%% file: main.tex
\documentclass[12pt]{article}

\usepackage{fullpage}

\usepackage{amsmath,amsfonts}

\usepackage{adjustbox}

\usepackage{tipa}

\usepackage{url}

\usepackage{hyperref}

\usepackage{microtype}

\usepackage{natbib}

\usepackage{tikz}
\usepackage{setspace}
\usepackage{authblk}

\usepackage{pslatex}

\usepackage{lineno}

\title{Complexity counts: global and local perspectives on Indo-Aryan numeral systems}

\author{Chundra A.\ Cathcart}

\affil{Institute for the Interdisciplinary Study of Language Evolution\\
University of Zurich}

\affil{{\tt chundra.cathcart@uzh.ch}}

\affil{ORCID: 0000-0002-3066-4532}

\date{}

\begin{document}


\maketitle



\input{abstract}


\onehalfspacing

\input{introduction}

\input{background}

\input{results}

\input{discussion}

\input{methods}
All code used to generate metrics, create graphics, and carry out statistical analyses can be found at \url{https://github.com/chundrac/IA-numeral-complexity}.

\input{acknowledgements}
This work was supported by the NCCR Evolving Language (SNSF Agreement No. 51NF40\underline{\phantom{X}}180888). 
Swiss National Science Foundation grant No.\ 207573 is gratefully acknowledged as well. 
I am grateful to audiences at the Universities of Konstanz and Zurich (in particular Christin Beck, Miriam Butt, and Balthasar Bickel) for comments on an early version of this work. I thank Carmen Saldana for detailed comments on an earlier manuscript version, along with two anonymous reviewers whose insights and clarifications --- in particular on the diachronic background and cultural context surrounding South Asian numeral systems, as well as the typology of numeral systems and morphological complexity --- greatly improved the paper. All errors and infelicities are my own responsibility.

\bibliographystyle{chicago}
\bibliography{bibliography}

\end{document}

%% file: abstract.tex
\begin{abstract}

\noindent 
The numeral systems of Indo-Aryan languages such as Hindi, Gujarati, and Bengali are highly unusual in that unlike most numeral systems (e.g., those of English, Chinese, etc.), forms referring to 1--99 are highly non-transparent and cannot be constructed using straightforward rules for forming combinations of tens and digits. 
Importantly, this non-transparency is not, as in some languages tolerating a mild degree of irregularity (e.g., Spanish {\it catorce} `14', {\it cuarenta} `40', which are not transparent combinations of {\it cuatro} `4' and {\it diez} `10'), limited to the teens and decade forms. 
As an example, Hindi/Urdu {\it iky\=anve} `91' is not decomposable into the composite elements {\it ek} `one' and {\it nave} `ninety' in the way that its English counterpart is. 
This paper further clarifies the position of Indo-Aryan languages within the typology of numeral systems, 
and explores the linguistic and non-linguistic factors that may be responsible for the persistence of complex systems in these languages. 
Using data from multiple databases, we develop and employ a number of cross-linguistically applicable metrics to quantify the complexity of languages' numeral systems, and demonstrate that Indo-Aryan languages have decisively more complex numeral systems than the world's languages as a whole, though individual Indo-Aryan languages differ from each other in terms of the complexity of the patterns they display. We investigate the factors (e.g., religion, geographic isolation, etc.) that underlie complexity in numeral systems, with a focus on South Asia, in an attempt to develop an account of why complex numeral systems developed and persisted in certain Indo-Aryan languages but not elsewhere. 
Finally, we demonstrate that Indo-Aryan numeral systems adhere to certain general pressures toward efficient communication found cross-linguistically, despite their high complexity. 
We call for this somewhat overlooked dimension of complexity to be taken seriously when discussing general variation in numeral systems. 

\end{abstract}

%% file: introduction.tex
\pagebreak

\section{Introduction}
A hallmark of human languages is the presence of numeral systems comprising terms used to refer to different amounts of elements.\footnote{We use the term numeral to refer to a word form referring to some numerosity, not a visual symbol.} 
Numeral systems of the world's languages exhibit considerable diversity across a number of dimensions, and much attention has been paid to their variation along a number of dimensions across languages. 
One such dimension concerns the number of quantities that a system can express: some languages lack exact numerals completely \citep{frank2008number}, while other languages have only a restricted set of exact numerals \citep{pica2004exact}; most languages of the world, however, possess more extensive exact recursive numeral systems with compositionality, combining a limited number of elements in order to generate expressions referring to an extensive range of cardinalities. 
Languages of this sort build their systems around different arithmetic bases, with decimal systems being most common but hardly universal \citep{wals-131,comrie2022arithmetic}. 
The syntax of numeral systems varies across languages as well, involving different orders of arithmetic bases with respect to modifiers (e.g., Danish {\it fem-og-fyrre} vs. Swedish {\it fyrtio-fem} `45'), and interacts with different parts of the linguistic system in different ways \citep{caludeverkerk2016,allassonniere2020numeral}. 



The vast majority of research on cross-linguistic numeral systems has tended to focus on {\it what} numeral systems express (in terms of exact versus approximate numerosities) as well as the structural building blocks and rules 
composing and governing systems. 
These properties of numeral systems are argued to be highly germane to our understanding of the evolution of numeric cognition and numeral systems in human history: for instance, the fact that many languages have restricted systems bolsters the idea that exact numeral systems are a relatively recent phenomenon, and that early humans may have lacked the ability to discriminate between precise quantities \citep{nunez2017there}. 

This paper focuses on another dimension according to which numeral systems can vary, which until recently \citep[cf.][]{denic2024recursive,rubehn2025annotating,koile2025decimal} has not received much attention compared to others: 
the morphophonological formal aspect of {\it how} numeral systems can express numerosities, which can range from the highly irregular and opaque (e.g., English {\it eleven}, which lacks a clear relationship to the arithmetic base and additive modifier from which it is composed) to the regular and highly transparent (e.g., Japanese {\it ni-jū-roku} `26', literally `2-10-6', with a canonical decade form that can be decomposed into base and multiplier). 
We approach the highly understudied phenomenon of irregular allomorphy within numeral systems from a global perspective 
that 
casts 
into relief 
the Indo-Aryan subgroup of the Indo-European language family. 
Situated within a family known for higher levels of this type of complexity in its numeral systems \citep{koile2025decimal}, Indo-Aryan numerals exhibit possibly the most striking examples of morphological irregularity and opacity, despite these languages not being noted for their morphological irregularity in other domains (systematic quantitative studies of the complexity of inflectional paradigms across languages, e.g., \citealt{cotterell2019complexity,wu2019morphological,guzman2024analogical}, tell us little about the status of Indo-Aryan, as they either exclude these languages altogether or use incomplete data from a handful of them). 
Using 
databases of numeral terms, we demonstrate quantitatively  that South Asia and specifically Indo-Aryan languages are a hotbed of formal complexity in numeral systems. 
Using qualitative as well as quantitative evidence, we explore the historical dynamics involved in development, maintenance, and loss of this complexity. 
Finally, we show that Indo-Aryan numeral systems fit into generalizations regarding communicative efficiency of numeral systems, albeit as extreme outliers with higher complexity. 
We discuss what predictions can be made about the future of such systems and outline a more general research program addressing the evolution 
of complexity within numeral systems.


%% file: background.tex
\section{Background}

Different Indo-Aryan languages exhibit different degrees of irregularity in their numeral systems. The numeral systems of a single pair of Indo-Aryan languages will serve to illustrate an extreme case of this variation. Table \ref{tab:sanskrit} gives numerals 1–99 in Classical Sanskrit, an Old Indo-Aryan language; while Sanskrit is not the direct ancestor of all Indo-Aryan languages (on this relationship see \citealt{Emeneau1966}; \citealt{Masica1991}; \citealt{Kogan2005}; \citealt{Smith2017}), the Old Indo-Aryan varieties from which contemporary Indo-Aryan languages descend likely had numerals virtually identical to those of Sanskrit. 
Generally speaking, most Sanskrit numerals are transparently segmentable into base and modifier elements. 
Some idiosyncrasies can be identified: for instance, the modifier {\it \d{s}a\d{s}} `6' exhibits several allomorphs stemming from the operation of multiple sound changes (for a detailed overview, see \citealt[I 83--88, 223--4]{lipp2009indogermanischen}). The long {\it \={a}} of {\it ek\=ada\'sa} `11' is unexpected and likely due to the influence of the word for `12' \citep[171]{emmerick1992}. 
The digits place in {\it dvya\'s\={\i}ti} `82' and {\it trya\'s\={\i}ti} `83' is represented by different allomorphs than found elsewhere in the system, which are listed as exceptions in traditional Sanskrit grammars \citep{panini}. 
However, most other instances of allomorphy are due to regular phonological alternations that are found in all relevant conditioning environments, even outside of the numeral system (so-called sandhi rules, e.g., {\it a} $+$ {\it a} $\rightarrow$ {\it \={a}} as in {\it nava} $+$ {\it a\'s\={\i}ti} $\rightarrow$ {\it nav\=a\'s\={\i}ti} `89').\footnote{The forms in Table \ref{tab:sanskrit} are the standard citation forms found in grammars, but by no means constitute an exhaustive list of the numeral constructions available in the language, which included a fairly flexible and productive range of subtractive constructions, e.g., {\it try-\=una-tri\d{m}\'sat} `thirty diminished by three = 27' (\citealt[69]{macdonell1927}; \citealt[194--5]{emmerick1992}), as well as multi-word expressions \citep{petrocchi2022morphosyntax}. Crucially, these alternative strategies would be subject to the same sandhi rules (and hence no less transparent than) the forms in Table \ref{tab:sanskrit}.}

\begin{table}[]
    \centering
\adjustbox{max width=.8\textwidth}{

{ 
\begin{tabular}{l p{.18\linewidth} p{.18\linewidth} p{.18\linewidth} p{.18\linewidth} p{.18\linewidth}}
\hline
  & 0     & 1      & 2      & 3      & 4     \\
\hline
0 & ---   & eka & dva & tri & catur\\
10 & da\'{s}a & ek\={a}da\'{s}a & dv\={a}da\'{s}a & trayoda\'{s}a & caturda\'{s}a\\
20 & vi\d{m}\'sati & ekavi\d{m}\'sati & dv\=avi\d{m}\'sati & trayovi\d{m}\'sati & caturvi\d{m}\'sati\\
30 & tri\d{m}\'sat & ekatri\d{m}\'sat & dv\=atri\d{m}\'sat & trayastri\d{m}\'sat & catustri\d{m}\'sat\\
40 & catv\={a}ri\d{m}\'sat & ekacatv\={a}ri\d{m}\'sat & dv\=acatv\={a}ri\d{m}\'sat & traya\'{s}catv\={a}ri\d{m}\'sat & catu\'scatv\={a}ri\d{m}\'sat\\
50 & pa\~{n}c\={a}\'sat & ekapa\~{n}c\={a}\'sat & dv\=apa\~{n}c\={a}\'sat & traya\d{h}pa\~{n}c\={a}\'sat & catu\d{s}pa\~{n}c\={a}\'sat\\
60 & \d{s}a\d{s}\d{t}i & eka\d{s}a\d{s}\d{t}i & dv\=a\d{s}a\d{s}\d{t}i & traya\d{h}\d{s}a\d{s}\d{t}i & catur\d{s}a\d{s}\d{t}i\\
70 & saptati & ekasaptati & dv\=asaptati & traya\d{h}saptati & catu\d{h}saptati\\
80 & a\'s\={\i}ti & ek\=a\'s\={\i}ti & dvya\'s\={\i}ti & trya\'s\={\i}ti & catura\'s\={\i}ti\\
90 & navati & ekanavati & dv\=anavati & trayonavati & caturnavati\\
\hline


\hline
 & 5      & 6      & 7     & 8      & 9     \\
\hline
0 & pa\~{n}ca & \d{s}a\d{s} & sapta & a\d{s}\d{t}\={a} & nava\\
10 & pa\~{n}cada\'{s}a & \d{s}o\d{d}a\'{s}a & saptada\'{s}a & a\d{s}\d{t}\={a}da\'{s}a & navada\'{s}a\\
20 & pa\~{n}cavi\d{m}\'sati & \d{s}a\d{d}vi\d{m}\'sati & saptavi\d{m}\'sati & a\d{s}\d{t}\=avi\d{m}\'sati & navavi\d{m}\'sati\\
30 & pa\~{n}catri\d{m}\'sat & \d{s}a\d{t}tri\d{m}\'sat & saptatri\d{m}\'sat & a\d{s}\d{t}\=atri\d{m}\'sat & navatri\d{m}\'sat\\
40 & pa\~{n}cacatv\={a}ri\d{m}\'sat & \d{s}a\d{t}catv\={a}ri\d{m}\'sat & saptacatv\={a}ri\d{m}\'sat & a\d{s}\d{t}\=acatv\={a}ri\d{m}\'sat & navacatv\={a}ri\d{m}\'sat\\
50 & pa\~{n}capa\~{n}c\={a}\'sat & \d{s}a\d{t}pa\~{n}c\={a}\'sat & saptapa\~{n}c\={a}\'sat & a\d{s}\d{t}\=apa\~{n}c\={a}\'sat & navapa\~{n}c\={a}\'sat\\
60 & pa\~{n}ca\d{s}a\d{s}\d{t}i & \d{s}a\d{t}\d{s}a\d{s}\d{t}i & sapta\d{s}a\d{s}\d{t}i & a\d{s}\d{t}\=a\d{s}a\d{s}\d{t}i & nava\d{s}a\d{s}\d{t}i\\
70 & pa\~{n}casaptati & \d{s}a\d{t}saptati & saptasaptati & a\d{s}\d{t}\=asaptati & navasaptati\\
80 & pa\~{n}c\=a\'s\={\i}ti & \d{s}a\d{d}a\'s\={\i}ti & sapt\=a\'s\={\i}ti & a\d{s}\d{t}\=a\'s\={\i}ti & nav\=a\'s\={\i}ti\\
90 & pa\~{n}canavati & \d{s}a\d{n}\d{n}avati & saptanavati & a\d{s}\d{t}\=anavati & navanavati\\

\hline
\end{tabular}
}

}
    \caption{Sanskrit numerals 1--99 \citep{monier1872sanskrit,emmerick1992}. Row numbers indicate the value of the tens place and column numbers indicate the value of the digits place of individual numerals. Some variant forms are excluded due to reasons of space.}
    \label{tab:sanskrit}
\end{table}

Table \ref{tab:hindi} shows the state of affairs several millennia later in Hindi/Urdu,\footnote{We employ the label Hindi/Urdu to reflect the fact that Hindi and Urdu, despite their different orthographies and high-register vocabularies, are mutually intelligible to the same degree as Bosnian-Croatian-Montenegrin-Serbian.} one of many modern speech varieties descended from an Old Indo-Aryan variety similar to Sanskrit. 
The relatively (albeit not entirely) transparent system of Sanskrit has evolved into one that displays extreme irregularity and unpredictability (as well as considerable dialectal and inter-speaker variation; \citealt[62]{macgregor1972}). 
No economical set of rules explains the allomorphy displayed by the system. 
The syntax of the numerals exhibits unusual features, such as subtractive forms with 9 in the digits place that share the same base as forms from the following decade (e.g., {\IPA Untis} `29', {\IPA tis} `30'; cf. Latin {\it duodeviginti} `18'). 
The base {\IPA p@cAs} `50' bears almost no resemblance to the forms 51--8 --- and this is hardly an exhaustive list of peculiarities found in these forms. 
Anyone who has studied Hindi/Urdu or a related language will be familiar with the difficulty of learning such a system. 
This makes it all the more astonishing that in principle, a significant proportion of the world's population uses systems like Hindi/Urdu's in day-to-day life. 
All the same, this phenomenon is largely neglected, and is only discussed in depth in a handful of works apart from descriptive grammars. 
Book-length studies of numerals, even those with a focus on South Asia, devote no space to this issue or mention it only in passing \citep{hurford1987language,everett2017numbers,mamta2024typology}. 

\begin{table}[]
    \centering
\adjustbox{max width=.8\textwidth}{

{\large 
\begin{tabular}{l p{.17\linewidth} p{.17\linewidth} p{.17\linewidth} p{.17\linewidth} p{.17\linewidth}}
\hline
  & 0     & 1      & 2      & 3      & 4     \\
\hline
0 & --- & {\IPA ek} & {\IPA do} & {\IPA tin} & {\IPA cAr}\\
10 & {\IPA d@s} & {\IPA gjAr@} & {\IPA bAr@} & {\IPA ter@} & {\IPA cOd@}\\
20 & {\IPA bis} & {\IPA Ikkis} & {\IPA bAis} & {\IPA teis} & {\IPA cObis}\\
30 & {\IPA tis} & {\IPA Ik@ttis} & {\IPA b@ttis} & {\IPA t\~{\ae}tis} & {\IPA c\~{O}tis}\\
40 & {\IPA cAlis} & {\IPA IktAlis} & {\IPA b@jAlis} & {\IPA t\~{\ae}tAlis} & {\IPA c@VAlis}\\
50 & {\IPA p@cAs} & {\IPA IkjAV@n} & {\IPA bAV@n} & {\IPA tIrp@n} & {\IPA c@uV@n}\\
60 & {\IPA sA\:t\textsuperscript{h}} & {\IPA Iks@\:t\textsuperscript{h}} & {\IPA bAs@\:t\textsuperscript{h}} & {\IPA tIrs@\:t\textsuperscript{h}} & {\IPA c\~{O}s@\:t\textsuperscript{h}}\\
70 & {\IPA s@tt@r} & {\IPA Ikh@tt@r} & {\IPA b@h@tt@r} & {\IPA tIh@tt@r} & {\IPA cOh@tt@r}\\
80 & {\IPA @ssi} & {\IPA IkjAsi} & {\IPA b@jAsi} & {\IPA tIrAsi} & {\IPA cOrAsi}\\
90 & {\IPA n@Ve} & {\IPA IkjAnVe} & {\IPA bAnVe} & {\IPA tIrAnVe} & {\IPA cOrAnVe}\\
\hline


\hline
 & 5      & 6      & 7     & 8      & 9     \\
\hline
0 & {\IPA p\~{A}c} & {\IPA c\textsuperscript{h}E} & {\IPA sAt} & {\IPA A\:t\textsuperscript{h}} & {\IPA nO}\\
10 & {\IPA p@ndr@} & {\IPA sol@} & {\IPA s@tr@} & {\IPA @\:t\textsuperscript{h}Ar@} & {\IPA Unnis}\\
20 & {\IPA p@ccis} & {\IPA c\textsuperscript{h}@bbis} & {\IPA s@ttAis} & {\IPA @\:t\:tAis} & {\IPA Untis}\\
30 & {\IPA p\~{\ae}tis} & {\IPA c\textsuperscript{h}@ttis} & {\IPA s\~{\ae}tis} & {\IPA @\:rtis} & {\IPA UntAlis}\\
40 & {\IPA p\~{\ae}tAlis} & {\IPA c\textsuperscript{h}IjAlis} & {\IPA s\~{\ae}tAlis} & {\IPA @\:rtAlis} & {\IPA UncAs}\\
50 & {\IPA p@cp@n} & {\IPA c\textsuperscript{h}@pp@n} & {\IPA s@ttAV@n} & {\IPA @\:t\:t\textsuperscript{h}AV@n} & {\IPA Uns@\:t\textsuperscript{h}}\\
60 & {\IPA p\~{\ae}s@\:t\textsuperscript{h}} & {\IPA c\textsuperscript{h}IjAs@\:t\textsuperscript{h}} & {\IPA s@rs@\:t\textsuperscript{h}} & {\IPA @\:rs@\:th} & {\IPA Unh@tt@r}\\
70 & {\IPA p@ch@tt@r} & {\IPA c\textsuperscript{h}Ih@tt@r} & {\IPA s@th@tt@r} & {\IPA @\:t\textsuperscript{h}h@tt@r} & {\IPA UnjAsi}\\
80 & {\IPA p@cAsi} & {\IPA c\textsuperscript{h}IjAsi} & {\IPA s@ttAsi} & {\IPA @\:t\:t\textsuperscript{h}Asi} & {\IPA n@VAsi}\\
90 & {\IPA p@cAnVe} & {\IPA c\textsuperscript{h}IjAnVe} & {\IPA s@ttAnVe} & {\IPA @\:t\:t\textsuperscript{h}AnVe} & {\IPA nInjAnVe}\\
\hline
\end{tabular}
}

}
    \caption{Hindi/Urdu numerals 1--99 \citep{macgregor1972}. 
    Row numbers indicate the value of the tens place and column numbers indicate the value of the digits place of individual numerals.}
    \label{tab:hindi}
\end{table}

\citet{Bright1969,bright1972} provides an in-depth analysis of the Hindi numeral system, enumerating the generalizations that capture allomorphic variation within it. 
He argues that while it is difficult to come up with an economic set of rules to derive the words for numerals from 1--99, implicit rules may be available to Hindi speakers that help them account for the allomorphy within this system. 
\citet{berger1992modern} outlines the historical development of each individual numeral from 1 to 99 in several modern Indo-Aryan languages; importantly, he shows that while irregularity in such systems is due partly to regular sound change, which is capable of creating unpredictable allomorphy, 
it is also due to analogical changes such as contamination in specific contexts, a finding confirmed in subsequent historical linguistic work \citep{andrijanic2024hindi}. 
\citet{cathcart2017decomposability} uses a Bayesian computational modeling to classify Hindi/Urdu numerals 1--99 according to the phonological cues they display, exploring the role of factors like cardinality and frequency on classification accuracy and posterior uncertainty. 
\citet{schneider2020children} present a study of the acquisition of numerals by children learning several languages, including Hindi and Gujarati. They demonstrate that while mastery of the successor function (the principle that all natural numbers have successors) is needed to acquire numerals, Hindi and Gujarati speaking children struggle more in mastering it. 
Beyond these works, there has been little attention to this issue, and no major attempts to integrate this dimension of complexity into the broader study of complexity of numeral systems and how it impacts acquisition and usage.

There are a number of possible reasons that the issue of allomorphic complexity of numeral systems is largely overlooked, in comparison to more privileged topics (such as what numeral systems can express along with the structural patterns they display). One possibility, as noted above, is that the latter topics may be seen as lending themselves to inferences regarding the evolution of numeric cognition within the history of our species at a deep time scale, as mentioned previously. 
On the other hand, allomorphic complexity of the type discussed above is almost exclusive to Indo-Aryan languages, and very recent within the chronology of language evolution --- a highly local, phylogenetically restricted phenomenon that does not apply to most cross-linguistic numeral systems.\footnote{A similar line of argumentation is employed by \citet{everett2020native}, who cautions against making generalizations about the influence of numerical cognition on morphosyntactic number marking that hinge on trial number, which is restricted to a small number of phylogenetic lineages within the Austronesian family.} 
At the same time, geographically and phylogenetically restricted phenomena of this sort can serve as extreme examples of the degree of complexity that language users are able to contend with and willing to tolerate. 

Another possibility is that mutual (un)interpredictability of forms in the presence of a high degree of allomorphy, known as integrative complexity (\citealt{AckermanMalouf2013}, as opposed to enumerative complexity, which concerns the size of an inventory within a system), is generally thought to be the preserve of the study of inflectional paradigms, particularly the question of how speakers predict previously unseen inflected forms by leveraging information from relationships between forms they have previously encountered. 
In contrast, speakers of a language most likely will have learned forms 1--99, so their ability to realize and discriminate between them efficiently may be seen as less interesting to the study of morphological irregularity (this is essentially the point made by \citealt[304]{hurford1987language}). 
But given the rise of phonological cue-based models of lexical learning and access (the linear discriminative learning framework), particularly those that radically do away with the classical notion of the morpheme and treat the generation and comprehension of simplex and complex words as governed by the same underlying mechanism of mapping between form and meaning \citep{baayen2018inflectional,baayen2019discriminative}, it strikes us as worthwhile to investigate the integrative complexity of numeral systems along these lines, particularly since such models appear to be able to capture properties of lexical access as the metrics that they produce correlate strongly with behavioral measures from psycholinguistic experiments. 


We use a variety of metrics derived from quantitative and computational models in order to operationalize complexity in numeral systems, allowing us to situate Indo-Aryan numeral systems in a global as well as a South Asian context. We additionally assess the extent to which computational models of word production and comprehension can account for Indo-Aryan numeral systems. 


%% file: results.tex
\section{Results}


We use several quantitative and computationally derived metrics to characterize the complexity at the language level as well as at the level of individual forms. 
These operationalizations of complexity are described in detail in Section \ref{methods}, and consist of the following:

\begin{itemize}
    \item Minimum description length (MDL), the shortest set of combinable elements needed to generate numeral forms 1--99 in a given language (yielding a positive integerial value), as inferred by a Bayesian segmentation algorithm
    \item N-gram (specifically trigram) surprisal in context, representing the average unpredictability of a sequence of phonological segments or graphemes in a (held-out) numeral form, given the other words for 1--99 in the same language (yielding a positive real value)
    \item Accuracy of production using a linear discriminative learning (LDL) model, the ease of generating a numeral form on the basis of some semantic input, given the associations between semantic and phonological cues found in other words for 1--99 in the same language (yielding a value between 0 and 1, computed via the normalized Levenshtein distance between the ground truth and predicted forms)
    \item Accuracy of comprehension using a linear discriminative learning (LDL; \citealt{baayen2018inflectional}) model, the ease of classifying a numeral form on the basis of its phonological/orthographic form, given the associations between semantic and phonological cues found in other words for 1--99 in the same language (yielding a binary 0/1 value, depending on whether a form was incorrectly or correctly classified)
\end{itemize}
Of these metrics, the latter three quantify the complexity of individual forms within a language's numeral system, but can be averaged across forms to provide a language-level measure of complexity. MDL applies only at the language level.

We compute these four metrics for languages in two data sets, the global UniNum dataset (\citealt{ritchie2019unified}, containing orthographic transcriptions) and the region-specific South Asia Numerals Dataset (SAND; \citealt{mamta2024typology}, containing phonemic transcriptions). These metrics are highly correlated. We carry out principal component analysis on aggregated language-level metrics, taking the first component (PC1), which explains 99\% of variance in the data, to represent a latent complexity variable manifested in these correlated variables, with positive values indicating higher complexity. 

\subsection{South Asia is a hotbed of integrative complexity in numeral systems}

\begin{figure}
    \centering

    \adjustbox{max width=\textwidth}{
    \input{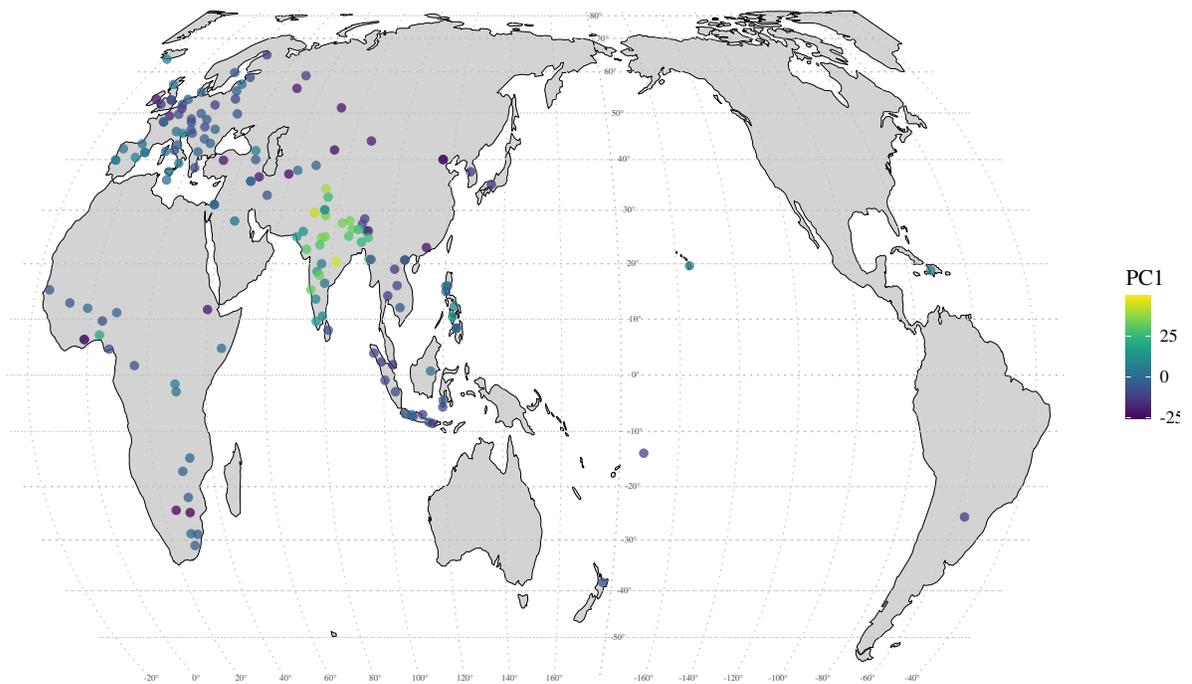}
    }

    \caption{Languages from the UniNum database. Colors represent the first component (PC1) of a principal components analysis of different variables used to operationalize the complexity of numeral systems at the language level (higher values represent higher complexity).}
    \label{fig:pc1.map}
\end{figure}

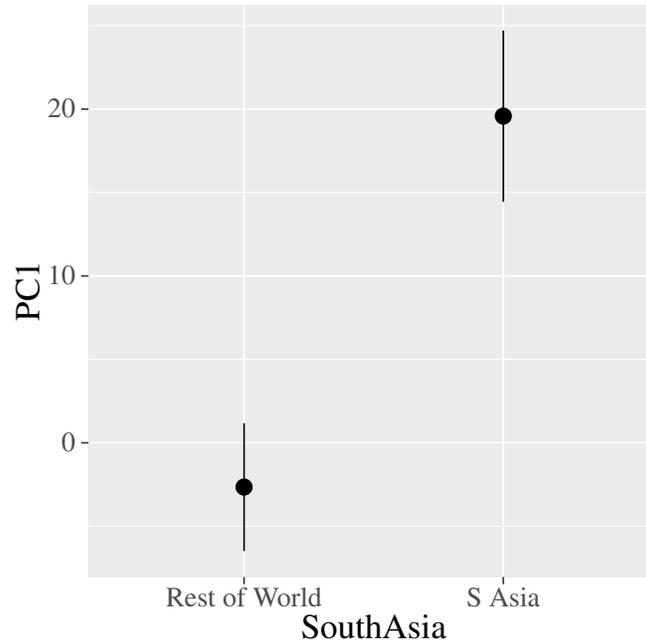
\begin{figure}
    \centering
    \adjustbox{max width=\textwidth}{
    \input{PC1-predictions-uninum}
    }

    \caption{Predicted PC1 (complexity) values from fitted hierarchical regression model for South Asian and non-South Asian languages in UniNum database.}
    \label{fig:marg.PC1}
\end{figure}

Figure \ref{fig:pc1.map} displays languages in the UniNum database \citep{ritchie2019unified} according to their geographic coordinates and their complexity levels (represented by PC1), with brighter colors indicating greater complexity. 
South Asia clearly emerges as a hotbed of high complexity in comparison to the rest of the world, as indicated by higher values of PC1. 
We fit a hierarchical regression model in which the response PC1 is a function of a predictor variable indicating whether a language is spoken in South Asia (comparison level) or not, with language family-level random intercepts. 
The prediction that South Asia has higher values of PC1 is confirmed; Figure \ref{fig:marg.PC1} shows predicted values for South Asian and non-South Asian languages.

\begin{figure}
    \centering

    \adjustbox{max width=.9\textwidth}{
    \input{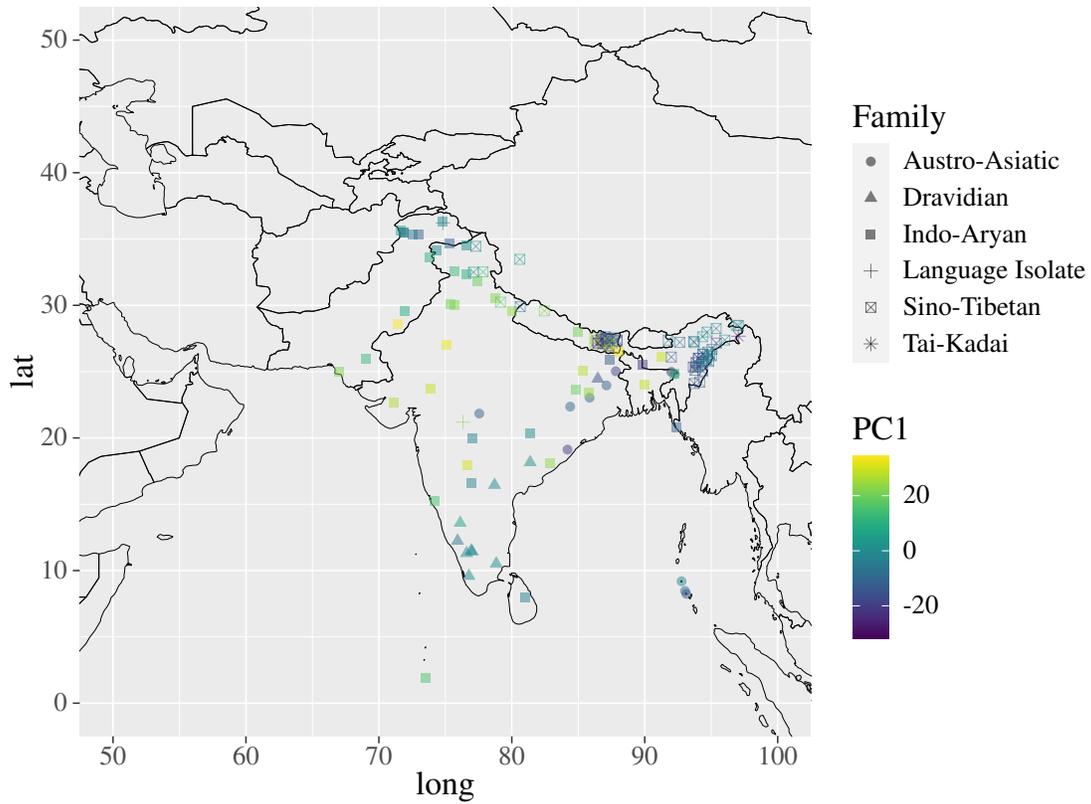}
    }
    
    \caption{Languages in the SAND database, plotted according to PC1 (color, representing complexity) and family (shape).}
    \label{fig:S-Asia-map}
\end{figure}

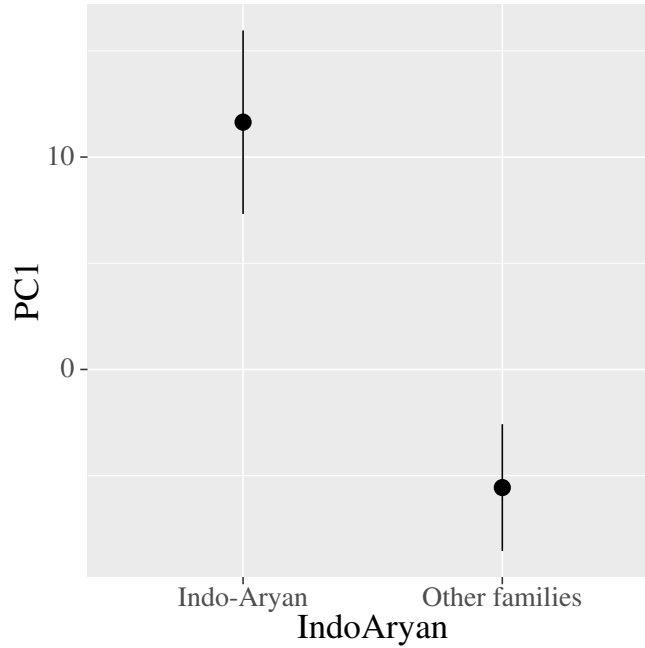
\begin{figure}
    \centering
    \adjustbox{max width=\textwidth}{
    \input{PC1-predictions-sand}
    }

    \caption{Predicted PC1 (complexity) values from fitted regression model for Indo-Aryan and non-Indo-Aryan languages in SAND database.}
    \label{fig:marg.SA}
\end{figure}

Through a closer inspection of the South Asian languages in the SAND database, we observe that this is a largely Indo-Aryan phenomenon, mostly absent from Dravidian and other families spoken in the region (Figure \ref{fig:S-Asia-map}). 
Non Indo-Aryan languages with higher complexity values have borrowed their numeral systems from Indo-Aryan languages. 
The considerably higher complexity level among Indo-Aryan versus other languages in South Asia is confirmed by a flat linear regression with family as a predictor and PC1 as a response; Figure \ref{fig:marg.SA} shows predicted values for Indo-Aryan and non-Indo-Aryan languages. 

This finding may not seem particularly striking to specialists who are familiar with Indo-Aryan numeral systems. Nonetheless, it serves as a methodologically rigorous confirmation of the intuition that South Asian (and specifically Indo-Aryan) numeral systems are more complex and irregular than those found elsewhere. 
Most Indo-Aryan languages spoken outside of South Asia, such as Romani \citep{matras2002romani} and Fiji Hindi \citep{moag1977fiji}, do not have systems of high complexity. 


\subsection{Peripheral Indo-Aryan languages with vigesimal systems lost or did not develop complexity, but other aspects of historical development remain unclear}


We take Indo-Aryan languages of South Asia as our focus and explore predictors of complexity within this group of languages, with an eye to making inferences about the history of numeral systems over the course of Indo-Aryan languages' descent and diversification from Old Indo-Aryan. 
As mentioned previously, there exist detailed studies of the sound and analogical changes affecting these systems \citep{berger1992modern}, but the problematic historical record --- not all numerals between 1 and 99 are attested in historical texts and most IA languages lack a continuous history of attestation --- has made it challenging to understand when exactly intense complexity took hold in these systems and under which extralinguistic circumstances. 
What we can infer from written records suggest that irregularity was not particularly extreme even in late Middle Indo-Aryan (ca.\ 1500 BP; \citealt{norman1992}), though this has changed by the early Modern period (ca.\ 1000 BP; \citealt{Baumann1975,strnad2013morphology}). 
\citet{andrijanic2024hindi} dates the advent of irregularity to after the Apabhra\d{m}\'sa stage (late 1st millennium CE). 
Still, some aspects of this dating are not totally straightforward, which is complicated particularly by details of the settlement of Sri Lanka and the Maldives. 
Insular Indo-Aryan spread to Sri Lanka prior in the latter part of the 1st millennium BCE, and further from Sri Lanka to the Maldives 
in the 1st millennium CE (\citealt[45]{Masica1991}; \citealt[18]{GnanadesikanDhivehi2017}). 
Sinhala, the main Indo-Aryan language of Sri Lanka (attested from 2200 BP onward), has a relatively transparent numeral system (possibly descending from a more irregular system), but Dhivehi, the language of the Maldives, has a system with irregularity comparable to mainland Indo-Aryan languages (along with multiple parallel systems: a largely obsolete, transparent duodecimal system and an innovative transparent decimal system; \citealt{Fritz2002,GnanadesikanDhivehi2017}). 
If the irregularity found in mainland Indo-Aryan languages is a fairly recent phenomenon, then the irregularity seen in Dhivehi (and of which older Sinhala varieties show traces; \citealt[244]{berger1992modern}; \citealt[119]{Geiger1938}) cannot be a shared genetic inheritance with these languages, since the insular Indo-Aryan languages were relatively isolated from the rest of Indo-Aryan during their history \citep{gair1982sinhala}. 
Additionally, while it is challenging to pinpoint divergent diachronic developments in the treatment of numeral terms that would be diagnostic of genetic sugrouping, Dhivehi differs from the majority of modern Indo-Aryan languages in its use of a subtractive form {\it onasatta} alongside a non-subtractive form {\it nav\=anavai} for 99 \citep[109]{GnanadesikanDhivehi2017}; however, this behavior is also found in geographic outliers like Nepali and Assamese (though apparently also in some dialects of Hindi; \citealt[274]{berger1992modern}), and is thus not a feature unique to the insular languages. 
Ultimately, many questions regarding the early development of Sinhala and Dhivehi are difficult to answer with precision; however, the complex decimal systems found in Dhivehi and older Sinhala appear to have included some inherited forms, but also borrowings from Middle Indo-Aryan as well as analogical and hybrid formations \citep[117--8]{Fritz2002}. 

Also unclear is why a high degree of irregularity was fostered in the Indo-Aryan languages, but not elsewhere. 
There is a tendency in linguistics research to ascribe all aberrant (from an Indo-European perspective) features found in Indo-Aryan to contact with languages from other families spoken in South Asia (e.g., Dravidian, Austroasiatic; see \citealt{Hock1993} for a criticism of this trend), but this is not possible here, for non-Indo-Aryan South Asian languages have much more transparent, regular numeral systems. 
Even more controversial are theories that relate some grammatical properties of Indo-Aryan languages to the rigid social hierarchy found for long periods of time in South Asia \citep{emeneau1974indian}; in line with this view, it is possible that such social structures are responsible for the maintenance of highly complex numeral systems, though there is no evidence for this idea and the underlying mechanisms are unclear. 
Other cultural (e.g., educational) institutions may be involved in the maintenance of complexity in the core area, which was lost in more peripheral areas. 
This behavior does not appear to be linked to any religious tradition, as it is found in languages with high proportions of Hindu and Muslim speakers (though not in the Dardic languages of the Northwest Indo-Aryan region, predominantly spoken by Muslims) and languages with a history of Buddhism (e.g., Dhivehi, prior to Islamization), as well as in documents associated with Jainism. 

\begin{figure}
    \centering

    \adjustbox{max width=.9\textwidth}{
    \input{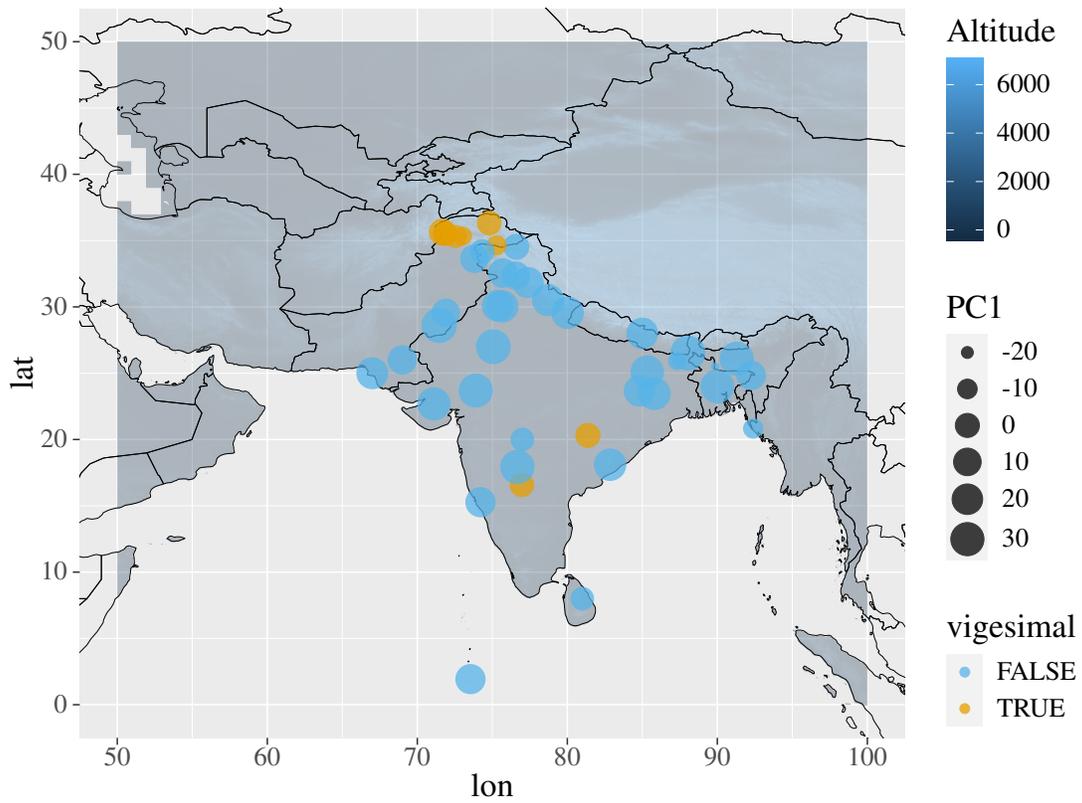}
    }
    
    \caption{Indo-Aryan languages in the SAND database, plotted according to PC1 (complexity, represented by size) and vigesimality (color) on an elevation map}
    \label{fig:IA-map}
\end{figure}

\begin{figure}
    \centering
    \begin{minipage}{.475\linewidth}
    \adjustbox{max width=\textwidth}{
    \input{elevation-plot}
    }
    \end{minipage}
     \hspace{.025\linewidth}
    \begin{minipage}{.475\linewidth}
    \adjustbox{max width=\textwidth}{
    \input{vigesimal-plot}
    }
    \end{minipage}
    
    \caption{PC1 (complexity) values of Indo-Aryan languages in the SAND database, plotted according to elevation values for coordinates associated with languages (left, with LOESS smooth) and vigesimality (right)}
    \label{fig:enter-label}
\end{figure}
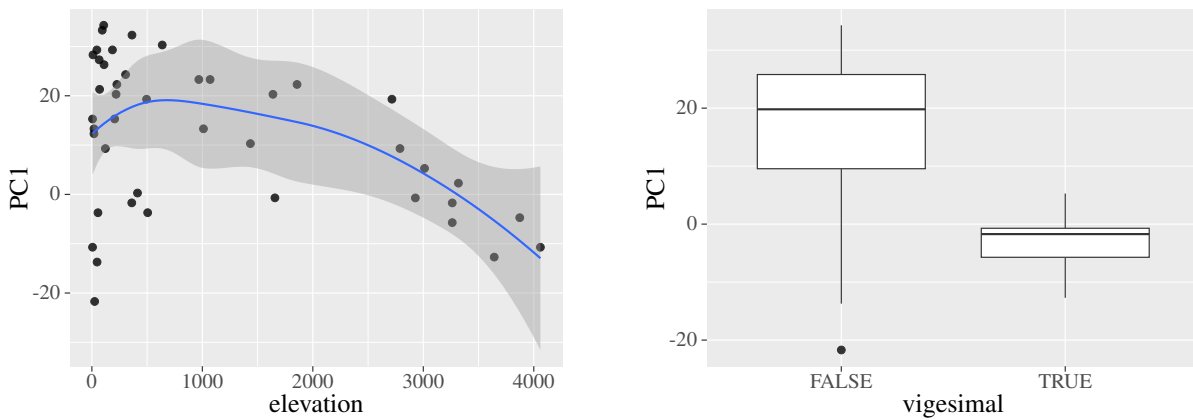

Observing the synchronic distribution of complexity across these languages' systems, some facts can be gleaned about the relationship between this variable and other covariates (Figure \ref{fig:IA-map}--\ref{fig:enter-label}). Complexity appears to have been preserved in the core Indo-Aryan area and lost in languages at higher altitudes 
(altitude can be construed in the context of South Asia as a proxy for social isolation from the main Indo-Aryan speech area), 
particularly those that developed vigesimality, a feature of the Hindu Kush linguistic area not originally present in Indo-Aryan languages \citep{weinreich2015not,liljegren2020hindu}. 
We fit three GAMs with PC1 as a response to (1) vigesimality, (2) a smooth function of elevation, and (3) both predictors. 
Analysis of deviance shows that a model containing both predictors shows improved fit over a model with only elevation as a predictor ($\chi^2(0.0001) = 548.52, p < 0.0001$), but not in comparison to a model with only vigesimality as a predictor ($\chi^2(1.1656) = 137.14, p = 0.4234$); 
this indicates that vigesimality is the key explanatory predictor of complexity among the variables we collected, though this could be an artifact of the low frequency and geographic concentration of systems with vigesimality. 

These findings allow us to sketch the following picture of Indo-Aryan numeral systems' development: complexity developed and was maintained in the core Indo-Aryan area, as well as in the geographically outlying insular Indo-Aryan languages. Complexity was either lost or prevented from developing in the higher-altitude languages of the Hindu Kush area, which shifted to a vigesimal system. 
The detailed role of contact (thought to affect properties of language involved in counting; e.g., numeral classifiers; \citealt{Grinevald2000,allassonniere2021expansion}) in fostering these patterns needs further clarification. Tentatively however, it seems that contact and social factors are ultimately responsible both for the maintenance of complexity in the core area as well as its loss in more peripheral areas. On one hand, the dense social networks found in the core Indo-Aryan area, largely free of internal natural barriers \citep[446]{Masica1991}, likely had a role in reinforcing this typologically unusual irregularity, as neighboring communities would require similar numeral forms in order to enable reliable communication (e.g., in the context of trade). Conversely, speech varieties outside of this network would be expected to lose, or never develop, this complexity, particularly those in pronounced contact with non-Indo-Aryan languages.

\subsection{Indo-Aryan systems are subject to some of the same principles of communicative efficiency found in other languages}

The above findings may be criticized as truistic on one hand and inconclusive on the other, in that many specialists are already aware of the relatively high irregularity of Indo-Aryan numeral systems and that while we have identified some covariates of complexity within South Asia, we have not reliably identified triggers that allowed this state of affairs to emerge and be preserved. 
Here, we highlight a striking finding that we believe is best arrived at using the quantitative metrics we employ --- that Indo-Aryan numeral systems, despite their complexity, follow principles of efficient communication related to those principles governing other systems.

Correlations between frequency and irregularity are well documented cross-linguistically; in noun and verb paradigms, more frequent terms tend to exhibit more unpredictable allomorphy and less frequent forms tend to be more regular and transparent, e.g., {\it go} $\sim$ {\it went} vs.\ {\it ambulate} $\sim$ {\it ambulated}. 
Distributional patterns of irregularity across paradigms are thought to facilitate efficient communication, not least because irregularity in more frequent paradigms makes individual inflected forms more discriminable \citep{Blevinsetal2017}. 
The same principle has been observed in numeral systems: English {\it twelve}, which is difficult to decompose into its component parts, is more frequent than transparent forms like {\it ninety-nine} \citep{Brysbaert2005}. 
If this is a universal principle underlying numeral systems, then it should apply in some manner to Indo-Aryan: specifically, we expect Indo-Aryan systems to exhibit higher predictability and lower complexity for items of higher cardinality, 
cardinality being shown to exhibit a negative correlation with frequency across a number of languages \citep{dehaene1992cross}. (Because we do not have reliable word frequencies for all languages in our sample, we use cardinality as a proxy for frequency in our main analyses, and confirm our findings for a handful of Indo-Aryan languages for which frequency data are available.)


We approach this question using 
the form-level 
metrics described above (n-gram surprisal, LDL production accuracy, LDL comprehension accuracy). 
We divide languages in the UniNum dataset into Indo-Aryan and non-Indo-Aryan groups, and fit generalized additive models (GAMs) with complexity metrics as responses which include by-group main effects and smooth terms representing the influence of cardinality on complexity, along with by-language random intercepts and slopes. We use a Gaussian log link function for surprisal, and a Binomial logit link function for LDL production accuracy and LDL comprehension accuracy. 

Figure \ref{fig:GAMs} shows marginal smooths for Indo-Aryan and non-Indo-Aryan languages from a hierarchical GAM relating different form-level complexity measure responses to the predictor cardinality. 
Smooth terms display some degree of non-monotonicity, which may be an artifact of the number of spline knots chosen (we used the default basis dimension of $k=10$); additionally, the slight rise in complexity after the cardinality value of $50$ may be partly related to the fact that for some IA languages, numerals 51--8 (e.g., Hindi/Urdu {\IPA IkjAv@n} `51') bear no formal resemblance to the numeral 50 (Hindi/Urdu {\IPA p@cAs}). All the same, for all smooths, elements with higher cardinality tend to have lower predicted complexity values than those of lower cardinality, as borne out by the marginal slopes for each model, which are negative for both groups for surprisal and production error and positive for comprehension accuracy (95\% confidence intervals for 
surprisal: IA $=[-0.01803, -0.01574]$, non-IA $=[-0.00923, -0.00828]$; 
production error: IA $=[-0.00583, -0.00362]$, non-IA $=[-0.00540, -0.00436]$; 
comprehension accuracy: IA $=[0.002502, 0.00518]$, non-IA $=[0.000561, 0.00102]$). 
Marginal effects of Indo-Aryan on complexity are consistently higher than for the non-Indo-Aryan group (95\% confidence intervals for 
surprisal: IA $=[1.929, 1.954]$, non-IA $=[0.856, 0.867]$; 
production error: IA $=[0.462, 0.497]$, non-IA $=[0.170, 0.181]$; 
comprehension accuracy: IA $=[0.521, 0.555]$, non-IA $=[0.919,0.927]$). 
Figure \ref{fig:emille} shows normalized frequency measures plotted against complexity metrics for selected Indo-Aryan languages for which reliable corpora were available. 
Strikingly, the non-monotonic trend seen in Figure \ref{fig:GAMs} around the mid-range of cardinalities is absent here or at least less pronounced, indicating that frequency is an even better predictor of complexity than cardinality; incidentally, 50 and other multiples of ten have higher frequency than would be predicted solely on the basis of their cardinality. 

What this shows is that while Indo-Aryan numeral systems exhibit higher overall complexity than non-Indo-Aryan numeral systems, complexity decays as cardinality increases (and frequency decreases), indicating that the distribution of irregularity and complexity in these systems is shaped by usage-based pressures in the same way that is observed in other numeral systems. 
It is worth noting that the situations found within versus outside of Indo-Aryan are in some ways incomparable: 
languages like English contain a small number of irregular forms of lower cardinality, with the remainder of forms being completely transparent and predictable. 
On the other hand, languages like Hindi/Urdu exhibit a less clearly discretizable gradient of irregularity, ranging from complex, unpredictable forms to even less predictable ones. 
Still, it is striking that Indo-Aryan systems display a gradient of complexity that tapers off as cardinality approaches 100. 
In the sense that they exhibit a more continuous cline (rather than a binary opposition between regular and irregular), 
Indo-Aryan systems are more probative with respect to questions regarding the relationship between frequency and irregularity within numeral systems, and may lend themselves to more fine-grained questions regarding this relationship. 
As an example of such a question, we know that both regular and irregular sound changes (viz., analogical changes) have affected Indo-Aryan numeral systems during their development, but to what extent did pressure to keep forms of higher cardinality more transparent play an active in the evolution of these systems?
Computational models of sound change capable of generating forms that are expected under regular sound change alone \citep[e.g.,][]{marr2023large} may help us pin down the extent to which the inverse correlation between cardinality and complexity found in these systems is due to neutral processes, or motivated patterns of change.



\begin{figure}
    \centering
    \begin{minipage}{.3\linewidth}
    \adjustbox{max width=\linewidth}{
    \input{surprisal-by-cardinality}
    }
    \centering (a)
    \end{minipage}
     \hspace{.0045\linewidth}
    \begin{minipage}{.3\linewidth}
    \adjustbox{max width=\linewidth}{
    \input{comprehension-by-cardinality}
    }
    \centering (b)
    \end{minipage}
     \hspace{.0045\linewidth}
    \begin{minipage}{.3\linewidth}
    \adjustbox{max width=\linewidth}{
    \input{production-by-cardinality}
    }
    \centering (c)
    \end{minipage}
    \caption{Marginal smooths for Indo-Aryan and non-Indo-Aryan languages from a hierarchical GAM relating 
    (a) surprisal,
    (b) comprehension accuracy,
    and 
    (c) production error rate 
    to the predictor cardinality. }
    \label{fig:GAMs}
\end{figure}
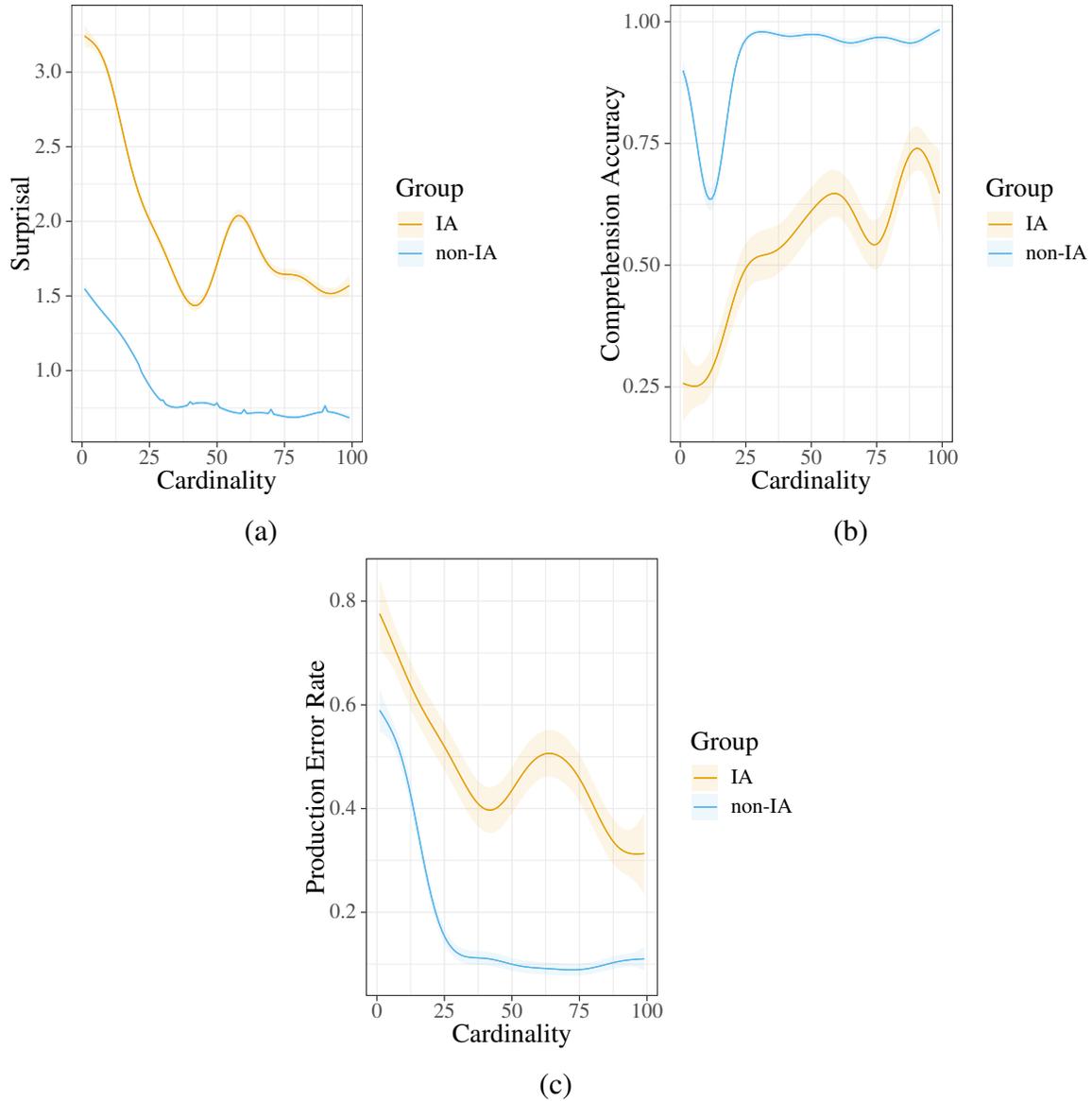



\begin{figure}
    \centering
    \begin{minipage}{.3\linewidth}
    \adjustbox{max width=\linewidth}{
    \input{surprisal-by-frequency}
    }
    \centering (a)
    \end{minipage}
     \hspace{.0045\linewidth}
    \begin{minipage}{.3\linewidth}
    \adjustbox{max width=\linewidth}{
    \input{accuracy-by-frequency}
    }
    \centering (b)
    \end{minipage}
     \hspace{.0045\linewidth}
    \begin{minipage}{.3\linewidth}
    \adjustbox{max width=\linewidth}{
    \input{production-by-frequency}
    }
    \centering (c)
    \end{minipage}
    \caption{Log normalized frequencies 
    from selected Indo-Aryan corpora 
    plotted against 
    (a) surprisal,
    (b) comprehension accuracy,
    and 
    (c) production error rate, with GAM smooths. }
    \label{fig:emille}
\end{figure}

%% file: PC1-predictions-uninum.tex
\begin{tikzpicture}[x=1pt,y=1pt]
\definecolor{fillColor}{RGB}{255,255,255}
\path[use as bounding box,fill=fillColor,fill opacity=0.00] (0,0) rectangle (252.94,252.94);
\begin{scope}
\path[clip] (  0.00,  0.00) rectangle (252.94,252.94);
\definecolor{drawColor}{RGB}{255,255,255}
\definecolor{fillColor}{RGB}{255,255,255}

\path[draw=drawColor,line width= 0.6pt,line join=round,line cap=round,fill=fillColor] (  0.00,  0.00) rectangle (252.94,252.94);
\end{scope}
\begin{scope}
\path[clip] ( 31.71, 30.69) rectangle (247.44,247.45);
\definecolor{fillColor}{gray}{0.92}

\path[fill=fillColor] ( 31.71, 30.69) rectangle (247.44,247.45);
\definecolor{drawColor}{RGB}{255,255,255}

\path[draw=drawColor,line width= 0.3pt,line join=round] ( 31.71, 49.89) --
	(247.44, 49.89);

\path[draw=drawColor,line width= 0.3pt,line join=round] ( 31.71,113.07) --
	(247.44,113.07);

\path[draw=drawColor,line width= 0.3pt,line join=round] ( 31.71,176.25) --
	(247.44,176.25);

\path[draw=drawColor,line width= 0.3pt,line join=round] ( 31.71,239.42) --
	(247.44,239.42);

\path[draw=drawColor,line width= 0.6pt,line join=round] ( 31.71, 81.48) --
	(247.44, 81.48);

\path[draw=drawColor,line width= 0.6pt,line join=round] ( 31.71,144.66) --
	(247.44,144.66);

\path[draw=drawColor,line width= 0.6pt,line join=round] ( 31.71,207.83) --
	(247.44,207.83);

\path[draw=drawColor,line width= 0.6pt,line join=round] ( 90.55, 30.69) --
	( 90.55,247.45);

\path[draw=drawColor,line width= 0.6pt,line join=round] (188.61, 30.69) --
	(188.61,247.45);
\definecolor{drawColor}{RGB}{0,0,0}

\path[draw=drawColor,line width= 0.6pt,line join=round] ( 90.55, 40.54) -- ( 90.55, 88.93);

\path[draw=drawColor,line width= 0.6pt,line join=round] (188.61,172.72) -- (188.61,237.59);
\definecolor{fillColor}{RGB}{0,0,0}

\path[draw=drawColor,line width= 0.8pt,line join=round,line cap=round,fill=fillColor] ( 90.55, 64.74) circle (  2.85);

\path[draw=drawColor,line width= 0.8pt,line join=round,line cap=round,fill=fillColor] (188.61,205.16) circle (  2.85);
\end{scope}
\begin{scope}
\path[clip] (  0.00,  0.00) rectangle (252.94,252.94);
\definecolor{drawColor}{gray}{0.30}

\node[text=drawColor,anchor=base east,inner sep=0pt, outer sep=0pt, scale=  0.88] at ( 26.76, 78.45) {0};

\node[text=drawColor,anchor=base east,inner sep=0pt, outer sep=0pt, scale=  0.88] at ( 26.76,141.63) {10};

\node[text=drawColor,anchor=base east,inner sep=0pt, outer sep=0pt, scale=  0.88] at ( 26.76,204.80) {20};
\end{scope}
\begin{scope}
\path[clip] (  0.00,  0.00) rectangle (252.94,252.94);
\definecolor{drawColor}{gray}{0.20}

\path[draw=drawColor,line width= 0.6pt,line join=round] ( 28.96, 81.48) --
	( 31.71, 81.48);

\path[draw=drawColor,line width= 0.6pt,line join=round] ( 28.96,144.66) --
	( 31.71,144.66);

\path[draw=drawColor,line width= 0.6pt,line join=round] ( 28.96,207.83) --
	( 31.71,207.83);
\end{scope}
\begin{scope}
\path[clip] (  0.00,  0.00) rectangle (252.94,252.94);
\definecolor{drawColor}{gray}{0.20}

\path[draw=drawColor,line width= 0.6pt,line join=round] ( 90.55, 27.94) --
	( 90.55, 30.69);

\path[draw=drawColor,line width= 0.6pt,line join=round] (188.61, 27.94) --
	(188.61, 30.69);
\end{scope}
\begin{scope}
\path[clip] (  0.00,  0.00) rectangle (252.94,252.94);
\definecolor{drawColor}{gray}{0.30}

\node[text=drawColor,anchor=base,inner sep=0pt, outer sep=0pt, scale=  0.88] at ( 90.55, 19.68) {Rest of World};

\node[text=drawColor,anchor=base,inner sep=0pt, outer sep=0pt, scale=  0.88] at (188.61, 19.68) {S Asia};
\end{scope}
\begin{scope}
\path[clip] (  0.00,  0.00) rectangle (252.94,252.94);
\definecolor{drawColor}{RGB}{0,0,0}

\node[text=drawColor,anchor=base,inner sep=0pt, outer sep=0pt, scale=  1.10] at (139.58,  7.64) {Region};
\end{scope}
\begin{scope}
\path[clip] (  0.00,  0.00) rectangle (252.94,252.94);
\definecolor{drawColor}{RGB}{0,0,0}

\node[text=drawColor,rotate= 90.00,anchor=base,inner sep=0pt, outer sep=0pt, scale=  1.10] at ( 13.08,139.07) {PC1};
\end{scope}
\end{tikzpicture}

%% file: PC1-predictions-sand.tex
\begin{tikzpicture}[x=1pt,y=1pt]
\definecolor{fillColor}{RGB}{255,255,255}
\path[use as bounding box,fill=fillColor,fill opacity=0.00] (0,0) rectangle (252.94,252.94);
\begin{scope}
\path[clip] (  0.00,  0.00) rectangle (252.94,252.94);
\definecolor{drawColor}{RGB}{255,255,255}
\definecolor{fillColor}{RGB}{255,255,255}

\path[draw=drawColor,line width= 0.6pt,line join=round,line cap=round,fill=fillColor] (  0.00,  0.00) rectangle (252.94,252.94);
\end{scope}
\begin{scope}
\path[clip] ( 31.71, 30.69) rectangle (247.44,247.45);
\definecolor{fillColor}{gray}{0.92}

\path[fill=fillColor] ( 31.71, 30.69) rectangle (247.44,247.45);
\definecolor{drawColor}{RGB}{255,255,255}

\path[draw=drawColor,line width= 0.3pt,line join=round] ( 31.71, 69.06) --
	(247.44, 69.06);

\path[draw=drawColor,line width= 0.3pt,line join=round] ( 31.71,149.46) --
	(247.44,149.46);

\path[draw=drawColor,line width= 0.3pt,line join=round] ( 31.71,229.86) --
	(247.44,229.86);

\path[draw=drawColor,line width= 0.6pt,line join=round] ( 31.71,109.26) --
	(247.44,109.26);

\path[draw=drawColor,line width= 0.6pt,line join=round] ( 31.71,189.66) --
	(247.44,189.66);

\path[draw=drawColor,line width= 0.6pt,line join=round] ( 90.55, 30.69) --
	( 90.55,247.45);

\path[draw=drawColor,line width= 0.6pt,line join=round] (188.61, 30.69) --
	(188.61,247.45);
\definecolor{drawColor}{RGB}{0,0,0}

\path[draw=drawColor,line width= 0.6pt,line join=round] (188.61, 40.54) -- (188.61, 88.54);

\path[draw=drawColor,line width= 0.6pt,line join=round] ( 90.55,168.14) -- ( 90.55,237.59);
\definecolor{fillColor}{RGB}{0,0,0}

\path[draw=drawColor,line width= 0.8pt,line join=round,line cap=round,fill=fillColor] (188.61, 64.54) circle (  2.85);

\path[draw=drawColor,line width= 0.8pt,line join=round,line cap=round,fill=fillColor] ( 90.55,202.87) circle (  2.85);
\end{scope}
\begin{scope}
\path[clip] (  0.00,  0.00) rectangle (252.94,252.94);
\definecolor{drawColor}{gray}{0.30}

\node[text=drawColor,anchor=base east,inner sep=0pt, outer sep=0pt, scale=  0.88] at ( 26.76,106.23) {0};

\node[text=drawColor,anchor=base east,inner sep=0pt, outer sep=0pt, scale=  0.88] at ( 26.76,186.63) {10};
\end{scope}
\begin{scope}
\path[clip] (  0.00,  0.00) rectangle (252.94,252.94);
\definecolor{drawColor}{gray}{0.20}

\path[draw=drawColor,line width= 0.6pt,line join=round] ( 28.96,109.26) --
	( 31.71,109.26);

\path[draw=drawColor,line width= 0.6pt,line join=round] ( 28.96,189.66) --
	( 31.71,189.66);
\end{scope}
\begin{scope}
\path[clip] (  0.00,  0.00) rectangle (252.94,252.94);
\definecolor{drawColor}{gray}{0.20}

\path[draw=drawColor,line width= 0.6pt,line join=round] ( 90.55, 27.94) --
	( 90.55, 30.69);

\path[draw=drawColor,line width= 0.6pt,line join=round] (188.61, 27.94) --
	(188.61, 30.69);
\end{scope}
\begin{scope}
\path[clip] (  0.00,  0.00) rectangle (252.94,252.94);
\definecolor{drawColor}{gray}{0.30}

\node[text=drawColor,anchor=base,inner sep=0pt, outer sep=0pt, scale=  0.88] at ( 90.55, 19.68) {Indo-Aryan};

\node[text=drawColor,anchor=base,inner sep=0pt, outer sep=0pt, scale=  0.88] at (188.61, 19.68) {Other families};
\end{scope}
\begin{scope}
\path[clip] (  0.00,  0.00) rectangle (252.94,252.94);
\definecolor{drawColor}{RGB}{0,0,0}

\node[text=drawColor,anchor=base,inner sep=0pt, outer sep=0pt, scale=  1.10] at (139.58,  7.64) {Family};
\end{scope}
\begin{scope}
\path[clip] (  0.00,  0.00) rectangle (252.94,252.94);
\definecolor{drawColor}{RGB}{0,0,0}

\node[text=drawColor,rotate= 90.00,anchor=base,inner sep=0pt, outer sep=0pt, scale=  1.10] at ( 13.08,139.07) {PC1};
\end{scope}
\end{tikzpicture}

%% file: elevation-plot.tex
\begin{tikzpicture}[x=1pt,y=1pt]
\definecolor{fillColor}{RGB}{255,255,255}
\path[use as bounding box,fill=fillColor,fill opacity=0.00] (0,0) rectangle (289.08,216.81);
\begin{scope}
\path[clip] (  0.00,  0.00) rectangle (289.08,216.81);
\definecolor{drawColor}{RGB}{255,255,255}
\definecolor{fillColor}{RGB}{255,255,255}

\path[draw=drawColor,line width= 0.6pt,line join=round,line cap=round,fill=fillColor] (  0.00,  0.00) rectangle (289.08,216.81);
\end{scope}
\begin{scope}
\path[clip] ( 34.64, 30.69) rectangle (283.58,211.31);
\definecolor{fillColor}{gray}{0.92}

\path[fill=fillColor] ( 34.64, 30.69) rectangle (283.58,211.31);
\definecolor{drawColor}{RGB}{255,255,255}

\path[draw=drawColor,line width= 0.3pt,line join=round] ( 34.64, 42.76) --
	(283.58, 42.76);

\path[draw=drawColor,line width= 0.3pt,line join=round] ( 34.64, 92.63) --
	(283.58, 92.63);

\path[draw=drawColor,line width= 0.3pt,line join=round] ( 34.64,142.51) --
	(283.58,142.51);

\path[draw=drawColor,line width= 0.3pt,line join=round] ( 34.64,192.39) --
	(283.58,192.39);

\path[draw=drawColor,line width= 0.3pt,line join=round] ( 73.54, 30.69) --
	( 73.54,211.31);

\path[draw=drawColor,line width= 0.3pt,line join=round] (129.36, 30.69) --
	(129.36,211.31);

\path[draw=drawColor,line width= 0.3pt,line join=round] (185.18, 30.69) --
	(185.18,211.31);

\path[draw=drawColor,line width= 0.3pt,line join=round] (241.00, 30.69) --
	(241.00,211.31);

\path[draw=drawColor,line width= 0.6pt,line join=round] ( 34.64, 67.70) --
	(283.58, 67.70);

\path[draw=drawColor,line width= 0.6pt,line join=round] ( 34.64,117.57) --
	(283.58,117.57);

\path[draw=drawColor,line width= 0.6pt,line join=round] ( 34.64,167.45) --
	(283.58,167.45);

\path[draw=drawColor,line width= 0.6pt,line join=round] ( 45.63, 30.69) --
	( 45.63,211.31);

\path[draw=drawColor,line width= 0.6pt,line join=round] (101.45, 30.69) --
	(101.45,211.31);

\path[draw=drawColor,line width= 0.6pt,line join=round] (157.27, 30.69) --
	(157.27,211.31);

\path[draw=drawColor,line width= 0.6pt,line join=round] (213.09, 30.69) --
	(213.09,211.31);

\path[draw=drawColor,line width= 0.6pt,line join=round] (268.92, 30.69) --
	(268.92,211.31);
\definecolor{drawColor}{RGB}{0,0,0}
\definecolor{fillColor}{RGB}{0,0,0}

\path[draw=drawColor,draw opacity=0.80,line width= 0.4pt,line join=round,line cap=round,fill=fillColor,fill opacity=0.80] ( 48.14,190.63) circle (  1.96);

\path[draw=drawColor,draw opacity=0.80,line width= 0.4pt,line join=round,line cap=round,fill=fillColor,fill opacity=0.80] ( 51.60,203.10) circle (  1.96);

\path[draw=drawColor,draw opacity=0.80,line width= 0.4pt,line join=round,line cap=round,fill=fillColor,fill opacity=0.80] ( 46.18,188.14) circle (  1.96);

\path[draw=drawColor,draw opacity=0.80,line width= 0.4pt,line join=round,line cap=round,fill=fillColor,fill opacity=0.80] (230.84,123.26) circle (  1.96);

\path[draw=drawColor,draw opacity=0.80,line width= 0.4pt,line join=round,line cap=round,fill=fillColor,fill opacity=0.80] ( 45.96,155.70) circle (  1.96);

\path[draw=drawColor,draw opacity=0.80,line width= 0.4pt,line join=round,line cap=round,fill=fillColor,fill opacity=0.80] (102.01,150.76) circle (  1.96);

\path[draw=drawColor,draw opacity=0.80,line width= 0.4pt,line join=round,line cap=round,fill=fillColor,fill opacity=0.80] (209.07,115.79) circle (  1.96);

\path[draw=drawColor,draw opacity=0.80,line width= 0.4pt,line join=round,line cap=round,fill=fillColor,fill opacity=0.80] (201.31,140.78) circle (  1.96);

\path[draw=drawColor,draw opacity=0.80,line width= 0.4pt,line join=round,line cap=round,fill=fillColor,fill opacity=0.80] (149.29,173.18) circle (  1.96);

\path[draw=drawColor,draw opacity=0.80,line width= 0.4pt,line join=round,line cap=round,fill=fillColor,fill opacity=0.80] ( 51.77,183.17) circle (  1.96);

\path[draw=drawColor,draw opacity=0.80,line width= 0.4pt,line join=round,line cap=round,fill=fillColor,fill opacity=0.80] (272.26, 90.85) circle (  1.96);

\path[draw=drawColor,draw opacity=0.80,line width= 0.4pt,line join=round,line cap=round,fill=fillColor,fill opacity=0.80] ( 47.02, 63.43) circle (  1.96);

\path[draw=drawColor,draw opacity=0.80,line width= 0.4pt,line join=round,line cap=round,fill=fillColor,fill opacity=0.80] ( 68.68,118.27) circle (  1.96);

\path[draw=drawColor,draw opacity=0.80,line width= 0.4pt,line join=round,line cap=round,fill=fillColor,fill opacity=0.80] ( 58.24,173.19) circle (  1.96);

\path[draw=drawColor,draw opacity=0.80,line width= 0.4pt,line join=round,line cap=round,fill=fillColor,fill opacity=0.80] (248.93, 85.87) circle (  1.96);

\path[draw=drawColor,draw opacity=0.80,line width= 0.4pt,line join=round,line cap=round,fill=fillColor,fill opacity=0.80] (213.71,130.73) circle (  1.96);

\path[draw=drawColor,draw opacity=0.80,line width= 0.4pt,line join=round,line cap=round,fill=fillColor,fill opacity=0.80] (138.12,115.83) circle (  1.96);

\path[draw=drawColor,draw opacity=0.80,line width= 0.4pt,line join=round,line cap=round,fill=fillColor,fill opacity=0.80] ( 49.31,185.65) circle (  1.96);

\path[draw=drawColor,draw opacity=0.80,line width= 0.4pt,line join=round,line cap=round,fill=fillColor,fill opacity=0.80] ( 57.12,155.73) circle (  1.96);

\path[draw=drawColor,draw opacity=0.80,line width= 0.4pt,line join=round,line cap=round,fill=fillColor,fill opacity=0.80] ( 48.25, 83.37) circle (  1.96);

\path[draw=drawColor,draw opacity=0.80,line width= 0.4pt,line join=round,line cap=round,fill=fillColor,fill opacity=0.80] (197.24,165.72) circle (  1.96);

\path[draw=drawColor,draw opacity=0.80,line width= 0.4pt,line join=round,line cap=round,fill=fillColor,fill opacity=0.80] ( 99.66,175.68) circle (  1.96);

\path[draw=drawColor,draw opacity=0.80,line width= 0.4pt,line join=round,line cap=round,fill=fillColor,fill opacity=0.80] ( 65.78,113.30) circle (  1.96);

\path[draw=drawColor,draw opacity=0.80,line width= 0.4pt,line join=round,line cap=round,fill=fillColor,fill opacity=0.80] ( 81.18,193.13) circle (  1.96);

\path[draw=drawColor,draw opacity=0.80,line width= 0.4pt,line join=round,line cap=round,fill=fillColor,fill opacity=0.80] ( 65.89,198.13) circle (  1.96);

\path[draw=drawColor,draw opacity=0.80,line width= 0.4pt,line join=round,line cap=round,fill=fillColor,fill opacity=0.80] ( 73.26,165.71) circle (  1.96);

\path[draw=drawColor,draw opacity=0.80,line width= 0.4pt,line join=round,line cap=round,fill=fillColor,fill opacity=0.80] (137.12,168.19) circle (  1.96);

\path[draw=drawColor,draw opacity=0.80,line width= 0.4pt,line join=round,line cap=round,fill=fillColor,fill opacity=0.80] (105.36,175.68) circle (  1.96);

\path[draw=drawColor,draw opacity=0.80,line width= 0.4pt,line join=round,line cap=round,fill=fillColor,fill opacity=0.80] (125.73,143.27) circle (  1.96);

\path[draw=drawColor,draw opacity=0.80,line width= 0.4pt,line join=round,line cap=round,fill=fillColor,fill opacity=0.80] ( 57.85,168.21) circle (  1.96);

\path[draw=drawColor,draw opacity=0.80,line width= 0.4pt,line join=round,line cap=round,fill=fillColor,fill opacity=0.80] (227.72,113.27) circle (  1.96);

\path[draw=drawColor,draw opacity=0.80,line width= 0.4pt,line join=round,line cap=round,fill=fillColor,fill opacity=0.80] (227.72,103.31) circle (  1.96);

\path[draw=drawColor,draw opacity=0.80,line width= 0.4pt,line join=round,line cap=round,fill=fillColor,fill opacity=0.80] ( 62.65,178.17) circle (  1.96);

\path[draw=drawColor,draw opacity=0.80,line width= 0.4pt,line join=round,line cap=round,fill=fillColor,fill opacity=0.80] ( 50.87,200.61) circle (  1.96);

\path[draw=drawColor,draw opacity=0.80,line width= 0.4pt,line join=round,line cap=round,fill=fillColor,fill opacity=0.80] ( 45.96, 90.86) circle (  1.96);

\path[draw=drawColor,draw opacity=0.80,line width= 0.4pt,line join=round,line cap=round,fill=fillColor,fill opacity=0.80] ( 73.82,108.33) circle (  1.96);

\path[draw=drawColor,draw opacity=0.80,line width= 0.4pt,line join=round,line cap=round,fill=fillColor,fill opacity=0.80] ( 46.69,148.24) circle (  1.96);

\path[draw=drawColor,draw opacity=0.80,line width= 0.4pt,line join=round,line cap=round,fill=fillColor,fill opacity=0.80] ( 48.70,108.31) circle (  1.96);

\path[draw=drawColor,draw opacity=0.80,line width= 0.4pt,line join=round,line cap=round,fill=fillColor,fill opacity=0.80] ( 52.44,140.76) circle (  1.96);

\path[draw=drawColor,draw opacity=0.80,line width= 0.4pt,line join=round,line cap=round,fill=fillColor,fill opacity=0.80] ( 46.69,150.74) circle (  1.96);

\path[draw=drawColor,draw opacity=0.80,line width= 0.4pt,line join=round,line cap=round,fill=fillColor,fill opacity=0.80] (261.83,105.80) circle (  1.96);

\path[draw=drawColor,draw opacity=0.80,line width= 0.4pt,line join=round,line cap=round,fill=fillColor,fill opacity=0.80] ( 49.53,170.70) circle (  1.96);

\path[draw=drawColor,draw opacity=0.80,line width= 0.4pt,line join=round,line cap=round,fill=fillColor,fill opacity=0.80] ( 56.01,190.65) circle (  1.96);
\definecolor{fillColor}{RGB}{153,153,153}

\path[fill=fillColor,fill opacity=0.40] ( 45.96,169.21) --
	( 48.82,167.74) --
	( 51.69,167.87) --
	( 54.55,169.69) --
	( 57.42,172.76) --
	( 60.28,176.26) --
	( 63.15,179.67) --
	( 66.01,182.84) --
	( 68.88,185.35) --
	( 71.74,187.00) --
	( 74.61,188.04) --
	( 77.47,188.74) --
	( 80.34,189.40) --
	( 83.20,190.19) --
	( 86.06,191.22) --
	( 88.93,192.44) --
	( 91.79,193.71) --
	( 94.66,194.85) --
	( 97.52,195.64) --
	(100.39,195.89) --
	(103.25,195.48) --
	(106.12,194.67) --
	(108.98,193.61) --
	(111.85,192.38) --
	(114.71,191.05) --
	(117.58,189.73) --
	(120.44,188.50) --
	(123.30,187.42) --
	(126.17,186.56) --
	(129.03,185.93) --
	(131.90,185.52) --
	(134.76,185.30) --
	(137.63,185.19) --
	(140.49,185.12) --
	(143.36,184.99) --
	(146.22,184.74) --
	(149.09,184.29) --
	(151.95,183.63) --
	(154.82,182.80) --
	(157.68,181.83) --
	(160.54,180.73) --
	(163.41,179.51) --
	(166.27,178.18) --
	(169.14,176.76) --
	(172.00,175.27) --
	(174.87,173.72) --
	(177.73,172.12) --
	(180.60,170.48) --
	(183.46,168.81) --
	(186.33,167.12) --
	(189.19,165.41) --
	(192.06,163.70) --
	(194.92,161.97) --
	(197.78,160.24) --
	(200.65,158.49) --
	(203.51,156.72) --
	(206.38,154.94) --
	(209.24,153.13) --
	(212.11,151.30) --
	(214.97,149.44) --
	(217.84,147.55) --
	(220.70,145.65) --
	(223.57,143.74) --
	(226.43,141.85) --
	(229.30,140.01) --
	(232.16,138.28) --
	(235.02,136.69) --
	(237.89,135.25) --
	(240.75,133.99) --
	(243.62,132.92) --
	(246.48,132.04) --
	(249.35,131.35) --
	(252.21,130.85) --
	(255.08,130.52) --
	(257.94,130.36) --
	(260.81,130.35) --
	(263.67,130.50) --
	(266.54,130.78) --
	(269.40,131.19) --
	(272.26,131.72) --
	(272.26, 38.90) --
	(269.40, 44.39) --
	(266.54, 49.69) --
	(263.67, 54.77) --
	(260.81, 59.63) --
	(257.94, 64.27) --
	(255.08, 68.66) --
	(252.21, 72.80) --
	(249.35, 76.68) --
	(246.48, 80.29) --
	(243.62, 83.64) --
	(240.75, 86.70) --
	(237.89, 89.50) --
	(235.02, 92.04) --
	(232.16, 94.33) --
	(229.30, 96.40) --
	(226.43, 98.29) --
	(223.57,100.04) --
	(220.70,101.70) --
	(217.84,103.30) --
	(214.97,104.83) --
	(212.11,106.31) --
	(209.24,107.73) --
	(206.38,109.10) --
	(203.51,110.42) --
	(200.65,111.67) --
	(197.78,112.84) --
	(194.92,113.95) --
	(192.06,114.98) --
	(189.19,115.92) --
	(186.33,116.79) --
	(183.46,117.58) --
	(180.60,118.29) --
	(177.73,118.94) --
	(174.87,119.53) --
	(172.00,120.07) --
	(169.14,120.57) --
	(166.27,121.05) --
	(163.41,121.51) --
	(160.54,121.98) --
	(157.68,122.46) --
	(154.82,122.96) --
	(151.95,123.50) --
	(149.09,124.10) --
	(146.22,124.86) --
	(143.36,125.81) --
	(140.49,126.89) --
	(137.63,128.01) --
	(134.76,129.08) --
	(131.90,130.03) --
	(129.03,130.78) --
	(126.17,131.29) --
	(123.30,131.55) --
	(120.44,131.58) --
	(117.58,131.43) --
	(114.71,131.17) --
	(111.85,130.88) --
	(108.98,130.65) --
	(106.12,130.57) --
	(103.25,130.71) --
	(100.39,131.21) --
	( 97.52,132.33) --
	( 94.66,133.93) --
	( 91.79,135.76) --
	( 88.93,137.58) --
	( 86.06,139.14) --
	( 83.20,140.27) --
	( 80.34,140.88) --
	( 77.47,141.03) --
	( 74.61,140.88) --
	( 71.74,140.64) --
	( 68.88,140.57) --
	( 66.01,140.88) --
	( 63.15,141.43) --
	( 60.28,141.90) --
	( 57.42,142.01) --
	( 54.55,141.24) --
	( 51.69,138.83) --
	( 48.82,134.25) --
	( 45.96,127.56) --
	cycle;

\path[] ( 45.96,169.21) --
	( 48.82,167.74) --
	( 51.69,167.87) --
	( 54.55,169.69) --
	( 57.42,172.76) --
	( 60.28,176.26) --
	( 63.15,179.67) --
	( 66.01,182.84) --
	( 68.88,185.35) --
	( 71.74,187.00) --
	( 74.61,188.04) --
	( 77.47,188.74) --
	( 80.34,189.40) --
	( 83.20,190.19) --
	( 86.06,191.22) --
	( 88.93,192.44) --
	( 91.79,193.71) --
	( 94.66,194.85) --
	( 97.52,195.64) --
	(100.39,195.89) --
	(103.25,195.48) --
	(106.12,194.67) --
	(108.98,193.61) --
	(111.85,192.38) --
	(114.71,191.05) --
	(117.58,189.73) --
	(120.44,188.50) --
	(123.30,187.42) --
	(126.17,186.56) --
	(129.03,185.93) --
	(131.90,185.52) --
	(134.76,185.30) --
	(137.63,185.19) --
	(140.49,185.12) --
	(143.36,184.99) --
	(146.22,184.74) --
	(149.09,184.29) --
	(151.95,183.63) --
	(154.82,182.80) --
	(157.68,181.83) --
	(160.54,180.73) --
	(163.41,179.51) --
	(166.27,178.18) --
	(169.14,176.76) --
	(172.00,175.27) --
	(174.87,173.72) --
	(177.73,172.12) --
	(180.60,170.48) --
	(183.46,168.81) --
	(186.33,167.12) --
	(189.19,165.41) --
	(192.06,163.70) --
	(194.92,161.97) --
	(197.78,160.24) --
	(200.65,158.49) --
	(203.51,156.72) --
	(206.38,154.94) --
	(209.24,153.13) --
	(212.11,151.30) --
	(214.97,149.44) --
	(217.84,147.55) --
	(220.70,145.65) --
	(223.57,143.74) --
	(226.43,141.85) --
	(229.30,140.01) --
	(232.16,138.28) --
	(235.02,136.69) --
	(237.89,135.25) --
	(240.75,133.99) --
	(243.62,132.92) --
	(246.48,132.04) --
	(249.35,131.35) --
	(252.21,130.85) --
	(255.08,130.52) --
	(257.94,130.36) --
	(260.81,130.35) --
	(263.67,130.50) --
	(266.54,130.78) --
	(269.40,131.19) --
	(272.26,131.72);

\path[] (272.26, 38.90) --
	(269.40, 44.39) --
	(266.54, 49.69) --
	(263.67, 54.77) --
	(260.81, 59.63) --
	(257.94, 64.27) --
	(255.08, 68.66) --
	(252.21, 72.80) --
	(249.35, 76.68) --
	(246.48, 80.29) --
	(243.62, 83.64) --
	(240.75, 86.70) --
	(237.89, 89.50) --
	(235.02, 92.04) --
	(232.16, 94.33) --
	(229.30, 96.40) --
	(226.43, 98.29) --
	(223.57,100.04) --
	(220.70,101.70) --
	(217.84,103.30) --
	(214.97,104.83) --
	(212.11,106.31) --
	(209.24,107.73) --
	(206.38,109.10) --
	(203.51,110.42) --
	(200.65,111.67) --
	(197.78,112.84) --
	(194.92,113.95) --
	(192.06,114.98) --
	(189.19,115.92) --
	(186.33,116.79) --
	(183.46,117.58) --
	(180.60,118.29) --
	(177.73,118.94) --
	(174.87,119.53) --
	(172.00,120.07) --
	(169.14,120.57) --
	(166.27,121.05) --
	(163.41,121.51) --
	(160.54,121.98) --
	(157.68,122.46) --
	(154.82,122.96) --
	(151.95,123.50) --
	(149.09,124.10) --
	(146.22,124.86) --
	(143.36,125.81) --
	(140.49,126.89) --
	(137.63,128.01) --
	(134.76,129.08) --
	(131.90,130.03) --
	(129.03,130.78) --
	(126.17,131.29) --
	(123.30,131.55) --
	(120.44,131.58) --
	(117.58,131.43) --
	(114.71,131.17) --
	(111.85,130.88) --
	(108.98,130.65) --
	(106.12,130.57) --
	(103.25,130.71) --
	(100.39,131.21) --
	( 97.52,132.33) --
	( 94.66,133.93) --
	( 91.79,135.76) --
	( 88.93,137.58) --
	( 86.06,139.14) --
	( 83.20,140.27) --
	( 80.34,140.88) --
	( 77.47,141.03) --
	( 74.61,140.88) --
	( 71.74,140.64) --
	( 68.88,140.57) --
	( 66.01,140.88) --
	( 63.15,141.43) --
	( 60.28,141.90) --
	( 57.42,142.01) --
	( 54.55,141.24) --
	( 51.69,138.83) --
	( 48.82,134.25) --
	( 45.96,127.56);
\definecolor{drawColor}{RGB}{51,102,255}

\path[draw=drawColor,line width= 1.1pt,line join=round] ( 45.96,148.39) --
	( 48.82,151.00) --
	( 51.69,153.35) --
	( 54.55,155.46) --
	( 57.42,157.38) --
	( 60.28,159.08) --
	( 63.15,160.55) --
	( 66.01,161.86) --
	( 68.88,162.96) --
	( 71.74,163.82) --
	( 74.61,164.46) --
	( 77.47,164.89) --
	( 80.34,165.14) --
	( 83.20,165.23) --
	( 86.06,165.18) --
	( 88.93,165.01) --
	( 91.79,164.74) --
	( 94.66,164.39) --
	( 97.52,163.99) --
	(100.39,163.55) --
	(103.25,163.09) --
	(106.12,162.62) --
	(108.98,162.13) --
	(111.85,161.63) --
	(114.71,161.11) --
	(117.58,160.58) --
	(120.44,160.04) --
	(123.30,159.49) --
	(126.17,158.93) --
	(129.03,158.35) --
	(131.90,157.78) --
	(134.76,157.19) --
	(137.63,156.60) --
	(140.49,156.00) --
	(143.36,155.40) --
	(146.22,154.80) --
	(149.09,154.19) --
	(151.95,153.56) --
	(154.82,152.88) --
	(157.68,152.14) --
	(160.54,151.35) --
	(163.41,150.51) --
	(166.27,149.61) --
	(169.14,148.67) --
	(172.00,147.67) --
	(174.87,146.62) --
	(177.73,145.53) --
	(180.60,144.38) --
	(183.46,143.19) --
	(186.33,141.95) --
	(189.19,140.67) --
	(192.06,139.34) --
	(194.92,137.96) --
	(197.78,136.54) --
	(200.65,135.08) --
	(203.51,133.57) --
	(206.38,132.02) --
	(209.24,130.43) --
	(212.11,128.80) --
	(214.97,127.13) --
	(217.84,125.42) --
	(220.70,123.67) --
	(223.57,121.89) --
	(226.43,120.07) --
	(229.30,118.21) --
	(232.16,116.30) --
	(235.02,114.36) --
	(237.89,112.38) --
	(240.75,110.35) --
	(243.62,108.28) --
	(246.48,106.17) --
	(249.35,104.02) --
	(252.21,101.82) --
	(255.08, 99.59) --
	(257.94, 97.31) --
	(260.81, 94.99) --
	(263.67, 92.63) --
	(266.54, 90.23) --
	(269.40, 87.79) --
	(272.26, 85.31);
\end{scope}
\begin{scope}
\path[clip] (  0.00,  0.00) rectangle (289.08,216.81);
\definecolor{drawColor}{gray}{0.30}

\node[text=drawColor,anchor=base east,inner sep=0pt, outer sep=0pt, scale=  0.88] at ( 29.69, 64.67) {-20};

\node[text=drawColor,anchor=base east,inner sep=0pt, outer sep=0pt, scale=  0.88] at ( 29.69,114.54) {0};

\node[text=drawColor,anchor=base east,inner sep=0pt, outer sep=0pt, scale=  0.88] at ( 29.69,164.42) {20};
\end{scope}
\begin{scope}
\path[clip] (  0.00,  0.00) rectangle (289.08,216.81);
\definecolor{drawColor}{gray}{0.20}

\path[draw=drawColor,line width= 0.6pt,line join=round] ( 31.89, 67.70) --
	( 34.64, 67.70);

\path[draw=drawColor,line width= 0.6pt,line join=round] ( 31.89,117.57) --
	( 34.64,117.57);

\path[draw=drawColor,line width= 0.6pt,line join=round] ( 31.89,167.45) --
	( 34.64,167.45);
\end{scope}
\begin{scope}
\path[clip] (  0.00,  0.00) rectangle (289.08,216.81);
\definecolor{drawColor}{gray}{0.20}

\path[draw=drawColor,line width= 0.6pt,line join=round] ( 45.63, 27.94) --
	( 45.63, 30.69);

\path[draw=drawColor,line width= 0.6pt,line join=round] (101.45, 27.94) --
	(101.45, 30.69);

\path[draw=drawColor,line width= 0.6pt,line join=round] (157.27, 27.94) --
	(157.27, 30.69);

\path[draw=drawColor,line width= 0.6pt,line join=round] (213.09, 27.94) --
	(213.09, 30.69);

\path[draw=drawColor,line width= 0.6pt,line join=round] (268.92, 27.94) --
	(268.92, 30.69);
\end{scope}
\begin{scope}
\path[clip] (  0.00,  0.00) rectangle (289.08,216.81);
\definecolor{drawColor}{gray}{0.30}

\node[text=drawColor,anchor=base,inner sep=0pt, outer sep=0pt, scale=  0.88] at ( 45.63, 19.68) {0};

\node[text=drawColor,anchor=base,inner sep=0pt, outer sep=0pt, scale=  0.88] at (101.45, 19.68) {1000};

\node[text=drawColor,anchor=base,inner sep=0pt, outer sep=0pt, scale=  0.88] at (157.27, 19.68) {2000};

\node[text=drawColor,anchor=base,inner sep=0pt, outer sep=0pt, scale=  0.88] at (213.09, 19.68) {3000};

\node[text=drawColor,anchor=base,inner sep=0pt, outer sep=0pt, scale=  0.88] at (268.92, 19.68) {4000};
\end{scope}
\begin{scope}
\path[clip] (  0.00,  0.00) rectangle (289.08,216.81);
\definecolor{drawColor}{RGB}{0,0,0}

\node[text=drawColor,anchor=base,inner sep=0pt, outer sep=0pt, scale=  1.10] at (159.11,  7.64) {elevation};
\end{scope}
\begin{scope}
\path[clip] (  0.00,  0.00) rectangle (289.08,216.81);
\definecolor{drawColor}{RGB}{0,0,0}

\node[text=drawColor,rotate= 90.00,anchor=base,inner sep=0pt, outer sep=0pt, scale=  1.10] at ( 13.08,121.00) {PC1};
\end{scope}
\end{tikzpicture}

%% file: vigesimal-plot.tex
\begin{tikzpicture}[x=1pt,y=1pt]
\definecolor{fillColor}{RGB}{255,255,255}
\path[use as bounding box,fill=fillColor,fill opacity=0.00] (0,0) rectangle (289.08,216.81);
\begin{scope}
\path[clip] (  0.00,  0.00) rectangle (289.08,216.81);
\definecolor{drawColor}{RGB}{255,255,255}
\definecolor{fillColor}{RGB}{255,255,255}

\path[draw=drawColor,line width= 0.6pt,line join=round,line cap=round,fill=fillColor] (  0.00,  0.00) rectangle (289.08,216.81);
\end{scope}
\begin{scope}
\path[clip] ( 34.64, 30.69) rectangle (283.58,211.31);
\definecolor{fillColor}{gray}{0.92}

\path[fill=fillColor] ( 34.64, 30.69) rectangle (283.58,211.31);
\definecolor{drawColor}{RGB}{255,255,255}

\path[draw=drawColor,line width= 0.3pt,line join=round] ( 34.64, 73.23) --
	(283.58, 73.23);

\path[draw=drawColor,line width= 0.3pt,line join=round] ( 34.64,131.87) --
	(283.58,131.87);

\path[draw=drawColor,line width= 0.3pt,line join=round] ( 34.64,190.51) --
	(283.58,190.51);

\path[draw=drawColor,line width= 0.6pt,line join=round] ( 34.64, 43.91) --
	(283.58, 43.91);

\path[draw=drawColor,line width= 0.6pt,line join=round] ( 34.64,102.55) --
	(283.58,102.55);

\path[draw=drawColor,line width= 0.6pt,line join=round] ( 34.64,161.19) --
	(283.58,161.19);

\path[draw=drawColor,line width= 0.6pt,line join=round] (102.54, 30.69) --
	(102.54,211.31);

\path[draw=drawColor,line width= 0.6pt,line join=round] (215.69, 30.69) --
	(215.69,211.31);
\definecolor{drawColor}{gray}{0.20}
\definecolor{fillColor}{gray}{0.20}

\path[draw=drawColor,line width= 0.4pt,line join=round,line cap=round,fill=fillColor] (102.54, 38.90) circle (  1.96);

\path[draw=drawColor,line width= 0.6pt,line join=round] (102.54,178.20) -- (102.54,203.10);

\path[draw=drawColor,line width= 0.6pt,line join=round] (102.54,130.56) -- (102.54, 62.34);
\definecolor{fillColor}{RGB}{255,255,255}

\path[draw=drawColor,line width= 0.6pt,fill=fillColor] ( 60.10,178.20) --
	( 60.10,130.56) --
	(144.97,130.56) --
	(144.97,178.20) --
	( 60.10,178.20) --
	cycle;

\path[draw=drawColor,line width= 1.1pt] ( 60.10,160.61) -- (144.97,160.61);

\path[draw=drawColor,line width= 0.6pt,line join=round] (215.69,100.45) -- (215.69,118.02);

\path[draw=drawColor,line width= 0.6pt,line join=round] (215.69, 85.78) -- (215.69, 65.28);

\path[draw=drawColor,line width= 0.6pt,fill=fillColor] (173.26,100.45) --
	(173.26, 85.78) --
	(258.12, 85.78) --
	(258.12,100.45) --
	(173.26,100.45) --
	cycle;

\path[draw=drawColor,line width= 1.1pt] (173.26, 97.50) -- (258.12, 97.50);
\end{scope}
\begin{scope}
\path[clip] (  0.00,  0.00) rectangle (289.08,216.81);
\definecolor{drawColor}{gray}{0.30}

\node[text=drawColor,anchor=base east,inner sep=0pt, outer sep=0pt, scale=  0.88] at ( 29.69, 40.88) {-20};

\node[text=drawColor,anchor=base east,inner sep=0pt, outer sep=0pt, scale=  0.88] at ( 29.69, 99.52) {0};

\node[text=drawColor,anchor=base east,inner sep=0pt, outer sep=0pt, scale=  0.88] at ( 29.69,158.16) {20};
\end{scope}
\begin{scope}
\path[clip] (  0.00,  0.00) rectangle (289.08,216.81);
\definecolor{drawColor}{gray}{0.20}

\path[draw=drawColor,line width= 0.6pt,line join=round] ( 31.89, 43.91) --
	( 34.64, 43.91);

\path[draw=drawColor,line width= 0.6pt,line join=round] ( 31.89,102.55) --
	( 34.64,102.55);

\path[draw=drawColor,line width= 0.6pt,line join=round] ( 31.89,161.19) --
	( 34.64,161.19);
\end{scope}
\begin{scope}
\path[clip] (  0.00,  0.00) rectangle (289.08,216.81);
\definecolor{drawColor}{gray}{0.20}

\path[draw=drawColor,line width= 0.6pt,line join=round] (102.54, 27.94) --
	(102.54, 30.69);

\path[draw=drawColor,line width= 0.6pt,line join=round] (215.69, 27.94) --
	(215.69, 30.69);
\end{scope}
\begin{scope}
\path[clip] (  0.00,  0.00) rectangle (289.08,216.81);
\definecolor{drawColor}{gray}{0.30}

\node[text=drawColor,anchor=base,inner sep=0pt, outer sep=0pt, scale=  0.88] at (102.54, 19.68) {FALSE};

\node[text=drawColor,anchor=base,inner sep=0pt, outer sep=0pt, scale=  0.88] at (215.69, 19.68) {TRUE};
\end{scope}
\begin{scope}
\path[clip] (  0.00,  0.00) rectangle (289.08,216.81);
\definecolor{drawColor}{RGB}{0,0,0}

\node[text=drawColor,anchor=base,inner sep=0pt, outer sep=0pt, scale=  1.10] at (159.11,  7.64) {vigesimal};
\end{scope}
\begin{scope}
\path[clip] (  0.00,  0.00) rectangle (289.08,216.81);
\definecolor{drawColor}{RGB}{0,0,0}

\node[text=drawColor,rotate= 90.00,anchor=base,inner sep=0pt, outer sep=0pt, scale=  1.10] at ( 13.08,121.00) {PC1};
\end{scope}
\end{tikzpicture}

%% file: surprisal-by-cardinality.tex
\begin{tikzpicture}[x=1pt,y=1pt]
\definecolor{fillColor}{RGB}{255,255,255}
\path[use as bounding box,fill=fillColor,fill opacity=0.00] (0,0) rectangle (252.94,252.94);
\begin{scope}
\path[clip] (  0.00,  0.00) rectangle (252.94,252.94);
\definecolor{drawColor}{RGB}{255,255,255}
\definecolor{fillColor}{RGB}{255,255,255}

\path[draw=drawColor,line width= 0.6pt,line join=round,line cap=round,fill=fillColor] (  0.00,  0.00) rectangle (252.94,252.94);
\end{scope}
\begin{scope}
\path[clip] ( 34.16, 30.69) rectangle (178.61,247.45);
\definecolor{fillColor}{RGB}{255,255,255}

\path[fill=fillColor] ( 34.16, 30.69) rectangle (178.61,247.45);
\definecolor{drawColor}{gray}{0.92}

\path[draw=drawColor,line width= 0.3pt,line join=round] ( 34.16, 47.58) --
	(178.61, 47.58);

\path[draw=drawColor,line width= 0.3pt,line join=round] ( 34.16, 84.62) --
	(178.61, 84.62);

\path[draw=drawColor,line width= 0.3pt,line join=round] ( 34.16,121.65) --
	(178.61,121.65);

\path[draw=drawColor,line width= 0.3pt,line join=round] ( 34.16,158.69) --
	(178.61,158.69);

\path[draw=drawColor,line width= 0.3pt,line join=round] ( 34.16,195.72) --
	(178.61,195.72);

\path[draw=drawColor,line width= 0.3pt,line join=round] ( 34.16,232.75) --
	(178.61,232.75);

\path[draw=drawColor,line width= 0.3pt,line join=round] ( 56.13, 30.69) --
	( 56.13,247.45);

\path[draw=drawColor,line width= 0.3pt,line join=round] ( 89.63, 30.69) --
	( 89.63,247.45);

\path[draw=drawColor,line width= 0.3pt,line join=round] (123.13, 30.69) --
	(123.13,247.45);

\path[draw=drawColor,line width= 0.3pt,line join=round] (156.63, 30.69) --
	(156.63,247.45);

\path[draw=drawColor,line width= 0.6pt,line join=round] ( 34.16, 66.10) --
	(178.61, 66.10);

\path[draw=drawColor,line width= 0.6pt,line join=round] ( 34.16,103.13) --
	(178.61,103.13);

\path[draw=drawColor,line width= 0.6pt,line join=round] ( 34.16,140.17) --
	(178.61,140.17);

\path[draw=drawColor,line width= 0.6pt,line join=round] ( 34.16,177.20) --
	(178.61,177.20);

\path[draw=drawColor,line width= 0.6pt,line join=round] ( 34.16,214.24) --
	(178.61,214.24);

\path[draw=drawColor,line width= 0.6pt,line join=round] ( 39.38, 30.69) --
	( 39.38,247.45);

\path[draw=drawColor,line width= 0.6pt,line join=round] ( 72.88, 30.69) --
	( 72.88,247.45);

\path[draw=drawColor,line width= 0.6pt,line join=round] (106.38, 30.69) --
	(106.38,247.45);

\path[draw=drawColor,line width= 0.6pt,line join=round] (139.88, 30.69) --
	(139.88,247.45);

\path[draw=drawColor,line width= 0.6pt,line join=round] (173.38, 30.69) --
	(173.38,247.45);
\definecolor{fillColor}{RGB}{230,159,0}

\path[fill=fillColor,fill opacity=0.10] ( 40.72,237.59) --
	( 42.06,235.76) --
	( 43.40,233.98) --
	( 44.74,232.22) --
	( 46.08,230.43) --
	( 47.42,228.48) --
	( 48.76,226.17) --
	( 50.10,223.34) --
	( 51.44,219.84) --
	( 52.78,215.65) --
	( 54.12,210.81) --
	( 55.46,205.43) --
	( 56.80,199.66) --
	( 58.14,193.70) --
	( 59.48,187.71) --
	( 60.82,181.86) --
	( 62.16,176.24) --
	( 63.50,170.94) --
	( 64.84,165.99) --
	( 66.18,161.41) --
	( 67.52,157.20) --
	( 68.86,153.35) --
	( 70.20,149.81) --
	( 71.54,146.56) --
	( 72.88,143.53) --
	( 74.22,140.66) --
	( 75.56,137.89) --
	( 76.90,135.13) --
	( 78.24,132.33) --
	( 79.58,129.43) --
	( 80.92,126.39) --
	( 82.26,123.23) --
	( 83.60,119.98) --
	( 84.94,116.70) --
	( 86.28,113.49) --
	( 87.62,110.45) --
	( 88.96,107.68) --
	( 90.30,105.28) --
	( 91.64,103.35) --
	( 92.98,101.95) --
	( 94.32,101.14) --
	( 95.66,100.95) --
	( 97.00,101.42) --
	( 98.34,102.56) --
	( 99.68,104.35) --
	(101.02,106.78) --
	(102.36,109.81) --
	(103.70,113.37) --
	(105.04,117.38) --
	(106.38,121.70) --
	(107.72,126.20) --
	(109.06,130.66) --
	(110.40,134.89) --
	(111.74,138.68) --
	(113.08,141.82) --
	(114.42,144.15) --
	(115.76,145.58) --
	(117.10,146.07) --
	(118.44,145.65) --
	(119.78,144.43) --
	(121.12,142.53) --
	(122.46,140.12) --
	(123.80,137.35) --
	(125.14,134.41) --
	(126.48,131.42) --
	(127.82,128.52) --
	(129.16,125.83) --
	(130.50,123.42) --
	(131.84,121.37) --
	(133.18,119.71) --
	(134.52,118.45) --
	(135.86,117.57) --
	(137.20,117.04) --
	(138.54,116.77) --
	(139.88,116.69) --
	(141.22,116.69) --
	(142.56,116.69) --
	(143.90,116.60) --
	(145.24,116.36) --
	(146.58,115.94) --
	(147.92,115.34) --
	(149.26,114.58) --
	(150.60,113.71) --
	(151.94,112.76) --
	(153.28,111.79) --
	(154.62,110.83) --
	(155.96,109.91) --
	(157.30,109.06) --
	(158.64,108.31) --
	(159.98,107.70) --
	(161.32,107.26) --
	(162.66,107.03) --
	(164.00,107.06) --
	(165.34,107.40) --
	(166.68,108.09) --
	(168.02,109.12) --
	(169.36,110.43) --
	(170.70,111.94) --
	(172.04,113.57) --
	(172.04,103.07) --
	(170.70,102.99) --
	(169.36,102.85) --
	(168.02,102.65) --
	(166.68,102.37) --
	(165.34,102.06) --
	(164.00,101.76) --
	(162.66,101.57) --
	(161.32,101.58) --
	(159.98,101.84) --
	(158.64,102.36) --
	(157.30,103.12) --
	(155.96,104.09) --
	(154.62,105.19) --
	(153.28,106.33) --
	(151.94,107.44) --
	(150.60,108.43) --
	(149.26,109.26) --
	(147.92,109.89) --
	(146.58,110.34) --
	(145.24,110.63) --
	(143.90,110.79) --
	(142.56,110.88) --
	(141.22,110.96) --
	(139.88,111.08) --
	(138.54,111.30) --
	(137.20,111.69) --
	(135.86,112.29) --
	(134.52,113.16) --
	(133.18,114.35) --
	(131.84,115.89) --
	(130.50,117.82) --
	(129.16,120.12) --
	(127.82,122.75) --
	(126.48,125.63) --
	(125.14,128.65) --
	(123.80,131.65) --
	(122.46,134.46) --
	(121.12,136.88) --
	(119.78,138.74) --
	(118.44,139.88) --
	(117.10,140.19) --
	(115.76,139.62) --
	(114.42,138.15) --
	(113.08,135.84) --
	(111.74,132.79) --
	(110.40,129.14) --
	(109.06,125.05) --
	(107.72,120.71) --
	(106.38,116.31) --
	(105.04,112.01) --
	(103.70,108.00) --
	(102.36,104.41) --
	(101.02,101.36) --
	( 99.68, 98.92) --
	( 98.34, 97.15) --
	( 97.00, 96.07) --
	( 95.66, 95.67) --
	( 94.32, 95.93) --
	( 92.98, 96.81) --
	( 91.64, 98.25) --
	( 90.30,100.17) --
	( 88.96,102.50) --
	( 87.62,105.16) --
	( 86.28,108.07) --
	( 84.94,111.14) --
	( 83.60,114.31) --
	( 82.26,117.51) --
	( 80.92,120.68) --
	( 79.58,123.78) --
	( 78.24,126.79) --
	( 76.90,129.69) --
	( 75.56,132.51) --
	( 74.22,135.30) --
	( 72.88,138.10) --
	( 71.54,141.02) --
	( 70.20,144.14) --
	( 68.86,147.55) --
	( 67.52,151.35) --
	( 66.18,155.56) --
	( 64.84,160.22) --
	( 63.50,165.31) --
	( 62.16,170.75) --
	( 60.82,176.46) --
	( 59.48,182.30) --
	( 58.14,188.15) --
	( 56.80,193.87) --
	( 55.46,199.36) --
	( 54.12,204.51) --
	( 52.78,209.25) --
	( 51.44,213.49) --
	( 50.10,217.19) --
	( 48.76,220.29) --
	( 47.42,222.75) --
	( 46.08,224.53) --
	( 44.74,225.67) --
	( 43.40,226.28) --
	( 42.06,226.54) --
	( 40.72,226.61) --
	cycle;

\path[] ( 40.72,237.59) --
	( 42.06,235.76) --
	( 43.40,233.98) --
	( 44.74,232.22) --
	( 46.08,230.43) --
	( 47.42,228.48) --
	( 48.76,226.17) --
	( 50.10,223.34) --
	( 51.44,219.84) --
	( 52.78,215.65) --
	( 54.12,210.81) --
	( 55.46,205.43) --
	( 56.80,199.66) --
	( 58.14,193.70) --
	( 59.48,187.71) --
	( 60.82,181.86) --
	( 62.16,176.24) --
	( 63.50,170.94) --
	( 64.84,165.99) --
	( 66.18,161.41) --
	( 67.52,157.20) --
	( 68.86,153.35) --
	( 70.20,149.81) --
	( 71.54,146.56) --
	( 72.88,143.53) --
	( 74.22,140.66) --
	( 75.56,137.89) --
	( 76.90,135.13) --
	( 78.24,132.33) --
	( 79.58,129.43) --
	( 80.92,126.39) --
	( 82.26,123.23) --
	( 83.60,119.98) --
	( 84.94,116.70) --
	( 86.28,113.49) --
	( 87.62,110.45) --
	( 88.96,107.68) --
	( 90.30,105.28) --
	( 91.64,103.35) --
	( 92.98,101.95) --
	( 94.32,101.14) --
	( 95.66,100.95) --
	( 97.00,101.42) --
	( 98.34,102.56) --
	( 99.68,104.35) --
	(101.02,106.78) --
	(102.36,109.81) --
	(103.70,113.37) --
	(105.04,117.38) --
	(106.38,121.70) --
	(107.72,126.20) --
	(109.06,130.66) --
	(110.40,134.89) --
	(111.74,138.68) --
	(113.08,141.82) --
	(114.42,144.15) --
	(115.76,145.58) --
	(117.10,146.07) --
	(118.44,145.65) --
	(119.78,144.43) --
	(121.12,142.53) --
	(122.46,140.12) --
	(123.80,137.35) --
	(125.14,134.41) --
	(126.48,131.42) --
	(127.82,128.52) --
	(129.16,125.83) --
	(130.50,123.42) --
	(131.84,121.37) --
	(133.18,119.71) --
	(134.52,118.45) --
	(135.86,117.57) --
	(137.20,117.04) --
	(138.54,116.77) --
	(139.88,116.69) --
	(141.22,116.69) --
	(142.56,116.69) --
	(143.90,116.60) --
	(145.24,116.36) --
	(146.58,115.94) --
	(147.92,115.34) --
	(149.26,114.58) --
	(150.60,113.71) --
	(151.94,112.76) --
	(153.28,111.79) --
	(154.62,110.83) --
	(155.96,109.91) --
	(157.30,109.06) --
	(158.64,108.31) --
	(159.98,107.70) --
	(161.32,107.26) --
	(162.66,107.03) --
	(164.00,107.06) --
	(165.34,107.40) --
	(166.68,108.09) --
	(168.02,109.12) --
	(169.36,110.43) --
	(170.70,111.94) --
	(172.04,113.57);

\path[] (172.04,103.07) --
	(170.70,102.99) --
	(169.36,102.85) --
	(168.02,102.65) --
	(166.68,102.37) --
	(165.34,102.06) --
	(164.00,101.76) --
	(162.66,101.57) --
	(161.32,101.58) --
	(159.98,101.84) --
	(158.64,102.36) --
	(157.30,103.12) --
	(155.96,104.09) --
	(154.62,105.19) --
	(153.28,106.33) --
	(151.94,107.44) --
	(150.60,108.43) --
	(149.26,109.26) --
	(147.92,109.89) --
	(146.58,110.34) --
	(145.24,110.63) --
	(143.90,110.79) --
	(142.56,110.88) --
	(141.22,110.96) --
	(139.88,111.08) --
	(138.54,111.30) --
	(137.20,111.69) --
	(135.86,112.29) --
	(134.52,113.16) --
	(133.18,114.35) --
	(131.84,115.89) --
	(130.50,117.82) --
	(129.16,120.12) --
	(127.82,122.75) --
	(126.48,125.63) --
	(125.14,128.65) --
	(123.80,131.65) --
	(122.46,134.46) --
	(121.12,136.88) --
	(119.78,138.74) --
	(118.44,139.88) --
	(117.10,140.19) --
	(115.76,139.62) --
	(114.42,138.15) --
	(113.08,135.84) --
	(111.74,132.79) --
	(110.40,129.14) --
	(109.06,125.05) --
	(107.72,120.71) --
	(106.38,116.31) --
	(105.04,112.01) --
	(103.70,108.00) --
	(102.36,104.41) --
	(101.02,101.36) --
	( 99.68, 98.92) --
	( 98.34, 97.15) --
	( 97.00, 96.07) --
	( 95.66, 95.67) --
	( 94.32, 95.93) --
	( 92.98, 96.81) --
	( 91.64, 98.25) --
	( 90.30,100.17) --
	( 88.96,102.50) --
	( 87.62,105.16) --
	( 86.28,108.07) --
	( 84.94,111.14) --
	( 83.60,114.31) --
	( 82.26,117.51) --
	( 80.92,120.68) --
	( 79.58,123.78) --
	( 78.24,126.79) --
	( 76.90,129.69) --
	( 75.56,132.51) --
	( 74.22,135.30) --
	( 72.88,138.10) --
	( 71.54,141.02) --
	( 70.20,144.14) --
	( 68.86,147.55) --
	( 67.52,151.35) --
	( 66.18,155.56) --
	( 64.84,160.22) --
	( 63.50,165.31) --
	( 62.16,170.75) --
	( 60.82,176.46) --
	( 59.48,182.30) --
	( 58.14,188.15) --
	( 56.80,193.87) --
	( 55.46,199.36) --
	( 54.12,204.51) --
	( 52.78,209.25) --
	( 51.44,213.49) --
	( 50.10,217.19) --
	( 48.76,220.29) --
	( 47.42,222.75) --
	( 46.08,224.53) --
	( 44.74,225.67) --
	( 43.40,226.28) --
	( 42.06,226.54) --
	( 40.72,226.61);
\definecolor{fillColor}{RGB}{86,180,233}

\path[fill=fillColor,fill opacity=0.10] ( 40.72,109.02) --
	( 42.06,106.78) --
	( 43.40,104.63) --
	( 44.74,102.61) --
	( 46.08,100.75) --
	( 47.42, 99.02) --
	( 48.76, 97.41) --
	( 50.10, 95.84) --
	( 51.44, 94.29) --
	( 52.78, 92.72) --
	( 54.12, 91.10) --
	( 55.46, 89.41) --
	( 56.80, 87.66) --
	( 58.14, 85.82) --
	( 59.48, 83.90) --
	( 60.82, 81.90) --
	( 62.16, 79.79) --
	( 63.50, 77.58) --
	( 64.84, 75.28) --
	( 66.18, 72.88) --
	( 67.52, 70.43) --
	( 68.86, 66.71) --
	( 70.20, 64.26) --
	( 71.54, 61.90) --
	( 72.88, 59.66) --
	( 74.22, 57.58) --
	( 75.56, 55.71) --
	( 76.90, 54.05) --
	( 78.24, 52.64) --
	( 79.58, 52.65) --
	( 80.92, 50.53) --
	( 82.26, 49.82) --
	( 83.60, 49.34) --
	( 84.94, 49.07) --
	( 86.28, 48.98) --
	( 87.62, 49.05) --
	( 88.96, 49.26) --
	( 90.30, 49.57) --
	( 91.64, 49.95) --
	( 92.98, 51.78) --
	( 94.32, 50.73) --
	( 95.66, 51.06) --
	( 97.00, 51.28) --
	( 98.34, 51.39) --
	( 99.68, 51.36) --
	(101.02, 51.20) --
	(102.36, 50.90) --
	(103.70, 50.50) --
	(105.04, 50.02) --
	(106.38, 51.20) --
	(107.72, 48.94) --
	(109.06, 48.39) --
	(110.40, 47.86) --
	(111.74, 47.38) --
	(113.08, 46.96) --
	(114.42, 46.61) --
	(115.76, 46.34) --
	(117.10, 46.14) --
	(118.44, 46.04) --
	(119.78, 47.90) --
	(121.12, 46.03) --
	(122.46, 46.11) --
	(123.80, 46.21) --
	(125.14, 46.31) --
	(126.48, 46.38) --
	(127.82, 46.40) --
	(129.16, 46.35) --
	(130.50, 46.24) --
	(131.84, 46.06) --
	(133.18, 48.05) --
	(134.52, 45.55) --
	(135.86, 45.26) --
	(137.20, 44.97) --
	(138.54, 44.71) --
	(139.88, 44.47) --
	(141.22, 44.29) --
	(142.56, 44.16) --
	(143.90, 44.09) --
	(145.24, 44.10) --
	(146.58, 44.18) --
	(147.92, 44.33) --
	(149.26, 44.56) --
	(150.60, 44.86) --
	(151.94, 45.22) --
	(153.28, 45.61) --
	(154.62, 46.00) --
	(155.96, 46.36) --
	(157.30, 46.66) --
	(158.64, 46.85) --
	(159.98, 49.97) --
	(161.32, 46.88) --
	(162.66, 46.72) --
	(164.00, 46.46) --
	(165.34, 46.16) --
	(166.68, 45.85) --
	(168.02, 45.55) --
	(169.36, 45.27) --
	(170.70, 45.03) --
	(172.04, 44.80) --
	(172.04, 40.54) --
	(170.70, 41.33) --
	(169.36, 42.08) --
	(168.02, 42.78) --
	(166.68, 43.38) --
	(165.34, 43.85) --
	(164.00, 44.17) --
	(162.66, 44.36) --
	(161.32, 44.44) --
	(159.98, 47.28) --
	(158.64, 44.32) --
	(157.30, 44.15) --
	(155.96, 43.92) --
	(154.62, 43.64) --
	(153.28, 43.33) --
	(151.94, 42.99) --
	(150.60, 42.66) --
	(149.26, 42.34) --
	(147.92, 42.06) --
	(146.58, 41.85) --
	(145.24, 41.73) --
	(143.90, 41.70) --
	(142.56, 41.76) --
	(141.22, 41.92) --
	(139.88, 42.14) --
	(138.54, 42.42) --
	(137.20, 42.72) --
	(135.86, 43.02) --
	(134.52, 43.30) --
	(133.18, 45.64) --
	(131.84, 43.73) --
	(130.50, 43.87) --
	(129.16, 43.95) --
	(127.82, 43.99) --
	(126.48, 43.99) --
	(125.14, 43.96) --
	(123.80, 43.90) --
	(122.46, 43.84) --
	(121.12, 43.79) --
	(119.78, 45.57) --
	(118.44, 43.78) --
	(117.10, 43.85) --
	(115.76, 44.01) --
	(114.42, 44.26) --
	(113.08, 44.59) --
	(111.74, 45.02) --
	(110.40, 45.52) --
	(109.06, 46.07) --
	(107.72, 46.64) --
	(106.38, 48.83) --
	(105.04, 47.74) --
	(103.70, 48.19) --
	(102.36, 48.56) --
	(101.02, 48.81) --
	( 99.68, 48.95) --
	( 98.34, 48.97) --
	( 97.00, 48.87) --
	( 95.66, 48.67) --
	( 94.32, 48.39) --
	( 92.98, 49.43) --
	( 91.64, 47.69) --
	( 90.30, 47.34) --
	( 88.96, 47.02) --
	( 87.62, 46.80) --
	( 86.28, 46.69) --
	( 84.94, 46.75) --
	( 83.60, 47.01) --
	( 82.26, 47.49) --
	( 80.92, 48.21) --
	( 79.58, 50.31) --
	( 78.24, 50.38) --
	( 76.90, 51.83) --
	( 75.56, 53.49) --
	( 74.22, 55.36) --
	( 72.88, 57.40) --
	( 71.54, 59.58) --
	( 70.20, 61.88) --
	( 68.86, 64.26) --
	( 67.52, 67.91) --
	( 66.18, 70.37) --
	( 64.84, 72.79) --
	( 63.50, 75.16) --
	( 62.16, 77.43) --
	( 60.82, 79.58) --
	( 59.48, 81.59) --
	( 58.14, 83.46) --
	( 56.80, 85.21) --
	( 55.46, 86.86) --
	( 54.12, 88.46) --
	( 52.78, 90.05) --
	( 51.44, 91.64) --
	( 50.10, 93.27) --
	( 48.76, 94.93) --
	( 47.42, 96.61) --
	( 46.08, 98.26) --
	( 44.74, 99.85) --
	( 43.40,101.39) --
	( 42.06,102.87) --
	( 40.72,104.33) --
	cycle;

\path[] ( 40.72,109.02) --
	( 42.06,106.78) --
	( 43.40,104.63) --
	( 44.74,102.61) --
	( 46.08,100.75) --
	( 47.42, 99.02) --
	( 48.76, 97.41) --
	( 50.10, 95.84) --
	( 51.44, 94.29) --
	( 52.78, 92.72) --
	( 54.12, 91.10) --
	( 55.46, 89.41) --
	( 56.80, 87.66) --
	( 58.14, 85.82) --
	( 59.48, 83.90) --
	( 60.82, 81.90) --
	( 62.16, 79.79) --
	( 63.50, 77.58) --
	( 64.84, 75.28) --
	( 66.18, 72.88) --
	( 67.52, 70.43) --
	( 68.86, 66.71) --
	( 70.20, 64.26) --
	( 71.54, 61.90) --
	( 72.88, 59.66) --
	( 74.22, 57.58) --
	( 75.56, 55.71) --
	( 76.90, 54.05) --
	( 78.24, 52.64) --
	( 79.58, 52.65) --
	( 80.92, 50.53) --
	( 82.26, 49.82) --
	( 83.60, 49.34) --
	( 84.94, 49.07) --
	( 86.28, 48.98) --
	( 87.62, 49.05) --
	( 88.96, 49.26) --
	( 90.30, 49.57) --
	( 91.64, 49.95) --
	( 92.98, 51.78) --
	( 94.32, 50.73) --
	( 95.66, 51.06) --
	( 97.00, 51.28) --
	( 98.34, 51.39) --
	( 99.68, 51.36) --
	(101.02, 51.20) --
	(102.36, 50.90) --
	(103.70, 50.50) --
	(105.04, 50.02) --
	(106.38, 51.20) --
	(107.72, 48.94) --
	(109.06, 48.39) --
	(110.40, 47.86) --
	(111.74, 47.38) --
	(113.08, 46.96) --
	(114.42, 46.61) --
	(115.76, 46.34) --
	(117.10, 46.14) --
	(118.44, 46.04) --
	(119.78, 47.90) --
	(121.12, 46.03) --
	(122.46, 46.11) --
	(123.80, 46.21) --
	(125.14, 46.31) --
	(126.48, 46.38) --
	(127.82, 46.40) --
	(129.16, 46.35) --
	(130.50, 46.24) --
	(131.84, 46.06) --
	(133.18, 48.05) --
	(134.52, 45.55) --
	(135.86, 45.26) --
	(137.20, 44.97) --
	(138.54, 44.71) --
	(139.88, 44.47) --
	(141.22, 44.29) --
	(142.56, 44.16) --
	(143.90, 44.09) --
	(145.24, 44.10) --
	(146.58, 44.18) --
	(147.92, 44.33) --
	(149.26, 44.56) --
	(150.60, 44.86) --
	(151.94, 45.22) --
	(153.28, 45.61) --
	(154.62, 46.00) --
	(155.96, 46.36) --
	(157.30, 46.66) --
	(158.64, 46.85) --
	(159.98, 49.97) --
	(161.32, 46.88) --
	(162.66, 46.72) --
	(164.00, 46.46) --
	(165.34, 46.16) --
	(166.68, 45.85) --
	(168.02, 45.55) --
	(169.36, 45.27) --
	(170.70, 45.03) --
	(172.04, 44.80);

\path[] (172.04, 40.54) --
	(170.70, 41.33) --
	(169.36, 42.08) --
	(168.02, 42.78) --
	(166.68, 43.38) --
	(165.34, 43.85) --
	(164.00, 44.17) --
	(162.66, 44.36) --
	(161.32, 44.44) --
	(159.98, 47.28) --
	(158.64, 44.32) --
	(157.30, 44.15) --
	(155.96, 43.92) --
	(154.62, 43.64) --
	(153.28, 43.33) --
	(151.94, 42.99) --
	(150.60, 42.66) --
	(149.26, 42.34) --
	(147.92, 42.06) --
	(146.58, 41.85) --
	(145.24, 41.73) --
	(143.90, 41.70) --
	(142.56, 41.76) --
	(141.22, 41.92) --
	(139.88, 42.14) --
	(138.54, 42.42) --
	(137.20, 42.72) --
	(135.86, 43.02) --
	(134.52, 43.30) --
	(133.18, 45.64) --
	(131.84, 43.73) --
	(130.50, 43.87) --
	(129.16, 43.95) --
	(127.82, 43.99) --
	(126.48, 43.99) --
	(125.14, 43.96) --
	(123.80, 43.90) --
	(122.46, 43.84) --
	(121.12, 43.79) --
	(119.78, 45.57) --
	(118.44, 43.78) --
	(117.10, 43.85) --
	(115.76, 44.01) --
	(114.42, 44.26) --
	(113.08, 44.59) --
	(111.74, 45.02) --
	(110.40, 45.52) --
	(109.06, 46.07) --
	(107.72, 46.64) --
	(106.38, 48.83) --
	(105.04, 47.74) --
	(103.70, 48.19) --
	(102.36, 48.56) --
	(101.02, 48.81) --
	( 99.68, 48.95) --
	( 98.34, 48.97) --
	( 97.00, 48.87) --
	( 95.66, 48.67) --
	( 94.32, 48.39) --
	( 92.98, 49.43) --
	( 91.64, 47.69) --
	( 90.30, 47.34) --
	( 88.96, 47.02) --
	( 87.62, 46.80) --
	( 86.28, 46.69) --
	( 84.94, 46.75) --
	( 83.60, 47.01) --
	( 82.26, 47.49) --
	( 80.92, 48.21) --
	( 79.58, 50.31) --
	( 78.24, 50.38) --
	( 76.90, 51.83) --
	( 75.56, 53.49) --
	( 74.22, 55.36) --
	( 72.88, 57.40) --
	( 71.54, 59.58) --
	( 70.20, 61.88) --
	( 68.86, 64.26) --
	( 67.52, 67.91) --
	( 66.18, 70.37) --
	( 64.84, 72.79) --
	( 63.50, 75.16) --
	( 62.16, 77.43) --
	( 60.82, 79.58) --
	( 59.48, 81.59) --
	( 58.14, 83.46) --
	( 56.80, 85.21) --
	( 55.46, 86.86) --
	( 54.12, 88.46) --
	( 52.78, 90.05) --
	( 51.44, 91.64) --
	( 50.10, 93.27) --
	( 48.76, 94.93) --
	( 47.42, 96.61) --
	( 46.08, 98.26) --
	( 44.74, 99.85) --
	( 43.40,101.39) --
	( 42.06,102.87) --
	( 40.72,104.33);
\definecolor{drawColor}{RGB}{230,159,0}

\path[draw=drawColor,line width= 0.6pt,line join=round] ( 40.72,232.10) --
	( 42.06,231.15) --
	( 43.40,230.13) --
	( 44.74,228.95) --
	( 46.08,227.48) --
	( 47.42,225.61) --
	( 48.76,223.23) --
	( 50.10,220.26) --
	( 51.44,216.67) --
	( 52.78,212.45) --
	( 54.12,207.66) --
	( 55.46,202.39) --
	( 56.80,196.77) --
	( 58.14,190.92) --
	( 59.48,185.01) --
	( 60.82,179.16) --
	( 62.16,173.50) --
	( 63.50,168.12) --
	( 64.84,163.11) --
	( 66.18,158.49) --
	( 67.52,154.27) --
	( 68.86,150.45) --
	( 70.20,146.97) --
	( 71.54,143.79) --
	( 72.88,140.81) --
	( 74.22,137.98) --
	( 75.56,135.20) --
	( 76.90,132.41) --
	( 78.24,129.56) --
	( 79.58,126.60) --
	( 80.92,123.53) --
	( 82.26,120.37) --
	( 83.60,117.14) --
	( 84.94,113.92) --
	( 86.28,110.78) --
	( 87.62,107.81) --
	( 88.96,105.09) --
	( 90.30,102.73) --
	( 91.64,100.80) --
	( 92.98, 99.38) --
	( 94.32, 98.54) --
	( 95.66, 98.31) --
	( 97.00, 98.75) --
	( 98.34, 99.85) --
	( 99.68,101.63) --
	(101.02,104.07) --
	(102.36,107.11) --
	(103.70,110.69) --
	(105.04,114.69) --
	(106.38,119.01) --
	(107.72,123.45) --
	(109.06,127.86) --
	(110.40,132.02) --
	(111.74,135.73) --
	(113.08,138.83) --
	(114.42,141.15) --
	(115.76,142.60) --
	(117.10,143.13) --
	(118.44,142.77) --
	(119.78,141.59) --
	(121.12,139.71) --
	(122.46,137.29) --
	(123.80,134.50) --
	(125.14,131.53) --
	(126.48,128.52) --
	(127.82,125.63) --
	(129.16,122.97) --
	(130.50,120.62) --
	(131.84,118.63) --
	(133.18,117.03) --
	(134.52,115.80) --
	(135.86,114.93) --
	(137.20,114.36) --
	(138.54,114.04) --
	(139.88,113.89) --
	(141.22,113.83) --
	(142.56,113.79) --
	(143.90,113.69) --
	(145.24,113.49) --
	(146.58,113.14) --
	(147.92,112.62) --
	(149.26,111.92) --
	(150.60,111.07) --
	(151.94,110.10) --
	(153.28,109.06) --
	(154.62,108.01) --
	(155.96,107.00) --
	(157.30,106.09) --
	(158.64,105.34) --
	(159.98,104.77) --
	(161.32,104.42) --
	(162.66,104.30) --
	(164.00,104.41) --
	(165.34,104.73) --
	(166.68,105.23) --
	(168.02,105.88) --
	(169.36,106.64) --
	(170.70,107.47) --
	(172.04,108.32);
\definecolor{drawColor}{RGB}{86,180,233}

\path[draw=drawColor,line width= 0.6pt,line join=round] ( 40.72,106.68) --
	( 42.06,104.83) --
	( 43.40,103.01) --
	( 44.74,101.23) --
	( 46.08, 99.50) --
	( 47.42, 97.82) --
	( 48.76, 96.17) --
	( 50.10, 94.56) --
	( 51.44, 92.97) --
	( 52.78, 91.38) --
	( 54.12, 89.78) --
	( 55.46, 88.14) --
	( 56.80, 86.43) --
	( 58.14, 84.64) --
	( 59.48, 82.75) --
	( 60.82, 80.74) --
	( 62.16, 78.61) --
	( 63.50, 76.37) --
	( 64.84, 74.03) --
	( 66.18, 71.62) --
	( 67.52, 69.17) --
	( 68.86, 65.48) --
	( 70.20, 63.07) --
	( 71.54, 60.74) --
	( 72.88, 58.53) --
	( 74.22, 56.47) --
	( 75.56, 54.60) --
	( 76.90, 52.94) --
	( 78.24, 51.51) --
	( 79.58, 51.48) --
	( 80.92, 49.37) --
	( 82.26, 48.66) --
	( 83.60, 48.18) --
	( 84.94, 47.91) --
	( 86.28, 47.83) --
	( 87.62, 47.92) --
	( 88.96, 48.14) --
	( 90.30, 48.45) --
	( 91.64, 48.82) --
	( 92.98, 50.61) --
	( 94.32, 49.56) --
	( 95.66, 49.86) --
	( 97.00, 50.08) --
	( 98.34, 50.18) --
	( 99.68, 50.16) --
	(101.02, 50.00) --
	(102.36, 49.73) --
	(103.70, 49.35) --
	(105.04, 48.88) --
	(106.38, 50.01) --
	(107.72, 47.79) --
	(109.06, 47.23) --
	(110.40, 46.69) --
	(111.74, 46.20) --
	(113.08, 45.78) --
	(114.42, 45.43) --
	(115.76, 45.17) --
	(117.10, 45.00) --
	(118.44, 44.91) --
	(119.78, 46.73) --
	(121.12, 44.91) --
	(122.46, 44.98) --
	(123.80, 45.06) --
	(125.14, 45.13) --
	(126.48, 45.18) --
	(127.82, 45.19) --
	(129.16, 45.15) --
	(130.50, 45.05) --
	(131.84, 44.89) --
	(133.18, 46.84) --
	(134.52, 44.43) --
	(135.86, 44.14) --
	(137.20, 43.85) --
	(138.54, 43.56) --
	(139.88, 43.31) --
	(141.22, 43.10) --
	(142.56, 42.96) --
	(143.90, 42.90) --
	(145.24, 42.91) --
	(146.58, 43.01) --
	(147.92, 43.20) --
	(149.26, 43.45) --
	(150.60, 43.76) --
	(151.94, 44.10) --
	(153.28, 44.47) --
	(154.62, 44.82) --
	(155.96, 45.14) --
	(157.30, 45.40) --
	(158.64, 45.59) --
	(159.98, 48.62) --
	(161.32, 45.66) --
	(162.66, 45.54) --
	(164.00, 45.32) --
	(165.34, 45.00) --
	(166.68, 44.61) --
	(168.02, 44.16) --
	(169.36, 43.68) --
	(170.70, 43.18) --
	(172.04, 42.67);
\definecolor{drawColor}{RGB}{230,159,0}

\path[draw=drawColor,line width= 0.6pt,line join=round] ( 40.72,232.10) --
	( 42.06,231.15) --
	( 43.40,230.13) --
	( 44.74,228.95) --
	( 46.08,227.48) --
	( 47.42,225.61) --
	( 48.76,223.23) --
	( 50.10,220.26) --
	( 51.44,216.67) --
	( 52.78,212.45) --
	( 54.12,207.66) --
	( 55.46,202.39) --
	( 56.80,196.77) --
	( 58.14,190.92) --
	( 59.48,185.01) --
	( 60.82,179.16) --
	( 62.16,173.50) --
	( 63.50,168.12) --
	( 64.84,163.11) --
	( 66.18,158.49) --
	( 67.52,154.27) --
	( 68.86,150.45) --
	( 70.20,146.97) --
	( 71.54,143.79) --
	( 72.88,140.81) --
	( 74.22,137.98) --
	( 75.56,135.20) --
	( 76.90,132.41) --
	( 78.24,129.56) --
	( 79.58,126.60) --
	( 80.92,123.53) --
	( 82.26,120.37) --
	( 83.60,117.14) --
	( 84.94,113.92) --
	( 86.28,110.78) --
	( 87.62,107.81) --
	( 88.96,105.09) --
	( 90.30,102.73) --
	( 91.64,100.80) --
	( 92.98, 99.38) --
	( 94.32, 98.54) --
	( 95.66, 98.31) --
	( 97.00, 98.75) --
	( 98.34, 99.85) --
	( 99.68,101.63) --
	(101.02,104.07) --
	(102.36,107.11) --
	(103.70,110.69) --
	(105.04,114.69) --
	(106.38,119.01) --
	(107.72,123.45) --
	(109.06,127.86) --
	(110.40,132.02) --
	(111.74,135.73) --
	(113.08,138.83) --
	(114.42,141.15) --
	(115.76,142.60) --
	(117.10,143.13) --
	(118.44,142.77) --
	(119.78,141.59) --
	(121.12,139.71) --
	(122.46,137.29) --
	(123.80,134.50) --
	(125.14,131.53) --
	(126.48,128.52) --
	(127.82,125.63) --
	(129.16,122.97) --
	(130.50,120.62) --
	(131.84,118.63) --
	(133.18,117.03) --
	(134.52,115.80) --
	(135.86,114.93) --
	(137.20,114.36) --
	(138.54,114.04) --
	(139.88,113.89) --
	(141.22,113.83) --
	(142.56,113.79) --
	(143.90,113.69) --
	(145.24,113.49) --
	(146.58,113.14) --
	(147.92,112.62) --
	(149.26,111.92) --
	(150.60,111.07) --
	(151.94,110.10) --
	(153.28,109.06) --
	(154.62,108.01) --
	(155.96,107.00) --
	(157.30,106.09) --
	(158.64,105.34) --
	(159.98,104.77) --
	(161.32,104.42) --
	(162.66,104.30) --
	(164.00,104.41) --
	(165.34,104.73) --
	(166.68,105.23) --
	(168.02,105.88) --
	(169.36,106.64) --
	(170.70,107.47) --
	(172.04,108.32);
\definecolor{drawColor}{RGB}{86,180,233}

\path[draw=drawColor,line width= 0.6pt,line join=round] ( 40.72,106.68) --
	( 42.06,104.83) --
	( 43.40,103.01) --
	( 44.74,101.23) --
	( 46.08, 99.50) --
	( 47.42, 97.82) --
	( 48.76, 96.17) --
	( 50.10, 94.56) --
	( 51.44, 92.97) --
	( 52.78, 91.38) --
	( 54.12, 89.78) --
	( 55.46, 88.14) --
	( 56.80, 86.43) --
	( 58.14, 84.64) --
	( 59.48, 82.75) --
	( 60.82, 80.74) --
	( 62.16, 78.61) --
	( 63.50, 76.37) --
	( 64.84, 74.03) --
	( 66.18, 71.62) --
	( 67.52, 69.17) --
	( 68.86, 65.48) --
	( 70.20, 63.07) --
	( 71.54, 60.74) --
	( 72.88, 58.53) --
	( 74.22, 56.47) --
	( 75.56, 54.60) --
	( 76.90, 52.94) --
	( 78.24, 51.51) --
	( 79.58, 51.48) --
	( 80.92, 49.37) --
	( 82.26, 48.66) --
	( 83.60, 48.18) --
	( 84.94, 47.91) --
	( 86.28, 47.83) --
	( 87.62, 47.92) --
	( 88.96, 48.14) --
	( 90.30, 48.45) --
	( 91.64, 48.82) --
	( 92.98, 50.61) --
	( 94.32, 49.56) --
	( 95.66, 49.86) --
	( 97.00, 50.08) --
	( 98.34, 50.18) --
	( 99.68, 50.16) --
	(101.02, 50.00) --
	(102.36, 49.73) --
	(103.70, 49.35) --
	(105.04, 48.88) --
	(106.38, 50.01) --
	(107.72, 47.79) --
	(109.06, 47.23) --
	(110.40, 46.69) --
	(111.74, 46.20) --
	(113.08, 45.78) --
	(114.42, 45.43) --
	(115.76, 45.17) --
	(117.10, 45.00) --
	(118.44, 44.91) --
	(119.78, 46.73) --
	(121.12, 44.91) --
	(122.46, 44.98) --
	(123.80, 45.06) --
	(125.14, 45.13) --
	(126.48, 45.18) --
	(127.82, 45.19) --
	(129.16, 45.15) --
	(130.50, 45.05) --
	(131.84, 44.89) --
	(133.18, 46.84) --
	(134.52, 44.43) --
	(135.86, 44.14) --
	(137.20, 43.85) --
	(138.54, 43.56) --
	(139.88, 43.31) --
	(141.22, 43.10) --
	(142.56, 42.96) --
	(143.90, 42.90) --
	(145.24, 42.91) --
	(146.58, 43.01) --
	(147.92, 43.20) --
	(149.26, 43.45) --
	(150.60, 43.76) --
	(151.94, 44.10) --
	(153.28, 44.47) --
	(154.62, 44.82) --
	(155.96, 45.14) --
	(157.30, 45.40) --
	(158.64, 45.59) --
	(159.98, 48.62) --
	(161.32, 45.66) --
	(162.66, 45.54) --
	(164.00, 45.32) --
	(165.34, 45.00) --
	(166.68, 44.61) --
	(168.02, 44.16) --
	(169.36, 43.68) --
	(170.70, 43.18) --
	(172.04, 42.67);
\definecolor{drawColor}{gray}{0.20}

\path[draw=drawColor,line width= 0.6pt,line join=round,line cap=round] ( 34.16, 30.69) rectangle (178.61,247.45);
\end{scope}
\begin{scope}
\path[clip] (  0.00,  0.00) rectangle (252.94,252.94);
\definecolor{drawColor}{gray}{0.30}

\node[text=drawColor,anchor=base east,inner sep=0pt, outer sep=0pt, scale=  0.88] at ( 29.21, 63.07) {1.0};

\node[text=drawColor,anchor=base east,inner sep=0pt, outer sep=0pt, scale=  0.88] at ( 29.21,100.10) {1.5};

\node[text=drawColor,anchor=base east,inner sep=0pt, outer sep=0pt, scale=  0.88] at ( 29.21,137.14) {2.0};

\node[text=drawColor,anchor=base east,inner sep=0pt, outer sep=0pt, scale=  0.88] at ( 29.21,174.17) {2.5};

\node[text=drawColor,anchor=base east,inner sep=0pt, outer sep=0pt, scale=  0.88] at ( 29.21,211.21) {3.0};
\end{scope}
\begin{scope}
\path[clip] (  0.00,  0.00) rectangle (252.94,252.94);
\definecolor{drawColor}{gray}{0.20}

\path[draw=drawColor,line width= 0.6pt,line join=round] ( 31.41, 66.10) --
	( 34.16, 66.10);

\path[draw=drawColor,line width= 0.6pt,line join=round] ( 31.41,103.13) --
	( 34.16,103.13);

\path[draw=drawColor,line width= 0.6pt,line join=round] ( 31.41,140.17) --
	( 34.16,140.17);

\path[draw=drawColor,line width= 0.6pt,line join=round] ( 31.41,177.20) --
	( 34.16,177.20);

\path[draw=drawColor,line width= 0.6pt,line join=round] ( 31.41,214.24) --
	( 34.16,214.24);
\end{scope}
\begin{scope}
\path[clip] (  0.00,  0.00) rectangle (252.94,252.94);
\definecolor{drawColor}{gray}{0.20}

\path[draw=drawColor,line width= 0.6pt,line join=round] ( 39.38, 27.94) --
	( 39.38, 30.69);

\path[draw=drawColor,line width= 0.6pt,line join=round] ( 72.88, 27.94) --
	( 72.88, 30.69);

\path[draw=drawColor,line width= 0.6pt,line join=round] (106.38, 27.94) --
	(106.38, 30.69);

\path[draw=drawColor,line width= 0.6pt,line join=round] (139.88, 27.94) --
	(139.88, 30.69);

\path[draw=drawColor,line width= 0.6pt,line join=round] (173.38, 27.94) --
	(173.38, 30.69);
\end{scope}
\begin{scope}
\path[clip] (  0.00,  0.00) rectangle (252.94,252.94);
\definecolor{drawColor}{gray}{0.30}

\node[text=drawColor,anchor=base,inner sep=0pt, outer sep=0pt, scale=  0.88] at ( 39.38, 19.68) {0};

\node[text=drawColor,anchor=base,inner sep=0pt, outer sep=0pt, scale=  0.88] at ( 72.88, 19.68) {25};

\node[text=drawColor,anchor=base,inner sep=0pt, outer sep=0pt, scale=  0.88] at (106.38, 19.68) {50};

\node[text=drawColor,anchor=base,inner sep=0pt, outer sep=0pt, scale=  0.88] at (139.88, 19.68) {75};

\node[text=drawColor,anchor=base,inner sep=0pt, outer sep=0pt, scale=  0.88] at (173.38, 19.68) {100};
\end{scope}
\begin{scope}
\path[clip] (  0.00,  0.00) rectangle (252.94,252.94);
\definecolor{drawColor}{RGB}{0,0,0}

\node[text=drawColor,anchor=base,inner sep=0pt, outer sep=0pt, scale=  1.10] at (106.38,  7.64) {Cardinality};
\end{scope}
\begin{scope}
\path[clip] (  0.00,  0.00) rectangle (252.94,252.94);
\definecolor{drawColor}{RGB}{0,0,0}

\node[text=drawColor,rotate= 90.00,anchor=base,inner sep=0pt, outer sep=0pt, scale=  1.10] at ( 13.08,139.07) {Surprisal};
\end{scope}
\begin{scope}
\path[clip] (  0.00,  0.00) rectangle (252.94,252.94);
\definecolor{fillColor}{RGB}{255,255,255}

\path[fill=fillColor] (189.61,111.50) rectangle (247.44,166.63);
\end{scope}
\begin{scope}
\path[clip] (  0.00,  0.00) rectangle (252.94,252.94);
\definecolor{drawColor}{RGB}{0,0,0}

\node[text=drawColor,anchor=base west,inner sep=0pt, outer sep=0pt, scale=  1.10] at (195.11,152.48) {Group};
\end{scope}
\begin{scope}
\path[clip] (  0.00,  0.00) rectangle (252.94,252.94);
\definecolor{fillColor}{RGB}{255,255,255}

\path[fill=fillColor] (195.11,131.46) rectangle (209.56,145.91);
\end{scope}
\begin{scope}
\path[clip] (  0.00,  0.00) rectangle (252.94,252.94);
\definecolor{fillColor}{RGB}{230,159,0}

\path[fill=fillColor,fill opacity=0.10] (195.82,132.17) rectangle (208.85,145.20);
\end{scope}
\begin{scope}
\path[clip] (  0.00,  0.00) rectangle (252.94,252.94);
\definecolor{drawColor}{RGB}{230,159,0}

\path[draw=drawColor,line width= 0.6pt,line join=round] (196.55,138.69) -- (208.12,138.69);
\end{scope}
\begin{scope}
\path[clip] (  0.00,  0.00) rectangle (252.94,252.94);
\definecolor{drawColor}{RGB}{230,159,0}

\path[draw=drawColor,line width= 0.6pt,line join=round] (196.55,138.69) -- (208.12,138.69);
\end{scope}
\begin{scope}
\path[clip] (  0.00,  0.00) rectangle (252.94,252.94);
\definecolor{fillColor}{RGB}{255,255,255}

\path[fill=fillColor] (195.11,117.00) rectangle (209.56,131.46);
\end{scope}
\begin{scope}
\path[clip] (  0.00,  0.00) rectangle (252.94,252.94);
\definecolor{fillColor}{RGB}{86,180,233}

\path[fill=fillColor,fill opacity=0.10] (195.82,117.72) rectangle (208.85,130.75);
\end{scope}
\begin{scope}
\path[clip] (  0.00,  0.00) rectangle (252.94,252.94);
\definecolor{drawColor}{RGB}{86,180,233}

\path[draw=drawColor,line width= 0.6pt,line join=round] (196.55,124.23) -- (208.12,124.23);
\end{scope}
\begin{scope}
\path[clip] (  0.00,  0.00) rectangle (252.94,252.94);
\definecolor{drawColor}{RGB}{86,180,233}

\path[draw=drawColor,line width= 0.6pt,line join=round] (196.55,124.23) -- (208.12,124.23);
\end{scope}
\begin{scope}
\path[clip] (  0.00,  0.00) rectangle (252.94,252.94);
\definecolor{drawColor}{RGB}{0,0,0}

\node[text=drawColor,anchor=base west,inner sep=0pt, outer sep=0pt, scale=  0.88] at (215.06,135.65) {IA};
\end{scope}
\begin{scope}
\path[clip] (  0.00,  0.00) rectangle (252.94,252.94);
\definecolor{drawColor}{RGB}{0,0,0}

\node[text=drawColor,anchor=base west,inner sep=0pt, outer sep=0pt, scale=  0.88] at (215.06,121.20) {non-IA};
\end{scope}
\end{tikzpicture}

%% file: comprehension-by-cardinality.tex
\begin{tikzpicture}[x=1pt,y=1pt]
\definecolor{fillColor}{RGB}{255,255,255}
\path[use as bounding box,fill=fillColor,fill opacity=0.00] (0,0) rectangle (252.94,252.94);
\begin{scope}
\path[clip] (  0.00,  0.00) rectangle (252.94,252.94);
\definecolor{drawColor}{RGB}{255,255,255}
\definecolor{fillColor}{RGB}{255,255,255}

\path[draw=drawColor,line width= 0.6pt,line join=round,line cap=round,fill=fillColor] (  0.00,  0.00) rectangle (252.94,252.94);
\end{scope}
\begin{scope}
\path[clip] ( 38.56, 30.69) rectangle (178.61,247.45);
\definecolor{fillColor}{RGB}{255,255,255}

\path[fill=fillColor] ( 38.56, 30.69) rectangle (178.61,247.45);
\definecolor{drawColor}{gray}{0.92}

\path[draw=drawColor,line width= 0.3pt,line join=round] ( 38.56, 88.20) --
	(178.61, 88.20);

\path[draw=drawColor,line width= 0.3pt,line join=round] ( 38.56,148.64) --
	(178.61,148.64);

\path[draw=drawColor,line width= 0.3pt,line join=round] ( 38.56,209.09) --
	(178.61,209.09);

\path[draw=drawColor,line width= 0.3pt,line join=round] ( 59.86, 30.69) --
	( 59.86,247.45);

\path[draw=drawColor,line width= 0.3pt,line join=round] ( 92.34, 30.69) --
	( 92.34,247.45);

\path[draw=drawColor,line width= 0.3pt,line join=round] (124.82, 30.69) --
	(124.82,247.45);

\path[draw=drawColor,line width= 0.3pt,line join=round] (157.30, 30.69) --
	(157.30,247.45);

\path[draw=drawColor,line width= 0.6pt,line join=round] ( 38.56, 57.97) --
	(178.61, 57.97);

\path[draw=drawColor,line width= 0.6pt,line join=round] ( 38.56,118.42) --
	(178.61,118.42);

\path[draw=drawColor,line width= 0.6pt,line join=round] ( 38.56,178.86) --
	(178.61,178.86);

\path[draw=drawColor,line width= 0.6pt,line join=round] ( 38.56,239.31) --
	(178.61,239.31);

\path[draw=drawColor,line width= 0.6pt,line join=round] ( 43.62, 30.69) --
	( 43.62,247.45);

\path[draw=drawColor,line width= 0.6pt,line join=round] ( 76.10, 30.69) --
	( 76.10,247.45);

\path[draw=drawColor,line width= 0.6pt,line join=round] (108.58, 30.69) --
	(108.58,247.45);

\path[draw=drawColor,line width= 0.6pt,line join=round] (141.06, 30.69) --
	(141.06,247.45);

\path[draw=drawColor,line width= 0.6pt,line join=round] (173.54, 30.69) --
	(173.54,247.45);
\definecolor{fillColor}{RGB}{230,159,0}

\path[fill=fillColor,fill opacity=0.10] ( 44.92, 79.02) --
	( 46.22, 75.94) --
	( 47.52, 73.17) --
	( 48.82, 70.88) --
	( 50.12, 69.28) --
	( 51.42, 68.50) --
	( 52.72, 68.58) --
	( 54.02, 69.42) --
	( 55.31, 70.88) --
	( 56.61, 72.83) --
	( 57.91, 75.21) --
	( 59.21, 77.96) --
	( 60.51, 81.09) --
	( 61.81, 84.58) --
	( 63.11, 88.46) --
	( 64.41, 92.73) --
	( 65.71, 97.34) --
	( 67.01,102.20) --
	( 68.31,107.17) --
	( 69.61,112.02) --
	( 70.91,116.57) --
	( 72.20,120.65) --
	( 73.50,124.14) --
	( 74.80,127.01) --
	( 76.10,129.29) --
	( 77.40,131.05) --
	( 78.70,132.40) --
	( 80.00,133.45) --
	( 81.30,134.29) --
	( 82.60,134.97) --
	( 83.90,135.53) --
	( 85.20,135.99) --
	( 86.50,136.38) --
	( 87.79,136.74) --
	( 89.09,137.12) --
	( 90.39,137.59) --
	( 91.69,138.20) --
	( 92.99,139.03) --
	( 94.29,140.10) --
	( 95.59,141.41) --
	( 96.89,142.92) --
	( 98.19,144.57) --
	( 99.49,146.27) --
	(100.79,147.97) --
	(102.09,149.61) --
	(103.38,151.18) --
	(104.68,152.69) --
	(105.98,154.17) --
	(107.28,155.64) --
	(108.58,157.14) --
	(109.88,158.66) --
	(111.18,160.17) --
	(112.48,161.62) --
	(113.78,162.94) --
	(115.08,164.06) --
	(116.38,164.92) --
	(117.68,165.49) --
	(118.98,165.75) --
	(120.27,165.71) --
	(121.57,165.37) --
	(122.87,164.74) --
	(124.17,163.82) --
	(125.47,162.59) --
	(126.77,161.04) --
	(128.07,159.14) --
	(129.37,156.90) --
	(130.67,154.37) --
	(131.97,151.64) --
	(133.27,148.86) --
	(134.57,146.20) --
	(135.86,143.88) --
	(137.16,142.08) --
	(138.46,140.97) --
	(139.76,140.64) --
	(141.06,141.15) --
	(142.36,142.47) --
	(143.66,144.54) --
	(144.96,147.29) --
	(146.26,150.61) --
	(147.56,154.41) --
	(148.86,158.58) --
	(150.16,163.00) --
	(151.46,167.54) --
	(152.75,172.03) --
	(154.05,176.27) --
	(155.35,180.04) --
	(156.65,183.17) --
	(157.95,185.50) --
	(159.25,186.95) --
	(160.55,187.49) --
	(161.85,187.17) --
	(163.15,186.10) --
	(164.45,184.47) --
	(165.75,182.59) --
	(167.05,180.74) --
	(168.34,179.15) --
	(169.64,177.88) --
	(170.94,176.93) --
	(172.24,176.24) --
	(172.24,131.89) --
	(170.94,139.30) --
	(169.64,146.16) --
	(168.34,152.21) --
	(167.05,157.19) --
	(165.75,160.94) --
	(164.45,163.48) --
	(163.15,164.94) --
	(161.85,165.50) --
	(160.55,165.27) --
	(159.25,164.32) --
	(157.95,162.69) --
	(156.65,160.38) --
	(155.35,157.42) --
	(154.05,153.82) --
	(152.75,149.64) --
	(151.46,144.96) --
	(150.16,139.96) --
	(148.86,134.89) --
	(147.56,130.03) --
	(146.26,125.65) --
	(144.96,121.98) --
	(143.66,119.18) --
	(142.36,117.31) --
	(141.06,116.36) --
	(139.76,116.26) --
	(138.46,116.88) --
	(137.16,118.11) --
	(135.86,119.82) --
	(134.57,121.91) --
	(133.27,124.28) --
	(131.97,126.86) --
	(130.67,129.55) --
	(129.37,132.23) --
	(128.07,134.80) --
	(126.77,137.12) --
	(125.47,139.10) --
	(124.17,140.65) --
	(122.87,141.71) --
	(121.57,142.29) --
	(120.27,142.41) --
	(118.98,142.15) --
	(117.68,141.58) --
	(116.38,140.80) --
	(115.08,139.86) --
	(113.78,138.81) --
	(112.48,137.66) --
	(111.18,136.40) --
	(109.88,135.03) --
	(108.58,133.51) --
	(107.28,131.87) --
	(105.98,130.10) --
	(104.68,128.28) --
	(103.38,126.44) --
	(102.09,124.66) --
	(100.79,122.96) --
	( 99.49,121.37) --
	( 98.19,119.91) --
	( 96.89,118.54) --
	( 95.59,117.25) --
	( 94.29,116.02) --
	( 92.99,114.87) --
	( 91.69,113.80) --
	( 90.39,112.87) --
	( 89.09,112.12) --
	( 87.79,111.57) --
	( 86.50,111.21) --
	( 85.20,111.00) --
	( 83.90,110.84) --
	( 82.60,110.62) --
	( 81.30,110.21) --
	( 80.00,109.46) --
	( 78.70,108.28) --
	( 77.40,106.62) --
	( 76.10,104.46) --
	( 74.80,101.84) --
	( 73.50, 98.81) --
	( 72.20, 95.42) --
	( 70.91, 91.74) --
	( 69.61, 87.79) --
	( 68.31, 83.62) --
	( 67.01, 79.27) --
	( 65.71, 74.80) --
	( 64.41, 70.32) --
	( 63.11, 65.97) --
	( 61.81, 61.92) --
	( 60.51, 58.33) --
	( 59.21, 55.33) --
	( 57.91, 52.96) --
	( 56.61, 51.22) --
	( 55.31, 50.06) --
	( 54.02, 49.32) --
	( 52.72, 48.79) --
	( 51.42, 48.22) --
	( 50.12, 47.38) --
	( 48.82, 46.17) --
	( 47.52, 44.58) --
	( 46.22, 42.68) --
	( 44.92, 40.54) --
	cycle;

\path[] ( 44.92, 79.02) --
	( 46.22, 75.94) --
	( 47.52, 73.17) --
	( 48.82, 70.88) --
	( 50.12, 69.28) --
	( 51.42, 68.50) --
	( 52.72, 68.58) --
	( 54.02, 69.42) --
	( 55.31, 70.88) --
	( 56.61, 72.83) --
	( 57.91, 75.21) --
	( 59.21, 77.96) --
	( 60.51, 81.09) --
	( 61.81, 84.58) --
	( 63.11, 88.46) --
	( 64.41, 92.73) --
	( 65.71, 97.34) --
	( 67.01,102.20) --
	( 68.31,107.17) --
	( 69.61,112.02) --
	( 70.91,116.57) --
	( 72.20,120.65) --
	( 73.50,124.14) --
	( 74.80,127.01) --
	( 76.10,129.29) --
	( 77.40,131.05) --
	( 78.70,132.40) --
	( 80.00,133.45) --
	( 81.30,134.29) --
	( 82.60,134.97) --
	( 83.90,135.53) --
	( 85.20,135.99) --
	( 86.50,136.38) --
	( 87.79,136.74) --
	( 89.09,137.12) --
	( 90.39,137.59) --
	( 91.69,138.20) --
	( 92.99,139.03) --
	( 94.29,140.10) --
	( 95.59,141.41) --
	( 96.89,142.92) --
	( 98.19,144.57) --
	( 99.49,146.27) --
	(100.79,147.97) --
	(102.09,149.61) --
	(103.38,151.18) --
	(104.68,152.69) --
	(105.98,154.17) --
	(107.28,155.64) --
	(108.58,157.14) --
	(109.88,158.66) --
	(111.18,160.17) --
	(112.48,161.62) --
	(113.78,162.94) --
	(115.08,164.06) --
	(116.38,164.92) --
	(117.68,165.49) --
	(118.98,165.75) --
	(120.27,165.71) --
	(121.57,165.37) --
	(122.87,164.74) --
	(124.17,163.82) --
	(125.47,162.59) --
	(126.77,161.04) --
	(128.07,159.14) --
	(129.37,156.90) --
	(130.67,154.37) --
	(131.97,151.64) --
	(133.27,148.86) --
	(134.57,146.20) --
	(135.86,143.88) --
	(137.16,142.08) --
	(138.46,140.97) --
	(139.76,140.64) --
	(141.06,141.15) --
	(142.36,142.47) --
	(143.66,144.54) --
	(144.96,147.29) --
	(146.26,150.61) --
	(147.56,154.41) --
	(148.86,158.58) --
	(150.16,163.00) --
	(151.46,167.54) --
	(152.75,172.03) --
	(154.05,176.27) --
	(155.35,180.04) --
	(156.65,183.17) --
	(157.95,185.50) --
	(159.25,186.95) --
	(160.55,187.49) --
	(161.85,187.17) --
	(163.15,186.10) --
	(164.45,184.47) --
	(165.75,182.59) --
	(167.05,180.74) --
	(168.34,179.15) --
	(169.64,177.88) --
	(170.94,176.93) --
	(172.24,176.24);

\path[] (172.24,131.89) --
	(170.94,139.30) --
	(169.64,146.16) --
	(168.34,152.21) --
	(167.05,157.19) --
	(165.75,160.94) --
	(164.45,163.48) --
	(163.15,164.94) --
	(161.85,165.50) --
	(160.55,165.27) --
	(159.25,164.32) --
	(157.95,162.69) --
	(156.65,160.38) --
	(155.35,157.42) --
	(154.05,153.82) --
	(152.75,149.64) --
	(151.46,144.96) --
	(150.16,139.96) --
	(148.86,134.89) --
	(147.56,130.03) --
	(146.26,125.65) --
	(144.96,121.98) --
	(143.66,119.18) --
	(142.36,117.31) --
	(141.06,116.36) --
	(139.76,116.26) --
	(138.46,116.88) --
	(137.16,118.11) --
	(135.86,119.82) --
	(134.57,121.91) --
	(133.27,124.28) --
	(131.97,126.86) --
	(130.67,129.55) --
	(129.37,132.23) --
	(128.07,134.80) --
	(126.77,137.12) --
	(125.47,139.10) --
	(124.17,140.65) --
	(122.87,141.71) --
	(121.57,142.29) --
	(120.27,142.41) --
	(118.98,142.15) --
	(117.68,141.58) --
	(116.38,140.80) --
	(115.08,139.86) --
	(113.78,138.81) --
	(112.48,137.66) --
	(111.18,136.40) --
	(109.88,135.03) --
	(108.58,133.51) --
	(107.28,131.87) --
	(105.98,130.10) --
	(104.68,128.28) --
	(103.38,126.44) --
	(102.09,124.66) --
	(100.79,122.96) --
	( 99.49,121.37) --
	( 98.19,119.91) --
	( 96.89,118.54) --
	( 95.59,117.25) --
	( 94.29,116.02) --
	( 92.99,114.87) --
	( 91.69,113.80) --
	( 90.39,112.87) --
	( 89.09,112.12) --
	( 87.79,111.57) --
	( 86.50,111.21) --
	( 85.20,111.00) --
	( 83.90,110.84) --
	( 82.60,110.62) --
	( 81.30,110.21) --
	( 80.00,109.46) --
	( 78.70,108.28) --
	( 77.40,106.62) --
	( 76.10,104.46) --
	( 74.80,101.84) --
	( 73.50, 98.81) --
	( 72.20, 95.42) --
	( 70.91, 91.74) --
	( 69.61, 87.79) --
	( 68.31, 83.62) --
	( 67.01, 79.27) --
	( 65.71, 74.80) --
	( 64.41, 70.32) --
	( 63.11, 65.97) --
	( 61.81, 61.92) --
	( 60.51, 58.33) --
	( 59.21, 55.33) --
	( 57.91, 52.96) --
	( 56.61, 51.22) --
	( 55.31, 50.06) --
	( 54.02, 49.32) --
	( 52.72, 48.79) --
	( 51.42, 48.22) --
	( 50.12, 47.38) --
	( 48.82, 46.17) --
	( 47.52, 44.58) --
	( 46.22, 42.68) --
	( 44.92, 40.54);
\definecolor{fillColor}{RGB}{86,180,233}

\path[fill=fillColor,fill opacity=0.10] ( 44.92,219.77) --
	( 46.22,215.62) --
	( 47.52,210.09) --
	( 48.82,203.14) --
	( 50.12,195.05) --
	( 51.42,186.38) --
	( 52.72,177.84) --
	( 54.02,170.13) --
	( 55.31,163.80) --
	( 56.61,159.30) --
	( 57.91,156.92) --
	( 59.21,156.85) --
	( 60.51,159.15) --
	( 61.81,163.74) --
	( 63.11,170.38) --
	( 64.41,178.64) --
	( 65.71,187.83) --
	( 67.01,197.18) --
	( 68.31,205.95) --
	( 69.61,213.57) --
	( 70.91,219.74) --
	( 72.20,224.47) --
	( 73.50,227.95) --
	( 74.80,230.46) --
	( 76.10,232.25) --
	( 77.40,233.52) --
	( 78.70,234.40) --
	( 80.00,235.00) --
	( 81.30,235.39) --
	( 82.60,235.62) --
	( 83.90,235.71) --
	( 85.20,235.70) --
	( 86.50,235.59) --
	( 87.79,235.41) --
	( 89.09,235.17) --
	( 90.39,234.89) --
	( 91.69,234.61) --
	( 92.99,234.36) --
	( 94.29,234.16) --
	( 95.59,234.02) --
	( 96.89,233.96) --
	( 98.19,233.96) --
	( 99.49,234.01) --
	(100.79,234.10) --
	(102.09,234.20) --
	(103.38,234.31) --
	(104.68,234.41) --
	(105.98,234.50) --
	(107.28,234.56) --
	(108.58,234.60) --
	(109.88,234.61) --
	(111.18,234.57) --
	(112.48,234.48) --
	(113.78,234.32) --
	(115.08,234.10) --
	(116.38,233.80) --
	(117.68,233.44) --
	(118.98,233.02) --
	(120.27,232.58) --
	(121.57,232.14) --
	(122.87,231.75) --
	(124.17,231.42) --
	(125.47,231.19) --
	(126.77,231.07) --
	(128.07,231.05) --
	(129.37,231.12) --
	(130.67,231.27) --
	(131.97,231.47) --
	(133.27,231.72) --
	(134.57,231.99) --
	(135.86,232.29) --
	(137.16,232.59) --
	(138.46,232.87) --
	(139.76,233.12) --
	(141.06,233.32) --
	(142.36,233.43) --
	(143.66,233.46) --
	(144.96,233.39) --
	(146.26,233.23) --
	(147.56,232.98) --
	(148.86,232.65) --
	(150.16,232.28) --
	(151.46,231.91) --
	(152.75,231.58) --
	(154.05,231.32) --
	(155.35,231.15) --
	(156.65,231.07) --
	(157.95,231.10) --
	(159.25,231.22) --
	(160.55,231.44) --
	(161.85,231.77) --
	(163.15,232.23) --
	(164.45,232.83) --
	(165.75,233.57) --
	(167.05,234.42) --
	(168.34,235.30) --
	(169.64,236.16) --
	(170.94,236.93) --
	(172.24,237.59) --
	(172.24,232.94) --
	(170.94,232.40) --
	(169.64,231.82) --
	(168.34,231.20) --
	(167.05,230.53) --
	(165.75,229.79) --
	(164.45,228.99) --
	(163.15,228.17) --
	(161.85,227.43) --
	(160.55,226.83) --
	(159.25,226.45) --
	(157.95,226.31) --
	(156.65,226.39) --
	(155.35,226.67) --
	(154.05,227.07) --
	(152.75,227.54) --
	(151.46,228.00) --
	(150.16,228.43) --
	(148.86,228.81) --
	(147.56,229.12) --
	(146.26,229.36) --
	(144.96,229.53) --
	(143.66,229.62) --
	(142.36,229.63) --
	(141.06,229.56) --
	(139.76,229.39) --
	(138.46,229.13) --
	(137.16,228.79) --
	(135.86,228.37) --
	(134.57,227.91) --
	(133.27,227.45) --
	(131.97,227.03) --
	(130.67,226.69) --
	(129.37,226.48) --
	(128.07,226.43) --
	(126.77,226.52) --
	(125.47,226.76) --
	(124.17,227.11) --
	(122.87,227.55) --
	(121.57,228.03) --
	(120.27,228.54) --
	(118.98,229.06) --
	(117.68,229.56) --
	(116.38,230.02) --
	(115.08,230.43) --
	(113.78,230.78) --
	(112.48,231.06) --
	(111.18,231.25) --
	(109.88,231.35) --
	(108.58,231.35) --
	(107.28,231.28) --
	(105.98,231.13) --
	(104.68,230.93) --
	(103.38,230.70) --
	(102.09,230.48) --
	(100.79,230.29) --
	( 99.49,230.17) --
	( 98.19,230.14) --
	( 96.89,230.20) --
	( 95.59,230.36) --
	( 94.29,230.60) --
	( 92.99,230.90) --
	( 91.69,231.24) --
	( 90.39,231.60) --
	( 89.09,231.95) --
	( 87.79,232.27) --
	( 86.50,232.54) --
	( 85.20,232.73) --
	( 83.90,232.82) --
	( 82.60,232.76) --
	( 81.30,232.54) --
	( 80.00,232.10) --
	( 78.70,231.37) --
	( 77.40,230.26) --
	( 76.10,228.62) --
	( 74.80,226.28) --
	( 73.50,223.02) --
	( 72.20,218.57) --
	( 70.91,212.75) --
	( 69.61,205.51) --
	( 68.31,197.03) --
	( 67.01,187.70) --
	( 65.71,178.09) --
	( 64.41,168.79) --
	( 63.11,160.41) --
	( 61.81,153.47) --
	( 60.51,148.43) --
	( 59.21,145.69) --
	( 57.91,145.50) --
	( 56.61,147.94) --
	( 55.31,152.88) --
	( 54.02,159.88) --
	( 52.72,168.27) --
	( 51.42,177.21) --
	( 50.12,185.93) --
	( 48.82,193.88) --
	( 47.52,200.75) --
	( 46.22,206.31) --
	( 44.92,210.39) --
	cycle;

\path[] ( 44.92,219.77) --
	( 46.22,215.62) --
	( 47.52,210.09) --
	( 48.82,203.14) --
	( 50.12,195.05) --
	( 51.42,186.38) --
	( 52.72,177.84) --
	( 54.02,170.13) --
	( 55.31,163.80) --
	( 56.61,159.30) --
	( 57.91,156.92) --
	( 59.21,156.85) --
	( 60.51,159.15) --
	( 61.81,163.74) --
	( 63.11,170.38) --
	( 64.41,178.64) --
	( 65.71,187.83) --
	( 67.01,197.18) --
	( 68.31,205.95) --
	( 69.61,213.57) --
	( 70.91,219.74) --
	( 72.20,224.47) --
	( 73.50,227.95) --
	( 74.80,230.46) --
	( 76.10,232.25) --
	( 77.40,233.52) --
	( 78.70,234.40) --
	( 80.00,235.00) --
	( 81.30,235.39) --
	( 82.60,235.62) --
	( 83.90,235.71) --
	( 85.20,235.70) --
	( 86.50,235.59) --
	( 87.79,235.41) --
	( 89.09,235.17) --
	( 90.39,234.89) --
	( 91.69,234.61) --
	( 92.99,234.36) --
	( 94.29,234.16) --
	( 95.59,234.02) --
	( 96.89,233.96) --
	( 98.19,233.96) --
	( 99.49,234.01) --
	(100.79,234.10) --
	(102.09,234.20) --
	(103.38,234.31) --
	(104.68,234.41) --
	(105.98,234.50) --
	(107.28,234.56) --
	(108.58,234.60) --
	(109.88,234.61) --
	(111.18,234.57) --
	(112.48,234.48) --
	(113.78,234.32) --
	(115.08,234.10) --
	(116.38,233.80) --
	(117.68,233.44) --
	(118.98,233.02) --
	(120.27,232.58) --
	(121.57,232.14) --
	(122.87,231.75) --
	(124.17,231.42) --
	(125.47,231.19) --
	(126.77,231.07) --
	(128.07,231.05) --
	(129.37,231.12) --
	(130.67,231.27) --
	(131.97,231.47) --
	(133.27,231.72) --
	(134.57,231.99) --
	(135.86,232.29) --
	(137.16,232.59) --
	(138.46,232.87) --
	(139.76,233.12) --
	(141.06,233.32) --
	(142.36,233.43) --
	(143.66,233.46) --
	(144.96,233.39) --
	(146.26,233.23) --
	(147.56,232.98) --
	(148.86,232.65) --
	(150.16,232.28) --
	(151.46,231.91) --
	(152.75,231.58) --
	(154.05,231.32) --
	(155.35,231.15) --
	(156.65,231.07) --
	(157.95,231.10) --
	(159.25,231.22) --
	(160.55,231.44) --
	(161.85,231.77) --
	(163.15,232.23) --
	(164.45,232.83) --
	(165.75,233.57) --
	(167.05,234.42) --
	(168.34,235.30) --
	(169.64,236.16) --
	(170.94,236.93) --
	(172.24,237.59);

\path[] (172.24,232.94) --
	(170.94,232.40) --
	(169.64,231.82) --
	(168.34,231.20) --
	(167.05,230.53) --
	(165.75,229.79) --
	(164.45,228.99) --
	(163.15,228.17) --
	(161.85,227.43) --
	(160.55,226.83) --
	(159.25,226.45) --
	(157.95,226.31) --
	(156.65,226.39) --
	(155.35,226.67) --
	(154.05,227.07) --
	(152.75,227.54) --
	(151.46,228.00) --
	(150.16,228.43) --
	(148.86,228.81) --
	(147.56,229.12) --
	(146.26,229.36) --
	(144.96,229.53) --
	(143.66,229.62) --
	(142.36,229.63) --
	(141.06,229.56) --
	(139.76,229.39) --
	(138.46,229.13) --
	(137.16,228.79) --
	(135.86,228.37) --
	(134.57,227.91) --
	(133.27,227.45) --
	(131.97,227.03) --
	(130.67,226.69) --
	(129.37,226.48) --
	(128.07,226.43) --
	(126.77,226.52) --
	(125.47,226.76) --
	(124.17,227.11) --
	(122.87,227.55) --
	(121.57,228.03) --
	(120.27,228.54) --
	(118.98,229.06) --
	(117.68,229.56) --
	(116.38,230.02) --
	(115.08,230.43) --
	(113.78,230.78) --
	(112.48,231.06) --
	(111.18,231.25) --
	(109.88,231.35) --
	(108.58,231.35) --
	(107.28,231.28) --
	(105.98,231.13) --
	(104.68,230.93) --
	(103.38,230.70) --
	(102.09,230.48) --
	(100.79,230.29) --
	( 99.49,230.17) --
	( 98.19,230.14) --
	( 96.89,230.20) --
	( 95.59,230.36) --
	( 94.29,230.60) --
	( 92.99,230.90) --
	( 91.69,231.24) --
	( 90.39,231.60) --
	( 89.09,231.95) --
	( 87.79,232.27) --
	( 86.50,232.54) --
	( 85.20,232.73) --
	( 83.90,232.82) --
	( 82.60,232.76) --
	( 81.30,232.54) --
	( 80.00,232.10) --
	( 78.70,231.37) --
	( 77.40,230.26) --
	( 76.10,228.62) --
	( 74.80,226.28) --
	( 73.50,223.02) --
	( 72.20,218.57) --
	( 70.91,212.75) --
	( 69.61,205.51) --
	( 68.31,197.03) --
	( 67.01,187.70) --
	( 65.71,178.09) --
	( 64.41,168.79) --
	( 63.11,160.41) --
	( 61.81,153.47) --
	( 60.51,148.43) --
	( 59.21,145.69) --
	( 57.91,145.50) --
	( 56.61,147.94) --
	( 55.31,152.88) --
	( 54.02,159.88) --
	( 52.72,168.27) --
	( 51.42,177.21) --
	( 50.12,185.93) --
	( 48.82,193.88) --
	( 47.52,200.75) --
	( 46.22,206.31) --
	( 44.92,210.39);
\definecolor{drawColor}{RGB}{230,159,0}

\path[draw=drawColor,line width= 0.6pt,line join=round] ( 44.92, 59.78) --
	( 46.22, 59.31) --
	( 47.52, 58.87) --
	( 48.82, 58.53) --
	( 50.12, 58.33) --
	( 51.42, 58.36) --
	( 52.72, 58.69) --
	( 54.02, 59.37) --
	( 55.31, 60.47) --
	( 56.61, 62.03) --
	( 57.91, 64.08) --
	( 59.21, 66.64) --
	( 60.51, 69.71) --
	( 61.81, 73.25) --
	( 63.11, 77.21) --
	( 64.41, 81.52) --
	( 65.71, 86.07) --
	( 67.01, 90.74) --
	( 68.31, 95.39) --
	( 69.61, 99.91) --
	( 70.91,104.15) --
	( 72.20,108.04) --
	( 73.50,111.47) --
	( 74.80,114.43) --
	( 76.10,116.87) --
	( 77.40,118.83) --
	( 78.70,120.34) --
	( 80.00,121.46) --
	( 81.30,122.25) --
	( 82.60,122.80) --
	( 83.90,123.18) --
	( 85.20,123.49) --
	( 86.50,123.79) --
	( 87.79,124.15) --
	( 89.09,124.62) --
	( 90.39,125.23) --
	( 91.69,126.00) --
	( 92.99,126.95) --
	( 94.29,128.06) --
	( 95.59,129.33) --
	( 96.89,130.73) --
	( 98.19,132.24) --
	( 99.49,133.82) --
	(100.79,135.46) --
	(102.09,137.13) --
	(103.38,138.81) --
	(104.68,140.48) --
	(105.98,142.13) --
	(107.28,143.75) --
	(108.58,145.33) --
	(109.88,146.84) --
	(111.18,148.29) --
	(112.48,149.64) --
	(113.78,150.87) --
	(115.08,151.96) --
	(116.38,152.86) --
	(117.68,153.54) --
	(118.98,153.95) --
	(120.27,154.06) --
	(121.57,153.83) --
	(122.87,153.22) --
	(124.17,152.23) --
	(125.47,150.85) --
	(126.77,149.08) --
	(128.07,146.97) --
	(129.37,144.57) --
	(130.67,141.96) --
	(131.97,139.25) --
	(133.27,136.57) --
	(134.57,134.06) --
	(135.86,131.85) --
	(137.16,130.09) --
	(138.46,128.93) --
	(139.76,128.45) --
	(141.06,128.76) --
	(142.36,129.89) --
	(143.66,131.86) --
	(144.96,134.63) --
	(146.26,138.13) --
	(147.56,142.22) --
	(148.86,146.73) --
	(150.16,151.48) --
	(151.46,156.25) --
	(152.75,160.83) --
	(154.05,165.05) --
	(155.35,168.73) --
	(156.65,171.77) --
	(157.95,174.09) --
	(159.25,175.63) --
	(160.55,176.38) --
	(161.85,176.34) --
	(163.15,175.52) --
	(164.45,173.98) --
	(165.75,171.76) --
	(167.05,168.96) --
	(168.34,165.68) --
	(169.64,162.02) --
	(170.94,158.12) --
	(172.24,154.06);
\definecolor{drawColor}{RGB}{86,180,233}

\path[draw=drawColor,line width= 0.6pt,line join=round] ( 44.92,215.08) --
	( 46.22,210.97) --
	( 47.52,205.42) --
	( 48.82,198.51) --
	( 50.12,190.49) --
	( 51.42,181.80) --
	( 52.72,173.06) --
	( 54.02,165.01) --
	( 55.31,158.34) --
	( 56.61,153.62) --
	( 57.91,151.21) --
	( 59.21,151.27) --
	( 60.51,153.79) --
	( 61.81,158.60) --
	( 63.11,165.40) --
	( 64.41,173.71) --
	( 65.71,182.96) --
	( 67.01,192.44) --
	( 68.31,201.49) --
	( 69.61,209.54) --
	( 70.91,216.25) --
	( 72.20,221.52) --
	( 73.50,225.48) --
	( 74.80,228.37) --
	( 76.10,230.44) --
	( 77.40,231.89) --
	( 78.70,232.88) --
	( 80.00,233.55) --
	( 81.30,233.97) --
	( 82.60,234.19) --
	( 83.90,234.26) --
	( 85.20,234.21) --
	( 86.50,234.06) --
	( 87.79,233.84) --
	( 89.09,233.56) --
	( 90.39,233.24) --
	( 91.69,232.93) --
	( 92.99,232.63) --
	( 94.29,232.38) --
	( 95.59,232.19) --
	( 96.89,232.08) --
	( 98.19,232.05) --
	( 99.49,232.09) --
	(100.79,232.19) --
	(102.09,232.34) --
	(103.38,232.50) --
	(104.68,232.67) --
	(105.98,232.81) --
	(107.28,232.92) --
	(108.58,232.98) --
	(109.88,232.98) --
	(111.18,232.91) --
	(112.48,232.77) --
	(113.78,232.55) --
	(115.08,232.26) --
	(116.38,231.91) --
	(117.68,231.50) --
	(118.98,231.04) --
	(120.27,230.56) --
	(121.57,230.09) --
	(122.87,229.65) --
	(124.17,229.27) --
	(125.47,228.98) --
	(126.77,228.80) --
	(128.07,228.74) --
	(129.37,228.80) --
	(130.67,228.98) --
	(131.97,229.25) --
	(133.27,229.58) --
	(134.57,229.95) --
	(135.86,230.33) --
	(137.16,230.69) --
	(138.46,231.00) --
	(139.76,231.26) --
	(141.06,231.44) --
	(142.36,231.53) --
	(143.66,231.54) --
	(144.96,231.46) --
	(146.26,231.30) --
	(147.56,231.05) --
	(148.86,230.73) --
	(150.16,230.36) --
	(151.46,229.96) --
	(152.75,229.56) --
	(154.05,229.20) --
	(155.35,228.91) --
	(156.65,228.73) --
	(157.95,228.70) --
	(159.25,228.83) --
	(160.55,229.13) --
	(161.85,229.60) --
	(163.15,230.20) --
	(164.45,230.91) --
	(165.75,231.68) --
	(167.05,232.47) --
	(168.34,233.25) --
	(169.64,233.99) --
	(170.94,234.67) --
	(172.24,235.27);
\definecolor{drawColor}{RGB}{230,159,0}

\path[draw=drawColor,line width= 0.6pt,line join=round] ( 44.92, 59.78) --
	( 46.22, 59.31) --
	( 47.52, 58.87) --
	( 48.82, 58.53) --
	( 50.12, 58.33) --
	( 51.42, 58.36) --
	( 52.72, 58.69) --
	( 54.02, 59.37) --
	( 55.31, 60.47) --
	( 56.61, 62.03) --
	( 57.91, 64.08) --
	( 59.21, 66.64) --
	( 60.51, 69.71) --
	( 61.81, 73.25) --
	( 63.11, 77.21) --
	( 64.41, 81.52) --
	( 65.71, 86.07) --
	( 67.01, 90.74) --
	( 68.31, 95.39) --
	( 69.61, 99.91) --
	( 70.91,104.15) --
	( 72.20,108.04) --
	( 73.50,111.47) --
	( 74.80,114.43) --
	( 76.10,116.87) --
	( 77.40,118.83) --
	( 78.70,120.34) --
	( 80.00,121.46) --
	( 81.30,122.25) --
	( 82.60,122.80) --
	( 83.90,123.18) --
	( 85.20,123.49) --
	( 86.50,123.79) --
	( 87.79,124.15) --
	( 89.09,124.62) --
	( 90.39,125.23) --
	( 91.69,126.00) --
	( 92.99,126.95) --
	( 94.29,128.06) --
	( 95.59,129.33) --
	( 96.89,130.73) --
	( 98.19,132.24) --
	( 99.49,133.82) --
	(100.79,135.46) --
	(102.09,137.13) --
	(103.38,138.81) --
	(104.68,140.48) --
	(105.98,142.13) --
	(107.28,143.75) --
	(108.58,145.33) --
	(109.88,146.84) --
	(111.18,148.29) --
	(112.48,149.64) --
	(113.78,150.87) --
	(115.08,151.96) --
	(116.38,152.86) --
	(117.68,153.54) --
	(118.98,153.95) --
	(120.27,154.06) --
	(121.57,153.83) --
	(122.87,153.22) --
	(124.17,152.23) --
	(125.47,150.85) --
	(126.77,149.08) --
	(128.07,146.97) --
	(129.37,144.57) --
	(130.67,141.96) --
	(131.97,139.25) --
	(133.27,136.57) --
	(134.57,134.06) --
	(135.86,131.85) --
	(137.16,130.09) --
	(138.46,128.93) --
	(139.76,128.45) --
	(141.06,128.76) --
	(142.36,129.89) --
	(143.66,131.86) --
	(144.96,134.63) --
	(146.26,138.13) --
	(147.56,142.22) --
	(148.86,146.73) --
	(150.16,151.48) --
	(151.46,156.25) --
	(152.75,160.83) --
	(154.05,165.05) --
	(155.35,168.73) --
	(156.65,171.77) --
	(157.95,174.09) --
	(159.25,175.63) --
	(160.55,176.38) --
	(161.85,176.34) --
	(163.15,175.52) --
	(164.45,173.98) --
	(165.75,171.76) --
	(167.05,168.96) --
	(168.34,165.68) --
	(169.64,162.02) --
	(170.94,158.12) --
	(172.24,154.06);
\definecolor{drawColor}{RGB}{86,180,233}

\path[draw=drawColor,line width= 0.6pt,line join=round] ( 44.92,215.08) --
	( 46.22,210.97) --
	( 47.52,205.42) --
	( 48.82,198.51) --
	( 50.12,190.49) --
	( 51.42,181.80) --
	( 52.72,173.06) --
	( 54.02,165.01) --
	( 55.31,158.34) --
	( 56.61,153.62) --
	( 57.91,151.21) --
	( 59.21,151.27) --
	( 60.51,153.79) --
	( 61.81,158.60) --
	( 63.11,165.40) --
	( 64.41,173.71) --
	( 65.71,182.96) --
	( 67.01,192.44) --
	( 68.31,201.49) --
	( 69.61,209.54) --
	( 70.91,216.25) --
	( 72.20,221.52) --
	( 73.50,225.48) --
	( 74.80,228.37) --
	( 76.10,230.44) --
	( 77.40,231.89) --
	( 78.70,232.88) --
	( 80.00,233.55) --
	( 81.30,233.97) --
	( 82.60,234.19) --
	( 83.90,234.26) --
	( 85.20,234.21) --
	( 86.50,234.06) --
	( 87.79,233.84) --
	( 89.09,233.56) --
	( 90.39,233.24) --
	( 91.69,232.93) --
	( 92.99,232.63) --
	( 94.29,232.38) --
	( 95.59,232.19) --
	( 96.89,232.08) --
	( 98.19,232.05) --
	( 99.49,232.09) --
	(100.79,232.19) --
	(102.09,232.34) --
	(103.38,232.50) --
	(104.68,232.67) --
	(105.98,232.81) --
	(107.28,232.92) --
	(108.58,232.98) --
	(109.88,232.98) --
	(111.18,232.91) --
	(112.48,232.77) --
	(113.78,232.55) --
	(115.08,232.26) --
	(116.38,231.91) --
	(117.68,231.50) --
	(118.98,231.04) --
	(120.27,230.56) --
	(121.57,230.09) --
	(122.87,229.65) --
	(124.17,229.27) --
	(125.47,228.98) --
	(126.77,228.80) --
	(128.07,228.74) --
	(129.37,228.80) --
	(130.67,228.98) --
	(131.97,229.25) --
	(133.27,229.58) --
	(134.57,229.95) --
	(135.86,230.33) --
	(137.16,230.69) --
	(138.46,231.00) --
	(139.76,231.26) --
	(141.06,231.44) --
	(142.36,231.53) --
	(143.66,231.54) --
	(144.96,231.46) --
	(146.26,231.30) --
	(147.56,231.05) --
	(148.86,230.73) --
	(150.16,230.36) --
	(151.46,229.96) --
	(152.75,229.56) --
	(154.05,229.20) --
	(155.35,228.91) --
	(156.65,228.73) --
	(157.95,228.70) --
	(159.25,228.83) --
	(160.55,229.13) --
	(161.85,229.60) --
	(163.15,230.20) --
	(164.45,230.91) --
	(165.75,231.68) --
	(167.05,232.47) --
	(168.34,233.25) --
	(169.64,233.99) --
	(170.94,234.67) --
	(172.24,235.27);
\definecolor{drawColor}{gray}{0.20}

\path[draw=drawColor,line width= 0.6pt,line join=round,line cap=round] ( 38.56, 30.69) rectangle (178.61,247.45);
\end{scope}
\begin{scope}
\path[clip] (  0.00,  0.00) rectangle (252.94,252.94);
\definecolor{drawColor}{gray}{0.30}

\node[text=drawColor,anchor=base east,inner sep=0pt, outer sep=0pt, scale=  0.88] at ( 33.61, 54.94) {0.25};

\node[text=drawColor,anchor=base east,inner sep=0pt, outer sep=0pt, scale=  0.88] at ( 33.61,115.39) {0.50};

\node[text=drawColor,anchor=base east,inner sep=0pt, outer sep=0pt, scale=  0.88] at ( 33.61,175.83) {0.75};

\node[text=drawColor,anchor=base east,inner sep=0pt, outer sep=0pt, scale=  0.88] at ( 33.61,236.28) {1.00};
\end{scope}
\begin{scope}
\path[clip] (  0.00,  0.00) rectangle (252.94,252.94);
\definecolor{drawColor}{gray}{0.20}

\path[draw=drawColor,line width= 0.6pt,line join=round] ( 35.81, 57.97) --
	( 38.56, 57.97);

\path[draw=drawColor,line width= 0.6pt,line join=round] ( 35.81,118.42) --
	( 38.56,118.42);

\path[draw=drawColor,line width= 0.6pt,line join=round] ( 35.81,178.86) --
	( 38.56,178.86);

\path[draw=drawColor,line width= 0.6pt,line join=round] ( 35.81,239.31) --
	( 38.56,239.31);
\end{scope}
\begin{scope}
\path[clip] (  0.00,  0.00) rectangle (252.94,252.94);
\definecolor{drawColor}{gray}{0.20}

\path[draw=drawColor,line width= 0.6pt,line join=round] ( 43.62, 27.94) --
	( 43.62, 30.69);

\path[draw=drawColor,line width= 0.6pt,line join=round] ( 76.10, 27.94) --
	( 76.10, 30.69);

\path[draw=drawColor,line width= 0.6pt,line join=round] (108.58, 27.94) --
	(108.58, 30.69);

\path[draw=drawColor,line width= 0.6pt,line join=round] (141.06, 27.94) --
	(141.06, 30.69);

\path[draw=drawColor,line width= 0.6pt,line join=round] (173.54, 27.94) --
	(173.54, 30.69);
\end{scope}
\begin{scope}
\path[clip] (  0.00,  0.00) rectangle (252.94,252.94);
\definecolor{drawColor}{gray}{0.30}

\node[text=drawColor,anchor=base,inner sep=0pt, outer sep=0pt, scale=  0.88] at ( 43.62, 19.68) {0};

\node[text=drawColor,anchor=base,inner sep=0pt, outer sep=0pt, scale=  0.88] at ( 76.10, 19.68) {25};

\node[text=drawColor,anchor=base,inner sep=0pt, outer sep=0pt, scale=  0.88] at (108.58, 19.68) {50};

\node[text=drawColor,anchor=base,inner sep=0pt, outer sep=0pt, scale=  0.88] at (141.06, 19.68) {75};

\node[text=drawColor,anchor=base,inner sep=0pt, outer sep=0pt, scale=  0.88] at (173.54, 19.68) {100};
\end{scope}
\begin{scope}
\path[clip] (  0.00,  0.00) rectangle (252.94,252.94);
\definecolor{drawColor}{RGB}{0,0,0}

\node[text=drawColor,anchor=base,inner sep=0pt, outer sep=0pt, scale=  1.10] at (108.58,  7.64) {Cardinality};
\end{scope}
\begin{scope}
\path[clip] (  0.00,  0.00) rectangle (252.94,252.94);
\definecolor{drawColor}{RGB}{0,0,0}

\node[text=drawColor,rotate= 90.00,anchor=base,inner sep=0pt, outer sep=0pt, scale=  1.10] at ( 13.08,139.07) {Comprehension Accuracy};
\end{scope}
\begin{scope}
\path[clip] (  0.00,  0.00) rectangle (252.94,252.94);
\definecolor{fillColor}{RGB}{255,255,255}

\path[fill=fillColor] (189.61,111.50) rectangle (247.44,166.63);
\end{scope}
\begin{scope}
\path[clip] (  0.00,  0.00) rectangle (252.94,252.94);
\definecolor{drawColor}{RGB}{0,0,0}

\node[text=drawColor,anchor=base west,inner sep=0pt, outer sep=0pt, scale=  1.10] at (195.11,152.48) {Group};
\end{scope}
\begin{scope}
\path[clip] (  0.00,  0.00) rectangle (252.94,252.94);
\definecolor{fillColor}{RGB}{255,255,255}

\path[fill=fillColor] (195.11,131.46) rectangle (209.56,145.91);
\end{scope}
\begin{scope}
\path[clip] (  0.00,  0.00) rectangle (252.94,252.94);
\definecolor{fillColor}{RGB}{230,159,0}

\path[fill=fillColor,fill opacity=0.10] (195.82,132.17) rectangle (208.85,145.20);
\end{scope}
\begin{scope}
\path[clip] (  0.00,  0.00) rectangle (252.94,252.94);
\definecolor{drawColor}{RGB}{230,159,0}

\path[draw=drawColor,line width= 0.6pt,line join=round] (196.55,138.69) -- (208.12,138.69);
\end{scope}
\begin{scope}
\path[clip] (  0.00,  0.00) rectangle (252.94,252.94);
\definecolor{drawColor}{RGB}{230,159,0}

\path[draw=drawColor,line width= 0.6pt,line join=round] (196.55,138.69) -- (208.12,138.69);
\end{scope}
\begin{scope}
\path[clip] (  0.00,  0.00) rectangle (252.94,252.94);
\definecolor{fillColor}{RGB}{255,255,255}

\path[fill=fillColor] (195.11,117.00) rectangle (209.56,131.46);
\end{scope}
\begin{scope}
\path[clip] (  0.00,  0.00) rectangle (252.94,252.94);
\definecolor{fillColor}{RGB}{86,180,233}

\path[fill=fillColor,fill opacity=0.10] (195.82,117.72) rectangle (208.85,130.75);
\end{scope}
\begin{scope}
\path[clip] (  0.00,  0.00) rectangle (252.94,252.94);
\definecolor{drawColor}{RGB}{86,180,233}

\path[draw=drawColor,line width= 0.6pt,line join=round] (196.55,124.23) -- (208.12,124.23);
\end{scope}
\begin{scope}
\path[clip] (  0.00,  0.00) rectangle (252.94,252.94);
\definecolor{drawColor}{RGB}{86,180,233}

\path[draw=drawColor,line width= 0.6pt,line join=round] (196.55,124.23) -- (208.12,124.23);
\end{scope}
\begin{scope}
\path[clip] (  0.00,  0.00) rectangle (252.94,252.94);
\definecolor{drawColor}{RGB}{0,0,0}

\node[text=drawColor,anchor=base west,inner sep=0pt, outer sep=0pt, scale=  0.88] at (215.06,135.65) {IA};
\end{scope}
\begin{scope}
\path[clip] (  0.00,  0.00) rectangle (252.94,252.94);
\definecolor{drawColor}{RGB}{0,0,0}

\node[text=drawColor,anchor=base west,inner sep=0pt, outer sep=0pt, scale=  0.88] at (215.06,121.20) {non-IA};
\end{scope}
\end{tikzpicture}

%% file: production-by-cardinality.tex
\begin{tikzpicture}[x=1pt,y=1pt]
\definecolor{fillColor}{RGB}{255,255,255}
\path[use as bounding box,fill=fillColor,fill opacity=0.00] (0,0) rectangle (252.94,252.94);
\begin{scope}
\path[clip] (  0.00,  0.00) rectangle (252.94,252.94);
\definecolor{drawColor}{RGB}{255,255,255}
\definecolor{fillColor}{RGB}{255,255,255}

\path[draw=drawColor,line width= 0.6pt,line join=round,line cap=round,fill=fillColor] (  0.00,  0.00) rectangle (252.94,252.94);
\end{scope}
\begin{scope}
\path[clip] ( 34.16, 30.69) rectangle (178.61,247.45);
\definecolor{fillColor}{RGB}{255,255,255}

\path[fill=fillColor] ( 34.16, 30.69) rectangle (178.61,247.45);
\definecolor{drawColor}{gray}{0.92}

\path[draw=drawColor,line width= 0.3pt,line join=round] ( 34.16, 46.28) --
	(178.61, 46.28);

\path[draw=drawColor,line width= 0.3pt,line join=round] ( 34.16, 97.75) --
	(178.61, 97.75);

\path[draw=drawColor,line width= 0.3pt,line join=round] ( 34.16,149.23) --
	(178.61,149.23);

\path[draw=drawColor,line width= 0.3pt,line join=round] ( 34.16,200.70) --
	(178.61,200.70);

\path[draw=drawColor,line width= 0.3pt,line join=round] ( 56.13, 30.69) --
	( 56.13,247.45);

\path[draw=drawColor,line width= 0.3pt,line join=round] ( 89.63, 30.69) --
	( 89.63,247.45);

\path[draw=drawColor,line width= 0.3pt,line join=round] (123.13, 30.69) --
	(123.13,247.45);

\path[draw=drawColor,line width= 0.3pt,line join=round] (156.63, 30.69) --
	(156.63,247.45);

\path[draw=drawColor,line width= 0.6pt,line join=round] ( 34.16, 72.02) --
	(178.61, 72.02);

\path[draw=drawColor,line width= 0.6pt,line join=round] ( 34.16,123.49) --
	(178.61,123.49);

\path[draw=drawColor,line width= 0.6pt,line join=round] ( 34.16,174.96) --
	(178.61,174.96);

\path[draw=drawColor,line width= 0.6pt,line join=round] ( 34.16,226.44) --
	(178.61,226.44);

\path[draw=drawColor,line width= 0.6pt,line join=round] ( 39.38, 30.69) --
	( 39.38,247.45);

\path[draw=drawColor,line width= 0.6pt,line join=round] ( 72.88, 30.69) --
	( 72.88,247.45);

\path[draw=drawColor,line width= 0.6pt,line join=round] (106.38, 30.69) --
	(106.38,247.45);

\path[draw=drawColor,line width= 0.6pt,line join=round] (139.88, 30.69) --
	(139.88,247.45);

\path[draw=drawColor,line width= 0.6pt,line join=round] (173.38, 30.69) --
	(173.38,247.45);
\definecolor{fillColor}{RGB}{230,159,0}

\path[fill=fillColor,fill opacity=0.10] ( 40.72,237.59) --
	( 42.06,233.39) --
	( 43.40,229.11) --
	( 44.74,224.85) --
	( 46.08,220.70) --
	( 47.42,216.77) --
	( 48.76,213.10) --
	( 50.10,209.71) --
	( 51.44,206.54) --
	( 52.78,203.53) --
	( 54.12,200.60) --
	( 55.46,197.71) --
	( 56.80,194.85) --
	( 58.14,192.03) --
	( 59.48,189.29) --
	( 60.82,186.66) --
	( 62.16,184.14) --
	( 63.50,181.76) --
	( 64.84,179.48) --
	( 66.18,177.27) --
	( 67.52,175.09) --
	( 68.86,172.90) --
	( 70.20,170.66) --
	( 71.54,168.36) --
	( 72.88,165.99) --
	( 74.22,163.57) --
	( 75.56,161.11) --
	( 76.90,158.63) --
	( 78.24,156.15) --
	( 79.58,153.67) --
	( 80.92,151.20) --
	( 82.26,148.76) --
	( 83.60,146.37) --
	( 84.94,144.05) --
	( 86.28,141.85) --
	( 87.62,139.82) --
	( 88.96,138.03) --
	( 90.30,136.53) --
	( 91.64,135.37) --
	( 92.98,134.56) --
	( 94.32,134.13) --
	( 95.66,134.07) --
	( 97.00,134.36) --
	( 98.34,134.99) --
	( 99.68,135.92) --
	(101.02,137.12) --
	(102.36,138.57) --
	(103.70,140.24) --
	(105.04,142.09) --
	(106.38,144.10) --
	(107.72,146.21) --
	(109.06,148.36) --
	(110.40,150.49) --
	(111.74,152.54) --
	(113.08,154.45) --
	(114.42,156.18) --
	(115.76,157.71) --
	(117.10,159.01) --
	(118.44,160.10) --
	(119.78,160.98) --
	(121.12,161.67) --
	(122.46,162.17) --
	(123.80,162.48) --
	(125.14,162.60) --
	(126.48,162.52) --
	(127.82,162.21) --
	(129.16,161.68) --
	(130.50,160.92) --
	(131.84,159.93) --
	(133.18,158.74) --
	(134.52,157.36) --
	(135.86,155.81) --
	(137.20,154.08) --
	(138.54,152.19) --
	(139.88,150.11) --
	(141.22,147.84) --
	(142.56,145.38) --
	(143.90,142.74) --
	(145.24,139.95) --
	(146.58,137.07) --
	(147.92,134.15) --
	(149.26,131.28) --
	(150.60,128.52) --
	(151.94,125.93) --
	(153.28,123.53) --
	(154.62,121.34) --
	(155.96,119.36) --
	(157.30,117.59) --
	(158.64,116.04) --
	(159.98,114.73) --
	(161.32,113.71) --
	(162.66,113.04) --
	(164.00,112.82) --
	(165.34,113.12) --
	(166.68,113.95) --
	(168.02,115.29) --
	(169.36,117.05) --
	(170.70,119.12) --
	(172.04,121.40) --
	(172.04, 80.90) --
	(170.70, 82.99) --
	(169.36, 84.91) --
	(168.02, 86.61) --
	(166.68, 88.04) --
	(165.34, 89.20) --
	(164.00, 90.12) --
	(162.66, 90.89) --
	(161.32, 91.66) --
	(159.98, 92.54) --
	(158.64, 93.65) --
	(157.30, 95.04) --
	(155.96, 96.73) --
	(154.62, 98.71) --
	(153.28,100.94) --
	(151.94,103.36) --
	(150.60,105.91) --
	(149.26,108.53) --
	(147.92,111.20) --
	(146.58,113.88) --
	(145.24,116.55) --
	(143.90,119.20) --
	(142.56,121.79) --
	(141.22,124.29) --
	(139.88,126.67) --
	(138.54,128.87) --
	(137.20,130.86) --
	(135.86,132.63) --
	(134.52,134.15) --
	(133.18,135.44) --
	(131.84,136.51) --
	(130.50,137.39) --
	(129.16,138.09) --
	(127.82,138.63) --
	(126.48,139.00) --
	(125.14,139.19) --
	(123.80,139.18) --
	(122.46,138.95) --
	(121.12,138.47) --
	(119.78,137.74) --
	(118.44,136.77) --
	(117.10,135.57) --
	(115.76,134.17) --
	(114.42,132.59) --
	(113.08,130.87) --
	(111.74,129.04) --
	(110.40,127.11) --
	(109.06,125.13) --
	(107.72,123.11) --
	(106.38,121.09) --
	(105.04,119.13) --
	(103.70,117.26) --
	(102.36,115.57) --
	(101.02,114.10) --
	( 99.68,112.91) --
	( 98.34,112.03) --
	( 97.00,111.50) --
	( 95.66,111.31) --
	( 94.32,111.47) --
	( 92.98,111.96) --
	( 91.64,112.75) --
	( 90.30,113.83) --
	( 88.96,115.17) --
	( 87.62,116.77) --
	( 86.28,118.60) --
	( 84.94,120.65) --
	( 83.60,122.90) --
	( 82.26,125.29) --
	( 80.92,127.78) --
	( 79.58,130.33) --
	( 78.24,132.88) --
	( 76.90,135.38) --
	( 75.56,137.81) --
	( 74.22,140.16) --
	( 72.88,142.45) --
	( 71.54,144.69) --
	( 70.20,146.93) --
	( 68.86,149.17) --
	( 67.52,151.46) --
	( 66.18,153.78) --
	( 64.84,156.15) --
	( 63.50,158.55) --
	( 62.16,160.99) --
	( 60.82,163.48) --
	( 59.48,166.04) --
	( 58.14,168.71) --
	( 56.80,171.52) --
	( 55.46,174.50) --
	( 54.12,177.62) --
	( 52.78,180.87) --
	( 51.44,184.16) --
	( 50.10,187.37) --
	( 48.76,190.40) --
	( 47.42,193.15) --
	( 46.08,195.57) --
	( 44.74,197.69) --
	( 43.40,199.55) --
	( 42.06,201.23) --
	( 40.72,202.80) --
	cycle;

\path[] ( 40.72,237.59) --
	( 42.06,233.39) --
	( 43.40,229.11) --
	( 44.74,224.85) --
	( 46.08,220.70) --
	( 47.42,216.77) --
	( 48.76,213.10) --
	( 50.10,209.71) --
	( 51.44,206.54) --
	( 52.78,203.53) --
	( 54.12,200.60) --
	( 55.46,197.71) --
	( 56.80,194.85) --
	( 58.14,192.03) --
	( 59.48,189.29) --
	( 60.82,186.66) --
	( 62.16,184.14) --
	( 63.50,181.76) --
	( 64.84,179.48) --
	( 66.18,177.27) --
	( 67.52,175.09) --
	( 68.86,172.90) --
	( 70.20,170.66) --
	( 71.54,168.36) --
	( 72.88,165.99) --
	( 74.22,163.57) --
	( 75.56,161.11) --
	( 76.90,158.63) --
	( 78.24,156.15) --
	( 79.58,153.67) --
	( 80.92,151.20) --
	( 82.26,148.76) --
	( 83.60,146.37) --
	( 84.94,144.05) --
	( 86.28,141.85) --
	( 87.62,139.82) --
	( 88.96,138.03) --
	( 90.30,136.53) --
	( 91.64,135.37) --
	( 92.98,134.56) --
	( 94.32,134.13) --
	( 95.66,134.07) --
	( 97.00,134.36) --
	( 98.34,134.99) --
	( 99.68,135.92) --
	(101.02,137.12) --
	(102.36,138.57) --
	(103.70,140.24) --
	(105.04,142.09) --
	(106.38,144.10) --
	(107.72,146.21) --
	(109.06,148.36) --
	(110.40,150.49) --
	(111.74,152.54) --
	(113.08,154.45) --
	(114.42,156.18) --
	(115.76,157.71) --
	(117.10,159.01) --
	(118.44,160.10) --
	(119.78,160.98) --
	(121.12,161.67) --
	(122.46,162.17) --
	(123.80,162.48) --
	(125.14,162.60) --
	(126.48,162.52) --
	(127.82,162.21) --
	(129.16,161.68) --
	(130.50,160.92) --
	(131.84,159.93) --
	(133.18,158.74) --
	(134.52,157.36) --
	(135.86,155.81) --
	(137.20,154.08) --
	(138.54,152.19) --
	(139.88,150.11) --
	(141.22,147.84) --
	(142.56,145.38) --
	(143.90,142.74) --
	(145.24,139.95) --
	(146.58,137.07) --
	(147.92,134.15) --
	(149.26,131.28) --
	(150.60,128.52) --
	(151.94,125.93) --
	(153.28,123.53) --
	(154.62,121.34) --
	(155.96,119.36) --
	(157.30,117.59) --
	(158.64,116.04) --
	(159.98,114.73) --
	(161.32,113.71) --
	(162.66,113.04) --
	(164.00,112.82) --
	(165.34,113.12) --
	(166.68,113.95) --
	(168.02,115.29) --
	(169.36,117.05) --
	(170.70,119.12) --
	(172.04,121.40);

\path[] (172.04, 80.90) --
	(170.70, 82.99) --
	(169.36, 84.91) --
	(168.02, 86.61) --
	(166.68, 88.04) --
	(165.34, 89.20) --
	(164.00, 90.12) --
	(162.66, 90.89) --
	(161.32, 91.66) --
	(159.98, 92.54) --
	(158.64, 93.65) --
	(157.30, 95.04) --
	(155.96, 96.73) --
	(154.62, 98.71) --
	(153.28,100.94) --
	(151.94,103.36) --
	(150.60,105.91) --
	(149.26,108.53) --
	(147.92,111.20) --
	(146.58,113.88) --
	(145.24,116.55) --
	(143.90,119.20) --
	(142.56,121.79) --
	(141.22,124.29) --
	(139.88,126.67) --
	(138.54,128.87) --
	(137.20,130.86) --
	(135.86,132.63) --
	(134.52,134.15) --
	(133.18,135.44) --
	(131.84,136.51) --
	(130.50,137.39) --
	(129.16,138.09) --
	(127.82,138.63) --
	(126.48,139.00) --
	(125.14,139.19) --
	(123.80,139.18) --
	(122.46,138.95) --
	(121.12,138.47) --
	(119.78,137.74) --
	(118.44,136.77) --
	(117.10,135.57) --
	(115.76,134.17) --
	(114.42,132.59) --
	(113.08,130.87) --
	(111.74,129.04) --
	(110.40,127.11) --
	(109.06,125.13) --
	(107.72,123.11) --
	(106.38,121.09) --
	(105.04,119.13) --
	(103.70,117.26) --
	(102.36,115.57) --
	(101.02,114.10) --
	( 99.68,112.91) --
	( 98.34,112.03) --
	( 97.00,111.50) --
	( 95.66,111.31) --
	( 94.32,111.47) --
	( 92.98,111.96) --
	( 91.64,112.75) --
	( 90.30,113.83) --
	( 88.96,115.17) --
	( 87.62,116.77) --
	( 86.28,118.60) --
	( 84.94,120.65) --
	( 83.60,122.90) --
	( 82.26,125.29) --
	( 80.92,127.78) --
	( 79.58,130.33) --
	( 78.24,132.88) --
	( 76.90,135.38) --
	( 75.56,137.81) --
	( 74.22,140.16) --
	( 72.88,142.45) --
	( 71.54,144.69) --
	( 70.20,146.93) --
	( 68.86,149.17) --
	( 67.52,151.46) --
	( 66.18,153.78) --
	( 64.84,156.15) --
	( 63.50,158.55) --
	( 62.16,160.99) --
	( 60.82,163.48) --
	( 59.48,166.04) --
	( 58.14,168.71) --
	( 56.80,171.52) --
	( 55.46,174.50) --
	( 54.12,177.62) --
	( 52.78,180.87) --
	( 51.44,184.16) --
	( 50.10,187.37) --
	( 48.76,190.40) --
	( 47.42,193.15) --
	( 46.08,195.57) --
	( 44.74,197.69) --
	( 43.40,199.55) --
	( 42.06,201.23) --
	( 40.72,202.80);
\definecolor{fillColor}{RGB}{86,180,233}

\path[fill=fillColor,fill opacity=0.10] ( 40.72,183.06) --
	( 42.06,179.57) --
	( 43.40,176.11) --
	( 44.74,172.75) --
	( 46.08,169.49) --
	( 47.42,166.30) --
	( 48.76,163.05) --
	( 50.10,159.51) --
	( 51.44,155.49) --
	( 52.78,150.85) --
	( 54.12,145.53) --
	( 55.46,139.56) --
	( 56.80,133.03) --
	( 58.14,126.11) --
	( 59.48,118.98) --
	( 60.82,111.83) --
	( 62.16,104.81) --
	( 63.50, 98.07) --
	( 64.84, 91.70) --
	( 66.18, 85.76) --
	( 67.52, 80.33) --
	( 68.86, 75.42) --
	( 70.20, 71.08) --
	( 71.54, 67.31) --
	( 72.88, 64.11) --
	( 74.22, 61.44) --
	( 75.56, 59.26) --
	( 76.90, 57.54) --
	( 78.24, 56.20) --
	( 79.58, 55.18) --
	( 80.92, 54.43) --
	( 82.26, 53.88) --
	( 83.60, 53.49) --
	( 84.94, 53.22) --
	( 86.28, 53.03) --
	( 87.62, 52.89) --
	( 88.96, 52.79) --
	( 90.30, 52.71) --
	( 91.64, 52.63) --
	( 92.98, 52.53) --
	( 94.32, 52.40) --
	( 95.66, 52.23) --
	( 97.00, 51.99) --
	( 98.34, 51.70) --
	( 99.68, 51.36) --
	(101.02, 50.97) --
	(102.36, 50.56) --
	(103.70, 50.15) --
	(105.04, 49.76) --
	(106.38, 49.39) --
	(107.72, 49.07) --
	(109.06, 48.78) --
	(110.40, 48.53) --
	(111.74, 48.31) --
	(113.08, 48.11) --
	(114.42, 47.93) --
	(115.76, 47.75) --
	(117.10, 47.60) --
	(118.44, 47.46) --
	(119.78, 47.34) --
	(121.12, 47.24) --
	(122.46, 47.16) --
	(123.80, 47.09) --
	(125.14, 47.02) --
	(126.48, 46.94) --
	(127.82, 46.85) --
	(129.16, 46.75) --
	(130.50, 46.64) --
	(131.84, 46.53) --
	(133.18, 46.43) --
	(134.52, 46.35) --
	(135.86, 46.32) --
	(137.20, 46.33) --
	(138.54, 46.40) --
	(139.88, 46.50) --
	(141.22, 46.65) --
	(142.56, 46.83) --
	(143.90, 47.03) --
	(145.24, 47.25) --
	(146.58, 47.49) --
	(147.92, 47.76) --
	(149.26, 48.07) --
	(150.60, 48.42) --
	(151.94, 48.81) --
	(153.28, 49.22) --
	(154.62, 49.64) --
	(155.96, 50.05) --
	(157.30, 50.41) --
	(158.64, 50.71) --
	(159.98, 50.95) --
	(161.32, 51.12) --
	(162.66, 51.27) --
	(164.00, 51.44) --
	(165.34, 51.68) --
	(166.68, 52.06) --
	(168.02, 52.57) --
	(169.36, 53.21) --
	(170.70, 53.95) --
	(172.04, 54.76) --
	(172.04, 43.09) --
	(170.70, 43.74) --
	(169.36, 44.31) --
	(168.02, 44.77) --
	(166.68, 45.08) --
	(165.34, 45.21) --
	(164.00, 45.18) --
	(162.66, 45.00) --
	(161.32, 44.75) --
	(159.98, 44.47) --
	(158.64, 44.18) --
	(157.30, 43.90) --
	(155.96, 43.63) --
	(154.62, 43.37) --
	(153.28, 43.10) --
	(151.94, 42.82) --
	(150.60, 42.51) --
	(149.26, 42.18) --
	(147.92, 41.85) --
	(146.58, 41.53) --
	(145.24, 41.25) --
	(143.90, 41.01) --
	(142.56, 40.82) --
	(141.22, 40.69) --
	(139.88, 40.60) --
	(138.54, 40.56) --
	(137.20, 40.54) --
	(135.86, 40.54) --
	(134.52, 40.55) --
	(133.18, 40.56) --
	(131.84, 40.59) --
	(130.50, 40.63) --
	(129.16, 40.68) --
	(127.82, 40.76) --
	(126.48, 40.85) --
	(125.14, 40.95) --
	(123.80, 41.06) --
	(122.46, 41.15) --
	(121.12, 41.24) --
	(119.78, 41.32) --
	(118.44, 41.38) --
	(117.10, 41.45) --
	(115.76, 41.54) --
	(114.42, 41.65) --
	(113.08, 41.80) --
	(111.74, 41.99) --
	(110.40, 42.21) --
	(109.06, 42.48) --
	(107.72, 42.76) --
	(106.38, 43.07) --
	(105.04, 43.38) --
	(103.70, 43.68) --
	(102.36, 43.99) --
	(101.02, 44.28) --
	( 99.68, 44.57) --
	( 98.34, 44.85) --
	( 97.00, 45.12) --
	( 95.66, 45.36) --
	( 94.32, 45.57) --
	( 92.98, 45.74) --
	( 91.64, 45.86) --
	( 90.30, 45.94) --
	( 88.96, 45.98) --
	( 87.62, 46.02) --
	( 86.28, 46.09) --
	( 84.94, 46.21) --
	( 83.60, 46.44) --
	( 82.26, 46.81) --
	( 80.92, 47.35) --
	( 79.58, 48.11) --
	( 78.24, 49.11) --
	( 76.90, 50.38) --
	( 75.56, 51.95) --
	( 74.22, 53.88) --
	( 72.88, 56.21) --
	( 71.54, 58.99) --
	( 70.20, 62.30) --
	( 68.86, 66.18) --
	( 67.52, 70.67) --
	( 66.18, 75.79) --
	( 64.84, 81.49) --
	( 63.50, 87.70) --
	( 62.16, 94.27) --
	( 60.82,101.05) --
	( 59.48,107.85) --
	( 58.14,114.52) --
	( 56.80,120.96) --
	( 55.46,127.08) --
	( 54.12,132.84) --
	( 52.78,138.19) --
	( 51.44,143.10) --
	( 50.10,147.52) --
	( 48.76,151.37) --
	( 47.42,154.56) --
	( 46.08,157.04) --
	( 44.74,158.83) --
	( 43.40,160.07) --
	( 42.06,160.94) --
	( 40.72,161.58) --
	cycle;

\path[] ( 40.72,183.06) --
	( 42.06,179.57) --
	( 43.40,176.11) --
	( 44.74,172.75) --
	( 46.08,169.49) --
	( 47.42,166.30) --
	( 48.76,163.05) --
	( 50.10,159.51) --
	( 51.44,155.49) --
	( 52.78,150.85) --
	( 54.12,145.53) --
	( 55.46,139.56) --
	( 56.80,133.03) --
	( 58.14,126.11) --
	( 59.48,118.98) --
	( 60.82,111.83) --
	( 62.16,104.81) --
	( 63.50, 98.07) --
	( 64.84, 91.70) --
	( 66.18, 85.76) --
	( 67.52, 80.33) --
	( 68.86, 75.42) --
	( 70.20, 71.08) --
	( 71.54, 67.31) --
	( 72.88, 64.11) --
	( 74.22, 61.44) --
	( 75.56, 59.26) --
	( 76.90, 57.54) --
	( 78.24, 56.20) --
	( 79.58, 55.18) --
	( 80.92, 54.43) --
	( 82.26, 53.88) --
	( 83.60, 53.49) --
	( 84.94, 53.22) --
	( 86.28, 53.03) --
	( 87.62, 52.89) --
	( 88.96, 52.79) --
	( 90.30, 52.71) --
	( 91.64, 52.63) --
	( 92.98, 52.53) --
	( 94.32, 52.40) --
	( 95.66, 52.23) --
	( 97.00, 51.99) --
	( 98.34, 51.70) --
	( 99.68, 51.36) --
	(101.02, 50.97) --
	(102.36, 50.56) --
	(103.70, 50.15) --
	(105.04, 49.76) --
	(106.38, 49.39) --
	(107.72, 49.07) --
	(109.06, 48.78) --
	(110.40, 48.53) --
	(111.74, 48.31) --
	(113.08, 48.11) --
	(114.42, 47.93) --
	(115.76, 47.75) --
	(117.10, 47.60) --
	(118.44, 47.46) --
	(119.78, 47.34) --
	(121.12, 47.24) --
	(122.46, 47.16) --
	(123.80, 47.09) --
	(125.14, 47.02) --
	(126.48, 46.94) --
	(127.82, 46.85) --
	(129.16, 46.75) --
	(130.50, 46.64) --
	(131.84, 46.53) --
	(133.18, 46.43) --
	(134.52, 46.35) --
	(135.86, 46.32) --
	(137.20, 46.33) --
	(138.54, 46.40) --
	(139.88, 46.50) --
	(141.22, 46.65) --
	(142.56, 46.83) --
	(143.90, 47.03) --
	(145.24, 47.25) --
	(146.58, 47.49) --
	(147.92, 47.76) --
	(149.26, 48.07) --
	(150.60, 48.42) --
	(151.94, 48.81) --
	(153.28, 49.22) --
	(154.62, 49.64) --
	(155.96, 50.05) --
	(157.30, 50.41) --
	(158.64, 50.71) --
	(159.98, 50.95) --
	(161.32, 51.12) --
	(162.66, 51.27) --
	(164.00, 51.44) --
	(165.34, 51.68) --
	(166.68, 52.06) --
	(168.02, 52.57) --
	(169.36, 53.21) --
	(170.70, 53.95) --
	(172.04, 54.76);

\path[] (172.04, 43.09) --
	(170.70, 43.74) --
	(169.36, 44.31) --
	(168.02, 44.77) --
	(166.68, 45.08) --
	(165.34, 45.21) --
	(164.00, 45.18) --
	(162.66, 45.00) --
	(161.32, 44.75) --
	(159.98, 44.47) --
	(158.64, 44.18) --
	(157.30, 43.90) --
	(155.96, 43.63) --
	(154.62, 43.37) --
	(153.28, 43.10) --
	(151.94, 42.82) --
	(150.60, 42.51) --
	(149.26, 42.18) --
	(147.92, 41.85) --
	(146.58, 41.53) --
	(145.24, 41.25) --
	(143.90, 41.01) --
	(142.56, 40.82) --
	(141.22, 40.69) --
	(139.88, 40.60) --
	(138.54, 40.56) --
	(137.20, 40.54) --
	(135.86, 40.54) --
	(134.52, 40.55) --
	(133.18, 40.56) --
	(131.84, 40.59) --
	(130.50, 40.63) --
	(129.16, 40.68) --
	(127.82, 40.76) --
	(126.48, 40.85) --
	(125.14, 40.95) --
	(123.80, 41.06) --
	(122.46, 41.15) --
	(121.12, 41.24) --
	(119.78, 41.32) --
	(118.44, 41.38) --
	(117.10, 41.45) --
	(115.76, 41.54) --
	(114.42, 41.65) --
	(113.08, 41.80) --
	(111.74, 41.99) --
	(110.40, 42.21) --
	(109.06, 42.48) --
	(107.72, 42.76) --
	(106.38, 43.07) --
	(105.04, 43.38) --
	(103.70, 43.68) --
	(102.36, 43.99) --
	(101.02, 44.28) --
	( 99.68, 44.57) --
	( 98.34, 44.85) --
	( 97.00, 45.12) --
	( 95.66, 45.36) --
	( 94.32, 45.57) --
	( 92.98, 45.74) --
	( 91.64, 45.86) --
	( 90.30, 45.94) --
	( 88.96, 45.98) --
	( 87.62, 46.02) --
	( 86.28, 46.09) --
	( 84.94, 46.21) --
	( 83.60, 46.44) --
	( 82.26, 46.81) --
	( 80.92, 47.35) --
	( 79.58, 48.11) --
	( 78.24, 49.11) --
	( 76.90, 50.38) --
	( 75.56, 51.95) --
	( 74.22, 53.88) --
	( 72.88, 56.21) --
	( 71.54, 58.99) --
	( 70.20, 62.30) --
	( 68.86, 66.18) --
	( 67.52, 70.67) --
	( 66.18, 75.79) --
	( 64.84, 81.49) --
	( 63.50, 87.70) --
	( 62.16, 94.27) --
	( 60.82,101.05) --
	( 59.48,107.85) --
	( 58.14,114.52) --
	( 56.80,120.96) --
	( 55.46,127.08) --
	( 54.12,132.84) --
	( 52.78,138.19) --
	( 51.44,143.10) --
	( 50.10,147.52) --
	( 48.76,151.37) --
	( 47.42,154.56) --
	( 46.08,157.04) --
	( 44.74,158.83) --
	( 43.40,160.07) --
	( 42.06,160.94) --
	( 40.72,161.58);
\definecolor{drawColor}{RGB}{230,159,0}

\path[draw=drawColor,line width= 0.6pt,line join=round] ( 40.72,220.20) --
	( 42.06,217.31) --
	( 43.40,214.33) --
	( 44.74,211.27) --
	( 46.08,208.14) --
	( 47.42,204.96) --
	( 48.76,201.75) --
	( 50.10,198.54) --
	( 51.44,195.35) --
	( 52.78,192.20) --
	( 54.12,189.11) --
	( 55.46,186.10) --
	( 56.80,183.19) --
	( 58.14,180.37) --
	( 59.48,177.67) --
	( 60.82,175.07) --
	( 62.16,172.57) --
	( 63.50,170.15) --
	( 64.84,167.81) --
	( 66.18,165.53) --
	( 67.52,163.28) --
	( 68.86,161.04) --
	( 70.20,158.79) --
	( 71.54,156.53) --
	( 72.88,154.22) --
	( 74.22,151.86) --
	( 75.56,149.46) --
	( 76.90,147.00) --
	( 78.24,144.51) --
	( 79.58,142.00) --
	( 80.92,139.49) --
	( 82.26,137.03) --
	( 83.60,134.63) --
	( 84.94,132.35) --
	( 86.28,130.22) --
	( 87.62,128.30) --
	( 88.96,126.60) --
	( 90.30,125.18) --
	( 91.64,124.06) --
	( 92.98,123.26) --
	( 94.32,122.80) --
	( 95.66,122.69) --
	( 97.00,122.93) --
	( 98.34,123.51) --
	( 99.68,124.41) --
	(101.02,125.61) --
	(102.36,127.07) --
	(103.70,128.75) --
	(105.04,130.61) --
	(106.38,132.60) --
	(107.72,134.66) --
	(109.06,136.74) --
	(110.40,138.80) --
	(111.74,140.79) --
	(113.08,142.66) --
	(114.42,144.39) --
	(115.76,145.94) --
	(117.10,147.29) --
	(118.44,148.43) --
	(119.78,149.36) --
	(121.12,150.07) --
	(122.46,150.56) --
	(123.80,150.83) --
	(125.14,150.90) --
	(126.48,150.76) --
	(127.82,150.42) --
	(129.16,149.89) --
	(130.50,149.15) --
	(131.84,148.22) --
	(133.18,147.09) --
	(134.52,145.76) --
	(135.86,144.22) --
	(137.20,142.47) --
	(138.54,140.53) --
	(139.88,138.39) --
	(141.22,136.07) --
	(142.56,133.59) --
	(143.90,130.97) --
	(145.24,128.25) --
	(146.58,125.47) --
	(147.92,122.68) --
	(149.26,119.91) --
	(150.60,117.22) --
	(151.94,114.64) --
	(153.28,112.24) --
	(154.62,110.03) --
	(155.96,108.05) --
	(157.30,106.31) --
	(158.64,104.84) --
	(159.98,103.63) --
	(161.32,102.68) --
	(162.66,101.97) --
	(164.00,101.47) --
	(165.34,101.16) --
	(166.68,101.00) --
	(168.02,100.95) --
	(169.36,100.98) --
	(170.70,101.06) --
	(172.04,101.15);
\definecolor{drawColor}{RGB}{86,180,233}

\path[draw=drawColor,line width= 0.6pt,line join=round] ( 40.72,172.32) --
	( 42.06,170.25) --
	( 43.40,168.09) --
	( 44.74,165.79) --
	( 46.08,163.27) --
	( 47.42,160.43) --
	( 48.76,157.21) --
	( 50.10,153.51) --
	( 51.44,149.30) --
	( 52.78,144.52) --
	( 54.12,139.18) --
	( 55.46,133.32) --
	( 56.80,127.00) --
	( 58.14,120.32) --
	( 59.48,113.41) --
	( 60.82,106.44) --
	( 62.16, 99.54) --
	( 63.50, 92.89) --
	( 64.84, 86.60) --
	( 66.18, 80.78) --
	( 67.52, 75.50) --
	( 68.86, 70.80) --
	( 70.20, 66.69) --
	( 71.54, 63.15) --
	( 72.88, 60.16) --
	( 74.22, 57.66) --
	( 75.56, 55.61) --
	( 76.90, 53.96) --
	( 78.24, 52.65) --
	( 79.58, 51.65) --
	( 80.92, 50.89) --
	( 82.26, 50.34) --
	( 83.60, 49.97) --
	( 84.94, 49.71) --
	( 86.28, 49.56) --
	( 87.62, 49.46) --
	( 88.96, 49.39) --
	( 90.30, 49.32) --
	( 91.64, 49.25) --
	( 92.98, 49.14) --
	( 94.32, 48.99) --
	( 95.66, 48.80) --
	( 97.00, 48.56) --
	( 98.34, 48.28) --
	( 99.68, 47.96) --
	(101.02, 47.63) --
	(102.36, 47.27) --
	(103.70, 46.92) --
	(105.04, 46.57) --
	(106.38, 46.23) --
	(107.72, 45.92) --
	(109.06, 45.63) --
	(110.40, 45.37) --
	(111.74, 45.15) --
	(113.08, 44.95) --
	(114.42, 44.79) --
	(115.76, 44.65) --
	(117.10, 44.53) --
	(118.44, 44.42) --
	(119.78, 44.33) --
	(121.12, 44.24) --
	(122.46, 44.16) --
	(123.80, 44.07) --
	(125.14, 43.99) --
	(126.48, 43.90) --
	(127.82, 43.81) --
	(129.16, 43.72) --
	(130.50, 43.63) --
	(131.84, 43.56) --
	(133.18, 43.49) --
	(134.52, 43.45) --
	(135.86, 43.43) --
	(137.20, 43.44) --
	(138.54, 43.48) --
	(139.88, 43.55) --
	(141.22, 43.67) --
	(142.56, 43.82) --
	(143.90, 44.02) --
	(145.24, 44.25) --
	(146.58, 44.51) --
	(147.92, 44.81) --
	(149.26, 45.13) --
	(150.60, 45.46) --
	(151.94, 45.81) --
	(153.28, 46.16) --
	(154.62, 46.51) --
	(155.96, 46.84) --
	(157.30, 47.16) --
	(158.64, 47.45) --
	(159.98, 47.71) --
	(161.32, 47.94) --
	(162.66, 48.14) --
	(164.00, 48.31) --
	(165.34, 48.45) --
	(166.68, 48.57) --
	(168.02, 48.67) --
	(169.36, 48.76) --
	(170.70, 48.84) --
	(172.04, 48.92);
\definecolor{drawColor}{RGB}{230,159,0}

\path[draw=drawColor,line width= 0.6pt,line join=round] ( 40.72,220.20) --
	( 42.06,217.31) --
	( 43.40,214.33) --
	( 44.74,211.27) --
	( 46.08,208.14) --
	( 47.42,204.96) --
	( 48.76,201.75) --
	( 50.10,198.54) --
	( 51.44,195.35) --
	( 52.78,192.20) --
	( 54.12,189.11) --
	( 55.46,186.10) --
	( 56.80,183.19) --
	( 58.14,180.37) --
	( 59.48,177.67) --
	( 60.82,175.07) --
	( 62.16,172.57) --
	( 63.50,170.15) --
	( 64.84,167.81) --
	( 66.18,165.53) --
	( 67.52,163.28) --
	( 68.86,161.04) --
	( 70.20,158.79) --
	( 71.54,156.53) --
	( 72.88,154.22) --
	( 74.22,151.86) --
	( 75.56,149.46) --
	( 76.90,147.00) --
	( 78.24,144.51) --
	( 79.58,142.00) --
	( 80.92,139.49) --
	( 82.26,137.03) --
	( 83.60,134.63) --
	( 84.94,132.35) --
	( 86.28,130.22) --
	( 87.62,128.30) --
	( 88.96,126.60) --
	( 90.30,125.18) --
	( 91.64,124.06) --
	( 92.98,123.26) --
	( 94.32,122.80) --
	( 95.66,122.69) --
	( 97.00,122.93) --
	( 98.34,123.51) --
	( 99.68,124.41) --
	(101.02,125.61) --
	(102.36,127.07) --
	(103.70,128.75) --
	(105.04,130.61) --
	(106.38,132.60) --
	(107.72,134.66) --
	(109.06,136.74) --
	(110.40,138.80) --
	(111.74,140.79) --
	(113.08,142.66) --
	(114.42,144.39) --
	(115.76,145.94) --
	(117.10,147.29) --
	(118.44,148.43) --
	(119.78,149.36) --
	(121.12,150.07) --
	(122.46,150.56) --
	(123.80,150.83) --
	(125.14,150.90) --
	(126.48,150.76) --
	(127.82,150.42) --
	(129.16,149.89) --
	(130.50,149.15) --
	(131.84,148.22) --
	(133.18,147.09) --
	(134.52,145.76) --
	(135.86,144.22) --
	(137.20,142.47) --
	(138.54,140.53) --
	(139.88,138.39) --
	(141.22,136.07) --
	(142.56,133.59) --
	(143.90,130.97) --
	(145.24,128.25) --
	(146.58,125.47) --
	(147.92,122.68) --
	(149.26,119.91) --
	(150.60,117.22) --
	(151.94,114.64) --
	(153.28,112.24) --
	(154.62,110.03) --
	(155.96,108.05) --
	(157.30,106.31) --
	(158.64,104.84) --
	(159.98,103.63) --
	(161.32,102.68) --
	(162.66,101.97) --
	(164.00,101.47) --
	(165.34,101.16) --
	(166.68,101.00) --
	(168.02,100.95) --
	(169.36,100.98) --
	(170.70,101.06) --
	(172.04,101.15);
\definecolor{drawColor}{RGB}{86,180,233}

\path[draw=drawColor,line width= 0.6pt,line join=round] ( 40.72,172.32) --
	( 42.06,170.25) --
	( 43.40,168.09) --
	( 44.74,165.79) --
	( 46.08,163.27) --
	( 47.42,160.43) --
	( 48.76,157.21) --
	( 50.10,153.51) --
	( 51.44,149.30) --
	( 52.78,144.52) --
	( 54.12,139.18) --
	( 55.46,133.32) --
	( 56.80,127.00) --
	( 58.14,120.32) --
	( 59.48,113.41) --
	( 60.82,106.44) --
	( 62.16, 99.54) --
	( 63.50, 92.89) --
	( 64.84, 86.60) --
	( 66.18, 80.78) --
	( 67.52, 75.50) --
	( 68.86, 70.80) --
	( 70.20, 66.69) --
	( 71.54, 63.15) --
	( 72.88, 60.16) --
	( 74.22, 57.66) --
	( 75.56, 55.61) --
	( 76.90, 53.96) --
	( 78.24, 52.65) --
	( 79.58, 51.65) --
	( 80.92, 50.89) --
	( 82.26, 50.34) --
	( 83.60, 49.97) --
	( 84.94, 49.71) --
	( 86.28, 49.56) --
	( 87.62, 49.46) --
	( 88.96, 49.39) --
	( 90.30, 49.32) --
	( 91.64, 49.25) --
	( 92.98, 49.14) --
	( 94.32, 48.99) --
	( 95.66, 48.80) --
	( 97.00, 48.56) --
	( 98.34, 48.28) --
	( 99.68, 47.96) --
	(101.02, 47.63) --
	(102.36, 47.27) --
	(103.70, 46.92) --
	(105.04, 46.57) --
	(106.38, 46.23) --
	(107.72, 45.92) --
	(109.06, 45.63) --
	(110.40, 45.37) --
	(111.74, 45.15) --
	(113.08, 44.95) --
	(114.42, 44.79) --
	(115.76, 44.65) --
	(117.10, 44.53) --
	(118.44, 44.42) --
	(119.78, 44.33) --
	(121.12, 44.24) --
	(122.46, 44.16) --
	(123.80, 44.07) --
	(125.14, 43.99) --
	(126.48, 43.90) --
	(127.82, 43.81) --
	(129.16, 43.72) --
	(130.50, 43.63) --
	(131.84, 43.56) --
	(133.18, 43.49) --
	(134.52, 43.45) --
	(135.86, 43.43) --
	(137.20, 43.44) --
	(138.54, 43.48) --
	(139.88, 43.55) --
	(141.22, 43.67) --
	(142.56, 43.82) --
	(143.90, 44.02) --
	(145.24, 44.25) --
	(146.58, 44.51) --
	(147.92, 44.81) --
	(149.26, 45.13) --
	(150.60, 45.46) --
	(151.94, 45.81) --
	(153.28, 46.16) --
	(154.62, 46.51) --
	(155.96, 46.84) --
	(157.30, 47.16) --
	(158.64, 47.45) --
	(159.98, 47.71) --
	(161.32, 47.94) --
	(162.66, 48.14) --
	(164.00, 48.31) --
	(165.34, 48.45) --
	(166.68, 48.57) --
	(168.02, 48.67) --
	(169.36, 48.76) --
	(170.70, 48.84) --
	(172.04, 48.92);
\definecolor{drawColor}{gray}{0.20}

\path[draw=drawColor,line width= 0.6pt,line join=round,line cap=round] ( 34.16, 30.69) rectangle (178.61,247.45);
\end{scope}
\begin{scope}
\path[clip] (  0.00,  0.00) rectangle (252.94,252.94);
\definecolor{drawColor}{gray}{0.30}

\node[text=drawColor,anchor=base east,inner sep=0pt, outer sep=0pt, scale=  0.88] at ( 29.21, 68.99) {0.2};

\node[text=drawColor,anchor=base east,inner sep=0pt, outer sep=0pt, scale=  0.88] at ( 29.21,120.46) {0.4};

\node[text=drawColor,anchor=base east,inner sep=0pt, outer sep=0pt, scale=  0.88] at ( 29.21,171.93) {0.6};

\node[text=drawColor,anchor=base east,inner sep=0pt, outer sep=0pt, scale=  0.88] at ( 29.21,223.41) {0.8};
\end{scope}
\begin{scope}
\path[clip] (  0.00,  0.00) rectangle (252.94,252.94);
\definecolor{drawColor}{gray}{0.20}

\path[draw=drawColor,line width= 0.6pt,line join=round] ( 31.41, 72.02) --
	( 34.16, 72.02);

\path[draw=drawColor,line width= 0.6pt,line join=round] ( 31.41,123.49) --
	( 34.16,123.49);

\path[draw=drawColor,line width= 0.6pt,line join=round] ( 31.41,174.96) --
	( 34.16,174.96);

\path[draw=drawColor,line width= 0.6pt,line join=round] ( 31.41,226.44) --
	( 34.16,226.44);
\end{scope}
\begin{scope}
\path[clip] (  0.00,  0.00) rectangle (252.94,252.94);
\definecolor{drawColor}{gray}{0.20}

\path[draw=drawColor,line width= 0.6pt,line join=round] ( 39.38, 27.94) --
	( 39.38, 30.69);

\path[draw=drawColor,line width= 0.6pt,line join=round] ( 72.88, 27.94) --
	( 72.88, 30.69);

\path[draw=drawColor,line width= 0.6pt,line join=round] (106.38, 27.94) --
	(106.38, 30.69);

\path[draw=drawColor,line width= 0.6pt,line join=round] (139.88, 27.94) --
	(139.88, 30.69);

\path[draw=drawColor,line width= 0.6pt,line join=round] (173.38, 27.94) --
	(173.38, 30.69);
\end{scope}
\begin{scope}
\path[clip] (  0.00,  0.00) rectangle (252.94,252.94);
\definecolor{drawColor}{gray}{0.30}

\node[text=drawColor,anchor=base,inner sep=0pt, outer sep=0pt, scale=  0.88] at ( 39.38, 19.68) {0};

\node[text=drawColor,anchor=base,inner sep=0pt, outer sep=0pt, scale=  0.88] at ( 72.88, 19.68) {25};

\node[text=drawColor,anchor=base,inner sep=0pt, outer sep=0pt, scale=  0.88] at (106.38, 19.68) {50};

\node[text=drawColor,anchor=base,inner sep=0pt, outer sep=0pt, scale=  0.88] at (139.88, 19.68) {75};

\node[text=drawColor,anchor=base,inner sep=0pt, outer sep=0pt, scale=  0.88] at (173.38, 19.68) {100};
\end{scope}
\begin{scope}
\path[clip] (  0.00,  0.00) rectangle (252.94,252.94);
\definecolor{drawColor}{RGB}{0,0,0}

\node[text=drawColor,anchor=base,inner sep=0pt, outer sep=0pt, scale=  1.10] at (106.38,  7.64) {Cardinality};
\end{scope}
\begin{scope}
\path[clip] (  0.00,  0.00) rectangle (252.94,252.94);
\definecolor{drawColor}{RGB}{0,0,0}

\node[text=drawColor,rotate= 90.00,anchor=base,inner sep=0pt, outer sep=0pt, scale=  1.10] at ( 13.08,139.07) {Production Error Rate};
\end{scope}
\begin{scope}
\path[clip] (  0.00,  0.00) rectangle (252.94,252.94);
\definecolor{fillColor}{RGB}{255,255,255}

\path[fill=fillColor] (189.61,111.50) rectangle (247.44,166.63);
\end{scope}
\begin{scope}
\path[clip] (  0.00,  0.00) rectangle (252.94,252.94);
\definecolor{drawColor}{RGB}{0,0,0}

\node[text=drawColor,anchor=base west,inner sep=0pt, outer sep=0pt, scale=  1.10] at (195.11,152.48) {Group};
\end{scope}
\begin{scope}
\path[clip] (  0.00,  0.00) rectangle (252.94,252.94);
\definecolor{fillColor}{RGB}{255,255,255}

\path[fill=fillColor] (195.11,131.46) rectangle (209.56,145.91);
\end{scope}
\begin{scope}
\path[clip] (  0.00,  0.00) rectangle (252.94,252.94);
\definecolor{fillColor}{RGB}{230,159,0}

\path[fill=fillColor,fill opacity=0.10] (195.82,132.17) rectangle (208.85,145.20);
\end{scope}
\begin{scope}
\path[clip] (  0.00,  0.00) rectangle (252.94,252.94);
\definecolor{drawColor}{RGB}{230,159,0}

\path[draw=drawColor,line width= 0.6pt,line join=round] (196.55,138.69) -- (208.12,138.69);
\end{scope}
\begin{scope}
\path[clip] (  0.00,  0.00) rectangle (252.94,252.94);
\definecolor{drawColor}{RGB}{230,159,0}

\path[draw=drawColor,line width= 0.6pt,line join=round] (196.55,138.69) -- (208.12,138.69);
\end{scope}
\begin{scope}
\path[clip] (  0.00,  0.00) rectangle (252.94,252.94);
\definecolor{fillColor}{RGB}{255,255,255}

\path[fill=fillColor] (195.11,117.00) rectangle (209.56,131.46);
\end{scope}
\begin{scope}
\path[clip] (  0.00,  0.00) rectangle (252.94,252.94);
\definecolor{fillColor}{RGB}{86,180,233}

\path[fill=fillColor,fill opacity=0.10] (195.82,117.72) rectangle (208.85,130.75);
\end{scope}
\begin{scope}
\path[clip] (  0.00,  0.00) rectangle (252.94,252.94);
\definecolor{drawColor}{RGB}{86,180,233}

\path[draw=drawColor,line width= 0.6pt,line join=round] (196.55,124.23) -- (208.12,124.23);
\end{scope}
\begin{scope}
\path[clip] (  0.00,  0.00) rectangle (252.94,252.94);
\definecolor{drawColor}{RGB}{86,180,233}

\path[draw=drawColor,line width= 0.6pt,line join=round] (196.55,124.23) -- (208.12,124.23);
\end{scope}
\begin{scope}
\path[clip] (  0.00,  0.00) rectangle (252.94,252.94);
\definecolor{drawColor}{RGB}{0,0,0}

\node[text=drawColor,anchor=base west,inner sep=0pt, outer sep=0pt, scale=  0.88] at (215.06,135.65) {IA};
\end{scope}
\begin{scope}
\path[clip] (  0.00,  0.00) rectangle (252.94,252.94);
\definecolor{drawColor}{RGB}{0,0,0}

\node[text=drawColor,anchor=base west,inner sep=0pt, outer sep=0pt, scale=  0.88] at (215.06,121.20) {non-IA};
\end{scope}
\end{tikzpicture}

%% file: discussion.tex
\section{Discussion and outlook}

The studies presented in the previous section highlighted a number of properties of Indo-Aryan numeral systems:
\begin{itemize}
\item that Indo-Aryan languages are more complex on average than numeral systems found elsewhere;
\item that while the role of specific sociocultural factors in mediating the maintenance of this complexity remains inconclusive, complexity was retained largely in the core Indo-Aryan area; and 
\item that Indo-Aryan systems, despite their higher overall complexity, bear signs of being structured to facilitate some degree of communicative efficiency. 
\end{itemize}
We used relatively simple metrics --- among which there was high agreement --- as a representation of irregularity, which we leave underspecified as a proxy for difficulty of production and/or processing. 
It is likely that more sophisticated models capable of capturing details of between-form relationships \cite[e.g.,][]{beniamine2021multiple,guzman2024analogical} will quantify complexity similarly across global samples of numeral systems. 

Our findings are reconcilable with recent results on the formal complexity of numeral systems, which apply different metrics to manually morphologically glossed data from a diverse range of languages. 
\citet{rubehn2025annotating}, working with terms for the numerals 1--40, find that two Indo-Aryan languages, Hindi and Assamese, have a higher number of numeral allomorphs than other languages, similar to our finding that Indo-Aryan numeral systems have higher minimum description length than others. 
Minimum description length also bears a resemblance to the notion of numeral lexicon size, used by \citet{denic2024recursive} to operationalize complexity; strangely, however, these authors do not take into account allomorphy within numeral systems, and hence arrive at much lower complexity values than ours. 
\citet{koile2025decimal}, restricting analyses to the numerals 1--30 plus the decades from 40 to 90, situate numeral systems according to their transparency (whether component forms can be identified within a form for a numeral term), canonicity (whether component values can be identified within the gloss of a numeral term), and compositionality (whether component glosses can be identified within the gloss of a numeral term). 
Of these, transparency, which detects allomorphy within numeral systems, is most translatable into the surprisal, production accuracy, and comprehension accuracy metrics used in this paper, which are however more flexible and granular than the quasi-binary classification used by the authors: their transparency criterion as stated would treat German {\it einundzwanzig} `21' as non-transparent, as it does not contain the string {\it eins} `1'. On the other hand, because {\it ein} is a frequent trigram across German numerals 1--99, occurring in all forms with 1 in the digits place, {\it einundzwanzig} 
emerges 
as a relatively predictable (i.e., regular and transparent) form according to the metrics we use in this paper. 
The metrics we use also tangentially capture some properties of non-canonicity, which involves fused, unanalyzable portmanteau morphemes and encompasses both decade forms like Spanish {\it cuarenta} `40' as well as certain irregular teen forms (e.g., {\it twelve}, Odia {\it \'soh\d{l}a} `16'). Forms like {\it twelve} and {\it \'soh\d{l}a} display high complexity under our metrics, as they are phonotactically unpredictable and do not display strong form-meaning cues corresponding to their underlying semantic representation; however, our methods do not actually distinguish between non-transparency and non-canonicity of this type, as the operationalization of canonicity employed by \citet{koile2025decimal} depends on manual morphological segmentation, which we eschew.\footnote{Apart from the MDL metric, which carries out automated segmentation with no glossing, our metrics generally do not rely explicitly on the classical construct of the morpheme. As such, they treat different types of unpredictable allomorphy and suppletion (e.g., segmentable {\it thir-teen} vs.\ non-segmentable {\it twelve}) as exhibiting the same type of irregularity, but also sidestep difficulties encountered in manual morphological glossing, namely the existence of competing valid analyses. As an example, it is not immediately clear why Odia {\it \'soh\d{l}a} `16' should be treated as an unsegmentable portmanteau form and not segmented {\it \'soh-\d{l}a}, since {\it \'soh-} and {\it -\d{l}a} are are etymologically related to other allomorphs for `6' and `10'. Our approach obviates decisions of this sort.} 
Non-canonical decades (e.g., {\it cuarenta}) would not stand out as irregular according to our metrics, as they are phonologically more predictable and more strongly associated with their underlying meanings, recurring in at least ten forms (additionally, when mapping between phonological and semantic cues, we assume holistic semantic representations for the portion of the meaning corresponding to the tens place, so in a sense the regularity of non-canonical decade forms is a model artifact). 
As it stands, however, repeated non-canonical forms are not particularly relevant to the focus of this paper, which concerns itself with the burden that forms in a numeral system place on a user due to their unpredictability. 

We believe that the inclusion of Indo-Aryan data in future research on cross-linguistic numeral systems will continue to offer insights into the nature of different types of complexity. 
We have shown that while Indo-Aryan numeral systems are more complex than others, they still seem to fall in line with some general properties of numeral systems, namely the frequency/regularity tradeoff, albeit as rather extreme outliers. 
A comparable observation can be gleaned from \citet{schneider2020children}, one of a very small number of cross-linguistic studies on numeral term acquisition including children learning an Indo-Aryan language: like learners of other languages, children acquiring Indo-Aryan languages make use of the successor function when learning productive counting, but show lower successor function performance than children speaking other languages; nonetheless, successor function performance remains a strong predictor of mastery of productive counting. 
In a similar vein, \citet{xu2020numeral} demonstrate that languages provide near-optimal solutions to tradeoffs between informativeness (whether numeral terms refer to individual quantities, weighted according to their need probability) and complexity (operationalized according to a rule-based framework), occupying a region below a Pareto frontier representing the optimal balance between these two variables. Although considerably higher in complexity than the most complex language included in analyses (Georgian), an Indo-Aryan language such as Hindi/Urdu would likely still be found in this region representing near-optimal tradeoffs, but would be an extreme outlier. 
Models of language learning and use can accommodate and should take into account such typological outliers (for a similar appeal made on the basis of Danish phonology, see \citealt{trecca2021danish}).








Many obstacles remain in the way of attempts to gain a conclusive understanding of the local historical contingencies that gave rise to the patterns we see in South Asia. Our analyses showed that they are not compatible with the idea that complexity is fostered and increased in more isolated regions (e.g., of higher altitude). 
At the same time, it is hard to identify the triggers directly responsible for fostering complexity in Indo-Aryan systems. 
It is worth noting that when compared to other ancient Indo-European languages (e.g., Latin, Ancient Greek, etc.), Old Indo-Aryan numerals exhibit a higher degree of allomorphy, both predictable and unpredictable. 
This is a consequence of sandhi, obligatory phonological alternations applying across morpheme and word boundaries. 
Similar sandhi rules are also found in Avestan (albeit less so across word boundaries), the oldest well-attested Iranian language, but middle and modern Iranian languages do not show high irregularity in numeral systems, apart from irregular teens and non-canonical decades; they generally exhibit transparent forms beyond 20 (e.g., Persian {\it se} `3', {\it sizdah} `13', but {\it bist o se} `23'), despite likely descending from an Old Iranian language like Avestan which would have had sandhi. 
However, from what we can glean from the textual record, Avestan does not represent complex numerals as compounds (as in Sanskrit), but with the enclitic coordinator {\it -ca} `and' 
(e.g., {\it pa\d{n}c\=a-ca v\={\i}sa\textsuperscript{i}ti} `25', lit.\ `five-and twenty'; \citealt{Skjærvø2007}) --- this would have drastically reduced the amount of allomorphic variation found in the system (in comparison to Old Indo-Aryan), as numeral bases would not be affected by sandhi following {\it -ca} plus a word boundary \citep[110--12]{Hoffmann-Forssman}. 
Had Old Indo-Aryan adopted a similar way of expressing numbers as its primary strategy (numeral constructions of this type appear to have been in use in Sanskrit, but not as frequently as single-word expressions; \citealt{petrocchi2022morphosyntax}), it is possible that its daughter languages would not have developed the degree of complexity that they display today. 
Other questions remain: did northwestern Indo-Aryan languages,\footnote{The languages in question belong to the so-called Dardic group. It is not entirely clear whether these languages form a legitimate subgroup defined by a shared history involving innovations not found in other Indo-Aryan languages. There is additionally debate as to whether Kashmiri, which exhibits a decimal numeral system of medium complexity (with more transparent morpheme boundaries than in e.g.\ Hindi, but with a degree of allomorphy as well as subtractive numerals), belongs to this group \citep{kogan2016}.} most of which have simpler, vigesimal systems, lose their irregularity, or never develop it in the first place? 
Ultimately, it will be difficult to advance these theories beyond the realm of speculation. 


All the more reason to focus on the present. 
Psycholinguistic and neurolinguistic experiments can shed light on the mechanisms involved in the production and comprehension of these words. 
Supplementing existing Indo-Aryan resources \citep{baker2002emille,nishioka2016development} with detailed naturalistic speech corpora will help to explore inter- and intra-individual variation in the realization of these forms, pointing to vulnerabilities in these systems. 
And what of the future? 
When will this roughly millennium-old state of affairs end, and which forms will serve as the locus of change, either via analogical remodeling or replacement (e.g., with English borrowings)? 
Though this system is a product of local historical and cultural idiosyncrasies, 
it is 
an extreme example of the degree of complexity that users are willing to tolerate. 
In a sense, the message of this paper to take Indo-Aryan languages into account in analyses of numeral systems can be reframed as a call to take integrative complexity into account. 
This dimension of complexity deserves to figure into discussions of the communicative pressures active in numeral systems, along with other widely discussed properties, as it provides a window into how culture-specific factors and general principles of efficient communication interact in the domain of numerical cognition.

%% file: methods.tex
\section{Materials and Methods}
\label{methods}

\subsection{Data}

Data for cross-linguistic numeral systems were taken from two sources.\footnote{The dataset of \citet{eugene_chan_2019_3475912} has broader cross-linguistic coverage than those we use in this paper, but does not contain all numerals in the range 1--99.} The first of these is UniNum \citep{ritchie2019unified}, a collection of numerals between 0 and 100 billion, provided by Google and language experts. This data set was curated for text-to-speech purposes, and contains data from 186 speech varieties. Representations of numeral word forms are orthographic, not phonemic. In some cases, this creates complications for the metrics we use; we exclude varieties with logographic writing systems, and convert transcriptions to a normalized decomposed format (Unicode NFD). In general, meaningful information can be extracted from orthographic representations even for languages that lack systematic orthographies. 
The second of these is the South Asian Numerals Database (SAND; \citealt{mamta_kumari_2023_10033151}; \citealt{sand}), a collection of numerals ranging from 1 to 10000000 from 131 speech varieties of South Asia, to which we added data from 8 additional Indo-Aryan languages \citep{GusainBagri2000,AhmedShina2019,BashirConnersSiraiki2019,GnanadesikanDhivehi2017,BaartKohistani1999,PetersenKalasha2015,GriersonTorwali1929,TrailLamani1970,LorimerDomaki1939}. Representations of numeral wordforms in SAND are phonemic, transcribed in the International Phonetic Alphabet (IPA). For additional languages, the systematic phonemic (but not necessarily IPA) representations provided in sources were used. Only data for the numerals 1--99 from both databases were used. Non-word characters like spaces and hyphens were removed from strings representing forms. 

For use in statistical analyses, elevation data from the World Climate data set \citep{hijmans2005very} were loaded using the function {\tt getData()} from the R package {\tt raster} (v.\ 3.0-7; \citealt{raster}). 
Linguistic areas (below the level of macroareas such as Eurasia, etc., that are already coded in UniNum's metadata) were extracted for each language from the AUTOTYP \citep{AUTOTYP} register. 

We extract numeral term frequencies for Indo-Aryan languages from the following corpora available through the Lancaster Corpus Query Processor (\citealt{hardie2012cqpweb}; \url{https://cqpweb.lancs.ac.uk/}). 
\begin{itemize}
    \item Emille-CIIL Assamese Corpus (\citealt{baker2002emille}; \url{https://cqpweb.lancs.ac.uk/asm_v2/})
    \item Emille Spoken Hindi: (\citealt{baker2002emille}; \url{https://cqpweb.lancs.ac.uk/emillehinsp/})
    \item Emille Hindi Webnews: (\citealt{baker2002emille}; \url{https://cqpweb.lancs.ac.uk/emillehinwbnws/})
    \item Emille Spoken Punjabi: (\citealt{baker2002emille}; \url{https://cqpweb.lancs.ac.uk/emillepunsp/})
    \item Emille Western Punjabi (Shahmukhi): (\citealt{baker2002emille}; \url{https://cqpweb.lancs.ac.uk/emillewpun/})
    \item Lancaster Urdu Web Corpus: (\citealt{jehangir2024design}; \url{https://cqpweb.lancs.ac.uk/luwc202108/})
\end{itemize}
We queried both numeric and orthographic representations of numeral terms, pooling these together. Because the corpora used are not lemmatized, it was not always possible to distinguish orthographic numeral terms from their homographs, or uses of numeral terms outside of the context of enumerating elements (e.g., Hindi {\it ek} `one' can be used to mark indefinite nouns). 
For comparability across corpora, we normalized frequencies within corpora by adding one to each frequency, dividing by the sum of all numeral frequencies in the corpus, and log transforming the resulting smoothed relative frequencies.

\subsection{Metrics}

All metrics were computed in Python (v.\ 3.8.10), relying primarily on the packages {\tt NumPy} (v.\ 1.24.4; \citealt{2020NumPy-Array}) and {\tt SciPy} (v.\ 1.10.1; \citealt{2020SciPy-NMeth}). 

\subsubsection{Minimum Description Length (MDL)}

In line with the information theoretic principle of minimum description length (MDL; \citealt{rissanen1983universal}), we seek the shortest set of combinable elements needed to generate all 99 numeral words of interest. 
We use a simplified version of models employed for morpheme and word segmentation \citep{goldsmith2001unsupervised,creutz2007unsupervised,goldwater2009bayesian} using expectation maximization (EM; \citealt{dempster1977maximum}) in order to segment each numeral form in each language into recurrent subword units such that the set of segmented units is minimized. 

For each language, we randomly initialize the segmentation of each numeral word form $w$. 
We do not allow numerals 1--9 to be segmented (as they tend to be simplex forms except in the case of systems with bases lower than 10, which are underrepresented in the data sample); we allow 11--19 and multiples of 10 to either be unsegmented (as in less transparent, more fusional forms like  English {\it twelve}, Spanish {\it cuarenta} `40') or have a single segmentation index $i \in \{2,...,|w|-1\}$, where $|w|$ represents word form length, and $i$ the index marking the start of the second subword unit (as in more transparent forms like German {\it acht-zehn} '18', Japanese {\it ni-jū} '20'). 
All other numerals are forced to have a single segmentation index (as above) or two segmentation indices $(i,j) \in \{2,...,|w|-1\}:i \neq j$. 
Segmented units are placed in a cache $\mathbb{W}$. 

For each EM iteration $t$, we randomly consider each word $w$ representing the numerals 10--99, removing the currently segmented units $\sigma(w)_z^{(t-1)}$ from the cache $\mathbb{W}$. 
We then choose a new segmentation $z^{(t)}$ (either no segmentation, one index, or two, depending on the conditions outlined above) 
equal to 
$\text{arg min}_{z^{(t)}} \sum_{s \in \mathbb{W} \cup \sigma(w)_{z^{(t)}}}|s|$ (i.e., the segmentation that minimizes the sum of the lengths of all segmented forms in the cache, plus those of the segmentation under consideration). 
We stop when the description length reaches a minimum, or after 1000 iterations. 
Because this version of the EM algorithm converges on local and not global optima, we run this procedure 10 times per language, storing the shortest description length across runs. 
We note that this procedure does not account for allophonic processes, and may infer different subword elements that are underlyingly the same according to standard phonological analyses due to allophony, or orthographic variation. 

This procedure yields a single MDL value for each language, representing the complexity of the language's numeral system as a whole.

\subsubsection{N-gram surprisal in context}

We additionally operationalize the complexity of numeral systems using n-gram (specifically segmental/grapheme trigram) continuation surprisal, representing the unpredictability of phoneme or grapheme in context, i.e., given the two previous phonemes/graphemes \citep{piantadosi2012communicative,dautriche2017words}, averaged within and across words. 
Numeral systems containing more recurrent, predictable elements are expected to exhibit lower surprisal. 
We compute the mean n-gram surprisal of individual numeral words conditioned on all other words in the system (training vs. held-out). 

For each language, we compute the held-out trigram surprisal (in nats) of each word form as follows. First, we prepend two instances of the beginning of string token {\tt $<$BOS$>$} and postpend one instance of the end of string token {\tt $<$EOS$>$} to each word form. We tabulate frequencies $c()$ of all trigram and bigram segment/grapheme sequences in the training data. For the word form under consideration, we compute the continuation surprisal of each segment/grapheme at index $i \in \{3,...,|w|\}$ ($3$ is the index of the first non-{\tt $<$BOS$>$} character, if two such characters are prepended to the word form)
$$s(w_i|w_{i-2...i-1}) = -\log \frac{c(w_{i-2...i}) + \alpha}{c(w_{i-2...i-1}) + V\alpha}$$
where $\alpha=0.1$ is an additive smoothing constant and $V$ is the number of segment types, and average these values. 

This procedure yields a surprisal value for each numeral in the system, conditioned on the other members of the system. 
These values can be averaged to represent the overall surprisal of the system at the language level. 

Additive smoothing can exhibit high sensitivity to values of the smoothing constant chosen. For this reason, we run sensitivity analyses, computing surprisals for forms in each data set for $\alpha \in \{0.0001,0.001,0.01,0.1,1\}$ and measuring correlations between surprisals under each pair of values of $\alpha$. Pairwise correlations are high (all greater than $0.864$ and most close to $1$), indicating that the main results we report are not artifacts of the smoothing constant chosen.

\subsubsection{Linear discriminative learning: production}


We use a simplified version of the linear mapping approach (proposed in \citealt{baayen2018inflectional} {\it et seq}) to model the production of numeral forms (e.g. {\it twelve}) given an underlying semantic representation (\{{\sc tens} $=1$, {\sc digits} $=2$\}). 
Unlike the original version of this model, numeral forms' meanings are represented by two one-hot vectors (comprising the tens- and digits-place value of a numeral) rather than continuous semantic vectors, which are concatenated together and make up rows of the meaning matrix $\boldsymbol{S}$. 
Numeral forms are represented by a vector containing the counts of the trigrams they contain, which make up rows of the word form matrix $\boldsymbol{W}$. 

Form generation given a semantic representation $\boldsymbol{s}_i$ proceeds as follows. 
We compute the least-squares solutions $\boldsymbol{\hat{\beta}_{sw}}$ and $\boldsymbol{\hat{\beta}_{ws}}$ that solve the equations $\boldsymbol{S}_{-i}\boldsymbol{\hat{\beta}_{sw}} = \boldsymbol{W}_{-i}$ and $\boldsymbol{W}_{-i}\boldsymbol{\hat{\beta}_{ws}} = \boldsymbol{S}_{-i}$, respectively. 

We then compute a vector of weights $\boldsymbol{s}_i^{\top} \boldsymbol{\hat{\beta}_{sw}}$, representing the association strengths of different trigrams (in reality their predicted counts, but in real-valued space) with the semantic representation $\boldsymbol{s}_i$. 
We decode the form via beam search: starting with trigrams beginning with the token {\tt $<$BOS$>$}, we consider the two trigrams with highest association strength, 
and proceed recursively in the same manner (selecting the two trigrams with highest association strength that form valid continuations of the previously selected trigram), stopping when the end-of-sequence token is reached. 
To avoid cycles, we rule out candidate trigrams during beam search if they have been selected more than two times during the previous two continuations. 
For each candidate form $\boldsymbol{\hat{w}}$, we compute the predicted meaning $\boldsymbol{\hat{w}}^{\top} \boldsymbol{\hat{\beta}_{ws}}$, choosing the candidate form that shows highest correlation (Pearson's $r$) with $\boldsymbol{s}_i$. 
We measure the error rate between the predicted form and the true form using the Levenshtein distance between the two forms divided by the length of the longer form.

\subsubsection{Linear discriminative learning: comprehension}

To model the comprehension or discrimination of the meaning of a given form, we train two multinomial logistic classifers on the word form matrix $\boldsymbol{W}_{-i}$ in order to predict the tens and digits label of $\boldsymbol{w}_i$ with maximum probability according to the classifiers. 
We treat classification accuracy as a binary variable valued $1$ if the tens and digits label are correctly predicted and $0$ otherwise.

\subsection{Modeling}

Statistical model fitting and interpretation were carried out in R (v.\ 4.4.3; \citealt{R}) using the packages {\tt lme4} (v.\ 0.25.0; \citealt{lme4}), {\tt mgcv} (v.\ 1.9-1; \citealt{mgcv}), and {\tt marginaleffects} (v.\ 1.1-37; \citealt{marginaleffects}). 

For smooth terms in GAMs, we used the default thin-plate regression splines implemented in the function {\tt s()} as the basis function. 
For language-level random slopes, 
we use smooth terms with the basis function {\tt s(..., bs='re')}. 
Models were fitted using maximum likelihood. 
Model comparison was carried out using the function {\tt anova.gam()}. 
We use the {\tt marginaleffects} functions {\tt plot\underline{\phantom{X}}predictions()} and {\tt plot\underline{\phantom{X}}slopes()} to generate graphics of predictions from predicted models and marginal slopes, respectively.


%% file: acknowledgements.tex
\section*{Acknowledgements}